%% file: matching_arXiv_v1.tex
\documentclass[a4paper,11pt]{article}
\pdfoutput=1 

\usepackage{jheppub} 

\usepackage[T1]{fontenc} 
\usepackage{arydshln}
\usepackage{slashed}
\usepackage{multirow}
\usepackage{braket}
\usepackage{units}

\usepackage{graphicx}
\usepackage{tikz}

\input{preamble.tex}

\newcommand{\be}{\begin{eqnarray}}
\newcommand{\ee}{\end{eqnarray}}

\title{One-loop Matching and Running with Covariant Derivative Expansion}

\author[a]{Brian Henning,}
\author[b]{Xiaochuan Lu,}
\author[c,d,e,f]{and Hitoshi Murayama}

\affiliation[a]{Department of Physics, Yale University, New Haven, Connecticut 06511, USA}
\affiliation[b]{Department of Physics, University of California, Davis, California 95616, USA}
\affiliation[c]{Department of Physics, University of California, Berkeley, California 94720, USA}
\affiliation[d]{Theoretical Physics Group, Lawrence Berkeley National Laboratory, Berkeley, California 94720, USA}
\affiliation[e]{Kavli Institute for the Physics and Mathematics of the Universe (WPI), Todai Institutes for Advanced Study, University of Tokyo, Kashiwa 277-8583, Japan}
\affiliation[f]{Center for Japanese Studies, University of California, Berkeley, CA 94720, USA}

\emailAdd{brian.henning@yale.edu}
\emailAdd{xclu@ucdavis.edu}
\emailAdd{hitoshi@berkeley.edu, hitoshi.murayama@ipmu.jp}

\preprint{IPMU16-0042}

\abstract{We develop tools for performing effective field theory (EFT) calculations in a manifestly gauge-covariant fashion. We clarify how functional methods account for one-loop diagrams resulting from the exchange of both heavy and light fields, as some confusion has recently arisen in the literature. To efficiently evaluate functional traces containing these ``mixed'' one-loop terms, we develop a new covariant derivative expansion (CDE) technique that is capable of evaluating a much wider class of traces than previous methods. The technique is detailed in an appendix, so that it can be read independently from the rest of this work. We review the well-known matching procedure to one-loop order with functional methods. What we add to this story is showing how to isolate one-loop terms coming from diagrams involving only heavy propagators from diagrams with mixed heavy and light propagators. This is done using a non-local effective action, which physically connects to the notion of ``integrating out'' heavy fields. Lastly, we show how to use a CDE to do running analyses in EFTs, \textit{i.e.} to obtain the anomalous dimension matrix. We demonstrate the methodologies by several explicit example calculations.}

\begin{document}
\maketitle
\flushbottom
\newpage

\section{Introduction}\label{sec:Introduction}

\input{sec_Introduction.tex}

\section{Matching by functional methods}\label{sec:MatchingFunctional}

\input{sec_MatchingFunctional.tex}

\section{Direct computation of $c_{i,\text{mixed}}^{(1)}(M)$ and a new CDE technique}\label{sec:MatchingDirect}

\input{sec_MatchingDirect.tex}

\section{Full example for matching: Yukawa theory with a heavy scalar}\label{sec:Yukawa}

\input{sec_Yukawa.tex}

\section{RG running by functional methods}\label{sec:Running}

\input{sec_Running.tex}

\section{Summary of results}\label{sec:Summary}

\input{sec_Summary.tex}

\acknowledgments
We are grateful to Daliang Li, Markus Luty, and Witold Skiba for conversations. We also thank Francesco del Aguila, Zoltan Kunszt, and Jose Santiago for correspondence. BH is supported in part by the U.S. DOE under the contract DE-FG02-92ER-40704. XL is supported by DOE grant DE-SC-000999. HM is supported in part by the U.S. DOE under Contract DE-AC03-76SF00098, in part by the NSF under grant PHY-1316783, in part by the JSPS Grant-in-Aid for Scientific Research (C) (No.~26400241), Scientific Research on Innovative Areas (No.~15H05887), and by WPI, MEXT, Japan.

\appendix

\section{Functional trace evaluation with a CDE}\label{app:Trace}

\input{app_Trace.tex}

\section{Derivation of the resolved functional determinant, Eq.~\eqref{eqn:determinantsresolved}}\label{app:DeriveResolve}
\input{app_DeriveResolve.tex}

\section{Local vs. non-local difference and the Wilsonian effective action}\label{app:Wilsonian}

\input{app_Wilsonian.tex}

\section{Details about the triplet scalar model}\label{app:TripletScalar}

\input{app_TripletScalar.tex}

\bibliography{./bibliography}
\bibliographystyle{JHEP}

\end{document}

%% file: preamble.tex
\def\d{\delta}

\def\g{\gamma}
\def\G{\Gamma}

\def\k{\kappa}
\def\l{\lambda}

\def\m{\mu}
\def\n{\nu}
\def\s{\sigma}

\def\ps{\psi}
\def\ph{\phi}
\def\Ph{\Phi}

\def\x{\xi}

\def\pd{\partial}

\def\scL{\mathcal{L}}

\def\scO{\mathcal{O}}

\makeatletter
\newsavebox\myboxA
\newsavebox\myboxB
\newlength\mylenA
\newcommand*\xoverline[2][0.75]{%
    \sbox{\myboxA}{$\m@th#2$}%
    \setbox\myboxB\null
    \ht\myboxB=\ht\myboxA%
    \dp\myboxB=\dp\myboxA%
    \wd\myboxB=#1\wd\myboxA
    \sbox\myboxB{$\m@th\overline{\copy\myboxB}$}
    \setlength\mylenA{\the\wd\myboxA}
    \addtolength\mylenA{-\the\wd\myboxB}%
    \ifdim\wd\myboxB<\wd\myboxA%
       \rlap{\hskip 0.5\mylenA\usebox\myboxB}{\usebox\myboxA}%
    \else
        \hskip -0.5\mylenA\rlap{\usebox\myboxA}{\hskip 0.5\mylenA\usebox\myboxB}%
    \fi}
\makeatother

\def\psb{\xoverline{\ps}}

\def\Gb{\xoverline{\Gamma}}

\def\pds{\slashed{\partial}}

\def\Tr{\text{Tr}}
\def\tr{\text{tr}}

\usetikzlibrary{positioning,decorations.pathreplacing,decorations.markings,shapes}
\usetikzlibrary{calc}
\usetikzlibrary{arrows,shapes,backgrounds}
\usetikzlibrary{decorations.pathmorphing}
\usetikzlibrary{decorations.markings}

\tikzset{ball/.style = {circle, draw, text centered, anchor=north, inner sep=0}}
\tikzset{ballc/.style = {circle, draw, text centered, inner sep=0}}


%% file: sec_Introduction.tex
The purpose of this paper primarily concerns the process of matching an ultraviolet (UV) theory onto an effective field theory (EFT). We review the proper matching procedure and present techniques for computing the Wilson coefficients up to one-loop level in a gauge-covariant fashion with functional methods, \textit{i.e.} using a covariant derivative expansion (CDE). As a byproduct, we also show how to use CDE to do running analyses in an EFT, namely to compute the anomalous dimension matrix of the Wilson coefficients.

\subsection{Review of matching to one-loop order}\label{subsec:MatchingReview}

Consider a theory, which we will call the UV theory, containing a heavy field $\Phi$ of mass $M$, and light fields denoted collectively by $\phi$. At low-energy scales, $E<M$, the only physically accessible degrees of freedom are those of the light fields $\phi$. So instead of working with the UV theory of both $\Phi$ and $\phi$, one can equivalently work with an EFT containing only $\phi$ by averaging over the short-distance physics.

Using Lagrangian language, one can split the UV Lagrangian into two parts: $\mathcal{L}_\phi$ that consists of only the light fields $\phi$, and $\mathcal{L}_\Phi$ that involves the heavy field $\Phi$:
\begin{equation}
{{\cal L}_{{\rm{UV}}}}\left( {\phi ,\Phi } \right) = {{\cal L}_\phi }\left( \phi  \right) + {{\cal L}_\Phi }\left( \phi, \Phi \right) . \label{eqn:LUV}
\end{equation}
At low energies, the EFT of $\phi$ is obtained essentially by replacing $\mathcal{L}_\Phi$ by a set of local effective operators $\mathcal{O}_i(\phi)$,
\begin{equation}
{{\cal L}_{{\rm{EFT}}}}\left( \phi  \right) = {{\cal L}_\phi }\left( \phi  \right) + \sum\limits_i {{c_i}\left(\mu\right){{\cal O}_i}\left( \phi  \right)} , \label{eqn:LEFT}
\end{equation}
with the Wilson coefficients $c_i\left(\mu\right)$ to be determined. The EFT is required to reproduce the physics predictions of the UV theory at low energy scales $E<M$. As physical observables are built from correlation functions, this is achieved by requiring that the one-light-particle irreducible (1LPI) diagrams computed from $\mathcal{L}_\text{EFT}$ and those computed from \(\scL_{\text{UV}}\) agree at the renormalization group (RG) scale $\mu=M$. Equivalently, in the functional approach we require that the 1LPI effective action \(\G_{\text{L}}[\ph]\)---the generating functional for 1LPI correlation functions---of each theory coincide at \(\m = M\). 
This is the well-known ``matching'' criterion, \textit{e.g.}~\cite{Georgi:1991ch,Georgi:1992xg}.

For weakly coupled theories, the matching of 1LPI diagrams between the EFT and the UV theory is done order-by-order in perturbation theory. For many practical applications, it is sufficient to perform the matching up to one-loop order. Restricting our attention to this case, the Wilson coefficients at $\mu=M$ can in general be decomposed into three parts:
\begin{equation}
{c_i}\left( M \right) = c_i^{\left( 0 \right)}\left( M \right) + c_{i,{\rm{heavy}}}^{\left( 1 \right)}\left( M \right) + c_{i,{\rm{mixed}}}^{\left( 1 \right)}\left( M \right) . \label{eqn:cMparts}
\end{equation}
Here the superscript denotes the loop order of the corresponding 1LPI diagrams in the UV theory. Specifically, $c_i^{(0)}(M)$ is tree-level, while $c_{i,\text{heavy}}^{(1)}$ and $c_{i,\text{mixed}}^{(1)}$ are one-loop size. In general there are two types of one-loop 1LPI diagrams in the UV theory: (1) those with only heavy field as propagators, and (2) those with mixed propagators of both light and heavy fields. $c_{i,\text{heavy}}^{(1)}$ and $c_{i,\text{mixed}}^{(1)}$ correspond to these two types of contributions, respectively.

Let us summarize the origin of the components in Eq.~\eqref{eqn:cMparts} and the relative ease in computing them:
\begin{description}
\item[$c_i^{(0)}(M)$:] Tree-level terms arise when the UV Lagrangian has a term linear in the heavy field $\Phi$, $\mathcal{L}_\text{UV} \supset \Phi B(\phi)$, with $B(\phi)$ built from light fields only. $c_i^{(0)}(M)$ is simply obtained by first solving the equation of motion of $\Phi$,
    \begin{equation}
    {\left. {\frac{{\delta {S_{{\rm{UV}}}}\left[ {\phi ,\Phi } \right]}}{{\delta \Phi }}} \right|_{{\Phi _c}\left[ \phi  \right]}} = 0 , \label{eqn:PhiEOM}
    \end{equation}
    and then inserting the solution $\Phi=\Phi_c\left[\phi\right]$ back into the UV Lagrangian in Eq.~\eqref{eqn:LUV}
    \begin{equation}
    \sum\limits_i {c_i^{\left( 0 \right)}\left( M \right){{\cal O}_i}\left( \phi  \right)}  = {{\cal L}_\Phi }\left( \phi, {\Phi _c}\left[ \phi  \right] \right) . \label{eqn:c0}
    \end{equation}
\item[$c_{i,\text{heavy}}^{(1)}(M)$:] One-loop contributions from 1LPI diagrams with only heavy fields as the propagators can always be computed by evaluating a functional determinant of the elliptic operator $D^2+M^2+U(x)$
    \begin{align}
    \int {{d^4}x\sum\limits_i {c_{i,{\rm{heavy}}}^{\left( 1 \right)}\left( M \right){{\cal O}_i}\left( \phi  \right)} }  \propto \log \det \left( { - \frac{{{\delta ^2}{S_{{\rm{UV}}}}}}{{\delta {\Phi ^2}}}} \right) \propto \log \det \left[ {{D^2} + {M^2} + U\left( x \right)} \right] , \label{eqn:c1heavy}
    \end{align}
    where $D_\mu$ is the covariant derivative, $U(x)$ depends only on the light fields $\phi$, and $S_\text{UV}=\int d^4x \mathcal{L}_\text{UV}$ is the action of the UV theory. In~\cite{Henning:2014wua}, building on techniques introduced in~\cite{Gaillard:1985,Cheyette:1987}, we showed how to evaluate such functional determinants in a manifestly gauge-covariant fashion, termed the covariant derivative expansion (CDE). Because of the simplicity of expressions gained through using the CDE, it was clear that universal results could be obtained in evaluating the functional determinant. The universal results of~\cite{Henning:2014wua}, and their generalization in~\cite{Drozd:2015rsp}, tremendously ease the computation of $c_{i,\text{heavy}}^{(1)}(M)$. These results have been successfully employed for a variety of matching analyses in the Standard Model EFT (SM EFT), \textit{e.g.}~\cite{Henning:2014wua,Drozd:2015rsp,Henning:2014gca,Drozd:2015kva,Chiang:2015ura,Huo:2015exa,Huo:2015nka}.
\item[$c_{i,\text{mixed}}^{(1)}(M)$:] One-loop contributions from 1LPI diagrams with mixed heavy and light propagators only arise when there is a non-zero tree-level piece, \textit{i.e.} if a certain \(c_j^{(0)}(M) \ne 0\). The calculation of these terms is generically fairly involved. A good portion of the present work is devoted to showing how to do this calculation in a relatively easier way, using functionals and evaluating them in a manifestly gauge-covariant fashion.
\end{description}

It was recently claimed~\cite{delAguila:2016zcb} that the CDE method cannot account for $c_{i,\text{mixed}}^{(1)}(M)$, due to the limitations of the nature of functional methods. In this paper, we show that this claim is not true and demonstrate how to compute $c_{i,\text{mixed}}^{(1)}(M)$ by functional methods. The functional traces encountered in this calculation are generally not of the form in Eq.~\eqref{eqn:c1heavy}, so the CDE results of~\cite{Gaillard:1985,Cheyette:1987,Henning:2014wua,Drozd:2015rsp} do not immediately apply. However, we present a new way of performing a covariant derivative expansion to evaluate a wide class of functional traces, which provides a way to use a CDE to directly extract $c_{i,\text{mixed}}^{(1)}(M)$.
Although we currently have no universal formula for $c_{i,\text{mixed}}^{(1)}(M)$ analogous to the formulas for $c_{i,\text{heavy}}^{(1)}(M)$ of~\cite{Henning:2014wua,Drozd:2015rsp}, the method we present is as systematic as Feynman diagram calculations, with the additional advantage of being manifestly gauge-covariant.

Even with the new CDE technique, the computation of $c_{i,\text{mixed}}^{(1)}(M)$ is generically fairly involved. Therefore, before diving into the discussion of its computation, we would like to review the practical relevance and the physical meaning of $c_{i,\text{mixed}}^{(1)}(M)$, so that we know when there is a true need for this effort.

\subsection{Practical relevance of $c_{i,\text{mixed}}^{(1)}(M)$}\label{subsec:PracticalRelevance}

Although $c_{i,\text{mixed}}^{(1)}(M)$ is a conceptually important piece in the matching at one-loop order, it is often not practically relevant. To see this, it is useful to consider the Wilson coefficients \(c_i(\m)\) at a scale \(\mu < M\) where physical observations are made. This requires RG evolving the Wilson coefficients from \(M\) to \(\mu\) using the anomalous dimension matrix \(\g_{ij}\), which is a function of marginal couplings in $\mathcal{L}_\phi(\phi)$ (\textit{e.g.} gauge, Yukawa, and the Higgs quartic couplings in the case of the Standard Model(SM)):
\begin{align}
{c_i}\left( \mu \right) &= {c_i}\left( M \right) + \frac{1}{{{{\left( {4\pi } \right)}^2}}}\sum\limits_j {{\gamma _{ij}}} {c_j}\left( M \right)\log \frac{\mu}{M} \nonumber \\
 &= c_i^{\left( 0 \right)}\left( M \right) + c_{i,{\rm{heavy}}}^{\left( 1 \right)}\left( M \right) + c_{i,{\rm{mixed}}}^{\left( 1 \right)}\left( M \right) + \frac{1}{{{{\left( {4\pi } \right)}^2}}}\sum\limits_j {{\gamma _{ij}}} c_j^{\left( 0 \right)}\left( M \right)\log \frac{\mu}{M} . \label{eqn:cvparts}
\end{align}
With Eq.~\eqref{eqn:cvparts} in mind, there are four cases in general to consider:
\begin{enumerate}
  \item There are no tree-level effects, \textit{i.e.} all $c_i^{(0)}(M)=0$. In this case, $c_{i,\text{mixed}}^{(1)}(M)$ vanishes. The only term to survive in Eq.~\eqref{eqn:cvparts} is $c_{i,\text{heavy}}^{(1)}(M)$, which is very straightforward to compute through Eq.~\eqref{eqn:c1heavy} and the universal results of evaluating the elliptic operator presented in~\cite{Henning:2014wua,Drozd:2015rsp}.
  \item There is a certain tree-level effect $c_j^{(0)}(M)\ne 0$ that results in a non-vanishing $c_{i,\text{mixed}}^{(1)}(M)$, but the tree-level piece of $c_i(M)$ also exists, \textit{i.e.} $c_i^{(0)}(M)\ne 0$. In this case, obviously the dominant contribution to Eq.~\eqref{eqn:cvparts} is $c_i^{(0)}(M)$, which can be easily computed through Eq.~\eqref{eqn:c0}.
  \item A certain tree-level effect $c_j^{(0)}(M)\ne 0$ yields a non-vanishing $c_{i,\text{mixed}}^{(1)}(M)$ while the tree-level piece $c_i^{(0)}(M)=0$, but the matching scale is much higher than the observation scale, \textit{i.e.} $M \gg \mu$. In this case, the RG running term dominates Eq.~\eqref{eqn:cvparts}. As the anomalous dimension matrix $\gamma_{ij}$ is inherent to the EFT and has nothing to do with the UV theory we start with, it can be computed separately once and for all and applied to any UV matching/running analysis. For example, in the case of the SM EFT, $\gamma_{ij}$ for dimension-six operators is known~\cite{Jenkins:2013zja,Jenkins:2013wua,Alonso:2013hga,Grojean:2013kd,Elias-Miro:2013gya,Elias-Miro:2013mua,Elias-Miro:2013eta}. Once $\gamma_{ij}$ is given, the RG running term in Eq.~\eqref{eqn:cvparts} is straightforward to compute.
  \item A certain tree-level effect $c_j^{(0)}(M)\ne 0$ yields a non-vanishing $c_{i,\text{mixed}}^{(1)}(M)$, the tree-level piece $c_i^{(0)}(M)=0$, and the matching scale is not much higher than the observation scale. It is only in this case that $c_{i,\text{mixed}}^{(1)}(M)$ could dominate Eq.~\eqref{eqn:cvparts}, and hence be of practical importance.
\end{enumerate}

In listing the four cases above, we have made the implicit assumption that experiments are only sensitive to the leading order term in Eq.~\eqref{eqn:cvparts}. We have in mind the SM, where present and near future precision Higgs and electroweak measurements will only be able to resolve the leading contribution to Wilson coefficients in the SM EFT. Such precision measurements have sensitivity up to the per mille level---enough to probe one-loop effects. However, existing measurements already constrain deviations from SM processes to be relatively small; in other words, we already know that the \(c_i(\mu=v)\) in the SM EFT cannot lead to \(\scO(1)\) deviations. Therefore, assuming some \(c_i(v)\) is non-zero due to new physics, it is unlikely that present and next-generation experiments will have the ability to resolve subleading terms in \(c_i(v)\).

\subsection{Physical meaning of $c_{i,\text{mixed}}^{(1)}(M)$}\label{subsec:PhysicalMeaning}

Apart from the practical relevance, it is also useful to examine the physical meaning of $c_{i,\text{mixed}}^{(1)}(M)$ more closely, to better understand on what we will spend so much effort later in this paper. The short answer is: $c_{i,\text{mixed}}^{(1)}(M)$ does not have a physical meaning on its own, in the sense that it is not renormalization scheme independent by itself. As we elaborate below, changing renormalization scheme amounts to a reallocation between $c_{i,\text{mixed}}^{(1)}(M)$ and the one-loop order part of the ``mapping'' of Wilson coefficients onto physical observables.\footnote{We adopt the language of~\cite{Henning:2014wua} where ``mapping'' refers to determining physical observables as functions of the EFT parameters.} The discussion in this subsection might be a bit subtle and digressive; readers mainly interested in the computational technique are welcome to skip ahead.

From general renormalized perturbation theory, we know that any renormalized coupling $c_i$ is understood to be accompanied by a counterterm $\delta c_i$. While computing a physical observable $T_i$ at one-loop order, we choose the value of $\delta c_i$ to cancel the divergences, and hence arrive at a well-defined expression of $T_i$ in terms of $c_i$. However, as $\delta c_i$ is only required to cancel the divergent part of the loop diagram, its finite part is essentially free to choose. Different choices of this finite part are referred to as different ``renormalization schemes''. Choosing different renormalization schemes can be understood as choosing different definitions of the renormalized coupling $c_i$, as this is just a re-splitting between $c_i$ and $\delta c_i$:
\begin{equation}
c_i + \delta c_i = c'_i + \delta c'_i . \label{eqn:csplit}
\end{equation}
As $\delta c'_i$ can only differ from $\delta c_i$ by a finite piece, this renormalization scheme change amounts to a change of $c_i$ by a one-loop finite piece
\begin{equation}
c'_i = c_i + \left( \delta c_i - \delta c'_i \right) . \label{eqn:cshift}
\end{equation}
While the coupling $c_i$ itself is subject to a renormalization scheme choice, the physical observable $T_i$ is not, as the relation between $T_i$ and $c'_i$ will be changed consistently according to the new choice $\delta c'_i$.

Because of its computational convenience, nowadays the most widely used renormalization scheme is the $\overline{\text{MS}}$ scheme under dimensional regularization. In this scheme, we make a somewhat universal choice for all $\delta c_i$, such that all the $1/\epsilon -\gamma +\log(4\pi)$ pieces in the loop expression are cancelled. Namely,
\begin{equation}
\delta {c_i} \propto {a_i}\frac{1}{{{{\left( {4\pi } \right)}^2}}}\left( {\frac{1}{\epsilon} - \gamma  + \log(4\pi) } \right) , \label{eqn:deltac1}
\end{equation}
where $a_i$ are certain appropriate coefficients. However, it is important to remember that one has the freedom to use a different renormalization scheme, such as
\begin{equation}
\delta {c'_i} \propto {a_i}\frac{1}{{{{\left( {4\pi } \right)}^2}}}\left( {\frac{1}{\epsilon} - \gamma  + \log(4\pi) + b_i } \right) , \label{eqn:deltac2}
\end{equation}
where $b_i$ can be different for each individual $c_i$.

To see what the above discussion implies for $c_{i,\text{mixed}}^{(1)}(M)$, first note that $c_{i,\text{mixed}}^{(1)}(M)$ is a one-loop sized piece in $c_i(M)$ in Eq.~\eqref{eqn:cMparts}. Therefore, one can change it at will by choosing a different renormalization scheme as going from Eq.~\eqref{eqn:deltac1} to Eq.~\eqref{eqn:deltac2}. One can even make $c_{i,\text{mixed}}^{(1)}(M)$ zero by choosing the appropriate values of $b_i$ in Eq.~\eqref{eqn:deltac2}. So we see that $c_{i,\text{mixed}}^{(1)}(M)$ itself does not have a physical meaning. On the other hand, physical observables will not be affected by the redefinition of $c_i$ through Eq.~\eqref{eqn:cshift}. For example, suppose that we map our Wilson coefficients, up to one-loop order, onto a low energy physical observable $T_i$ that directly corresponds to the effective operator $\mathcal{O}_i$. The generic expression of $T_i$ is
\begin{equation}
{T_i} = {c_i}\left( v \right) + \frac{1}{{{{\left( {4\pi } \right)}^2}}}\sum\limits_j {{\lambda _{ij}}{c_j}\left( v \right)} , \label{eqn:Ti}
\end{equation}
where the second piece is the one-loop order calculation of $T_i$ in terms of $c_i(v)$ (``mapping''), with $\lambda_{ij}$ denoting certain functions of the low energy couplings contained in $\mathcal{L}_\phi(\phi)$. If $c_{i,\text{mixed}}^{(1)}(M) \ne 0$, this implies a certain $c_j^{(0)}(M)\ne 0$ (with either $j=i$ or $j\ne i$). If this happens, however, the corresponding $\lambda_{ij}$ in Eq.~\eqref{eqn:Ti} must also be nonzero; namely, the second piece in Eq.~\eqref{eqn:Ti} has a one-to-one correspondence with the contributions to $c_{i,\text{mixed}}^{(1)}(M) \ne 0$. Now it is easy to see that a change in renormalization scheme will result in a change of \textit{both} $c_{i,\text{mixed}}^{(1)}(M) \ne 0$ \textit{and} the one-loop mapping term in a consistent way such that $T_i$ is unchanged.

The important conclusion is that it is not physically meaningful to include $c_{i,\text{mixed}}^{(1)}(M)$ in matching \textit{unless} we also include the one-loop piece in mapping.\footnote{The only exception to this is that sometimes a certain combination of $c_{i,\text{mixed}}^{(1)}$ might be protected by a symmetry in the theory, and the scheme dependence cancels out in this combination. In this case, it is physically meaningful to compute this combination of $c_{i,\text{mixed}}^{(1)}$.} This is the reason why we dropped $c_{i,\text{mixed}}^{(1)}(M)$ in~\cite{Henning:2014wua}: it was technically consistent as the mapping analysis was done at leading order.\footnote{One may wonder if the whole scheme dependence issue we discussed in this section also exists for the other one-loop piece $c_{i,\text{heavy}}^{(1)}$. The answer is yes. Generically, $c_{i,\text{heavy}}^{(1)}$ is not scheme independent either. But in a large class of models without tree-level Wilson coefficients, \textit{i.e.} all $c_i^{(0)}=0$, $c_{i,\text{heavy}}^{(1)}$ is scheme independent since it is the leading order piece, and the counterterm $\delta c_i$ starts from two-loop order. This makes it useful to provide a tool for computing $c_{i,\text{heavy}}^{(1)}$ alone.} Nevertheless, as experimental sensitivity increases, next-to-leading order mapping analyses may be important (\textit{e.g.}~\cite{Hartmann:2015oia,Grober:2015cwa,Ghezzi:2015vva,Chen:2013kfa,Dumont:2013wma,Zhang:2014rja,Zhang:2016omx,Bylund:2016phk,Mimasu:2015nqa,Gauld:2015lmb}), which makes it necessary to include the corresponding piece $c_{i,\text{mixed}}^{(1)}(M)$ in the matching step.\footnote{To be very explicit, we work under the reasonable assumption that next-to-leading order mapping analyses are done in the \(\overline{\text{MS}}\) scheme. In order to consistently use such results, the matching process needs to use the same scheme. Hence, we adopt \(\overline{\text{MS}}\) scheme throughout this work.} In this paper, we will show how to compute this piece with a CDE technique.

\subsection{Outline and summary of results}
For whatever reason, functional methods seem less commonly used in the EFT community in comparison to diagrammatic calculations. Moreover, while techniques for functional evaluation exist in the literature, they (unfortunately) are not standard material. The basic hurdle is sort of obvious: many textbooks explain how to evaluate functionals with constant field configurations (usually to obtain the Coleman-Weinberg effective potential), but do not cover the case of arbitrary field configurations. This latter situation is necessary if one wishes to compute, for example, Wilson coefficients of operators involving derivatives.

With this in mind, we provide many computational details in the spirit of pedagogy. While we hope this approach is helpful, it has contributed to the considerable length of this paper. Because of this length, here we provide an extended outline of the main ideas and results of this paper. We note that a ``short path'' through this paper, which hits the main and new ideas, is to read this introduction, sections~\ref{subsec:MatchingTraditional},~\ref{subsec:DirectGeneral}, and~\ref{subsec:RunningGeneral}, and appendix~\ref{app:Trace}.

As evaluation of functionals---in particular, functional determinants---pervades this work, let us first address this topic. A major contribution of the present work is showing how to evaluate a wide class of functional traces in a manifestly gauge-covariant fashion with a covariant derivative expansion. There seems to be some confusion in the literature as to what exactly is meant by a CDE, which we try to clear up in Sec.~\ref{subsec:CDEReview}. Essentially, a ``covariant derivative expansion'' is exactly what the name implies: if we can expand a function of the covariant derivative in a power series of \(D\) without breaking \(D_{\m}\) into its individual components, then this is a CDE by definition. In~\cite{Gaillard:1985}, a specific \textit{transformation} was introduced that \textit{enabled} functional determinants of the elliptic operator \(D^2 + m^2 + U(x)\) to be evaluated in a covariant derivative expansion. The \textit{two} steps of the transformation plus the expansion has gained the \textit{single} name ``the CDE'' in the literature. In many cases, the transformation introduced in~\cite{Gaillard:1985} \textit{is not needed} to develop a CDE. To avoid confusion, we will refer to covariant derivative expansions with the article ``a'' and not ``the''.

The ability to use a covariant derivative expansion to evaluate functionals tremendously improves calculations. In Appendix~\ref{app:Trace} we develop a method to evaluate arbitrary functional traces with a CDE. We separate this result into an appendix so that it stands alone and can be read separately from the rest of this work. The manipulations and techniques explained in this appendix are used over and over throughout this paper, and we encourage the interested reader to study it carefully.

Let us move onto the physics ideas contained in this work concerning EFT matching and running analyses. Section~\ref{sec:MatchingFunctional} begins with a basic review of functional methods. The material here is standard, but we include it to set the tone as well as clarify the equivalence between functional methods and Feynman diagrams, as some spurious claims have recently arisen in the literature.

We next move onto matching, the practical conclusion of which, \textit{e.g.}~\cite{Georgi:1991ch,Georgi:1992xg}, is that the one-light-particle-irreducible (1LPI) effective actions, \(\G_{\text{L}}[\ph]\), of the UV theory and the EFT are equated at the matching scale.\footnote{As a reminder, the 1LPI effective action diagrammatically consists of diagrams with only light fields as external legs that are one-particle-irreducible with respect to cutting \textit{light} field propagators. We use the subscript L to distinguish the 1LPI effective action \(\G_{\text{L}}\) from the 1PI effective action \(\G\).} In equations the matching condition reads,
\begin{equation}
\G_{\text{L,UV}}[\ph](\l,\m = M) - \G_{\text{L,EFT}}[\ph](c,\m = M) = 0,
\label{eq:match_1LPI_intro}
\end{equation}
where \(\l\) and \(c\) represent parameters of the UV theory and EFT, respectively, with the parameters of the EFT adjusted to solve the above equation. This equation is solved order-by-order in perturbation theory, \textit{i.e.} by first equating the tree-level components, \(\G_{\text{L,UV}}^{(0)} = \G_{\text{L,EFT}}^{(0)}\), then the one-loop-level terms, \(\G_{\text{L,UV}}^{(1)} = \G_{\text{L,EFT}}^{(1)}\), \textit{etc}.

The 1LPI correlation functions may be computed either with Feynman diagrams or with functional methods; in this work, we take the latter approach. The ``traditional'' solution through one-loop order is detailed in subsection~\ref{subsec:MatchingTraditional}. Here, ``traditional'' refers to the concept of first computing \(\G_{\text{L,UV}}\) and \(\G_{\text{L,EFT}}\) separately and then equating them as in Eq.~\eqref{eq:match_1LPI_intro}; this terminology is adopted as it is conceptually the most straightforward procedure for matching and is precisely how it is done when matching with Feynman diagrams.

However, the traditional procedure in the functional approach has the conceptual disadvantage that the origin of \(c_{i,\text{heavy}}^{(1)}(M)\) and \(c_{i,\text{light}}^{(1)}(M)\) are entangled, see Eq.~\eqref{eqn:c1Functional}. Clearly, it would be advantageous if we could isolate these terms individually. This is the task taken up in Section~\ref{sec:MatchingDirect}, with the answer given in Eqs.~\eqref{eqn:c1heavydirect} and~\eqref{eqn:c1mixeddirect}. The physical idea to arrive at these results is to understand what it means to integrate out \(\Ph\) from the UV theory using the path integral. To one-loop order this produces a \textit{non-local} effective action \(S_{\text{eff}}[\ph] = S_{\text{UV}}[\ph,\Ph_c[\ph]] + \frac{i}{2}\log\det (-\d^2S_{\text{UV}}[\ph,\Ph]/\d \Ph^2 |_{\Ph_c[\ph]})\), obtained by evaluating \(e^{iS_{\text{eff}}[\ph]} = \int \mathcal{D} \Ph e^{iS_{\text{UV}}[\ph,\Ph]}\) in the saddle-point approximation. If we tried to match this non-local action onto \(\G_{\text{L,EFT}}[\ph]\) by expanding \(S_{\text{eff}}[\ph]\) in a series of local operators we would get spurious results for the Wilson coefficients: \(c_i^{(0)}(M)\) and \(c_{i,\text{heavy}}^{(1)}(M)\) would be correctly identified but \(c_{i,\text{mixed}}^{(1)}(M)\) would be missed. This is not surprising: \(S_{\text{eff}}[\ph]\) is \textit{not} the same functional as \(\G_{\text{L,UV}}[\ph]\).

The crucial idea derived in Sec.~\ref{sec:MatchingDirect} is: \textit{we can use the non-local} \(S_{\text{eff}}[\ph]\) \textit{to compute} \(\G_{\text{L,UV}}[\ph]\). To one-loop order, it is clear that the UV 1LPI effective action contains \(S_{\text{eff}}[\ph]\) as well as the one-loop terms built from the tree-level terms in \(S_{\text{eff}}[\ph]\):
\begin{equation}
\G_{\text{L,UV}}[\ph] = S_{\text{eff}}[\ph] + \frac{i}{2} \log \det \left(-\frac{\d^2S_{\text{UV}}[\ph,\Ph_c[\ph]]}{\d\ph^2}\right).
\end{equation}
The  \(c_{i,\text{mixed}}^{(1)}(M)\) ``missing'' from \(S_{\text{eff}}[\ph]\) obviously must originate from the second term in the above equation.

Using the non-local action to isolate \(c_{i,\text{mixed}}^{(1)}(M)\) in the matching not only provides a better conceptual understanding, but also a route to an improved method for directly computing \(c_{i,\text{mixed}}^{(1)}(M)\). In particular, it provides a clear way to perform matching step in Eq.~\eqref{eq:match_1LPI_intro} \textit{before} computing both functionals individually. This computational technique is explained in the rest of Sec.~\ref{sec:MatchingDirect}. We highlight this technique by example in two different theories: the first is a simple toy theory of two scalars, the second is the more phenomenologically relevant (and more complicated) case of a heavy electroweak triplet scalar added to the SM.

In Sec.~\ref{sec:Yukawa} we demonstrate all aspects of matching in a toy Yukawa model with a heavy scalar. Specifically, we obtain the one-loop Wilson coefficients using both the ``traditional'' procedure of Sec.~\ref{subsec:MatchingTraditional} as well as the ``direct'' procedure of Sec.~\ref{sec:MatchingDirect}.

Section~\ref{sec:Running} concerns the process of renormalization group running in the EFT. We show how functional methods easily allow us to derive the RG equation for the Wilson coefficients. Since the relevant functional can be evaluated with a CDE, this provides an improved computational technique for obtaining the anomalous dimension matrix of an EFT.

We conclude in Sec.~\ref{sec:Summary}. Some detailed computations are relegated to appendices.

\begin{center}
\textbf{Note added}
\end{center}
During the preparation of this manuscript, the work~\cite{delAguila:2016zcb} was updated to roll-back its criticisms of functional methods. We wish to say that, except for a small confusion on functional methods, the physical ideas of matching are very well explained in~\cite{delAguila:2016zcb} and all calculations are completely correct.

The work~\cite{Boggia:2016asg} also appeared, which partially addresses the use of functional methods in matching analyses. While there is some conceptual overlap in~\cite{Boggia:2016asg} with the present work, the main purposes and computational techniques are quite different.

%% file: sec_MatchingFunctional.tex
The purpose of this section is to explain how to perform a matching analysis using functional methods, with an emphasis on computing the piece $c_{i,\text{mixed}}^{(1)}(M)$. For this purpose, several preparations are needed. We first review some basics about functional methods in Section~\ref{subsec:FunctionalReview}, to unambigously clarify that functional methods and Feynman diagrams are equivalent. We then discuss in Section~\ref{subsec:CDEReview} what the essence of a ``covariant derivative expansion'' is and what role it plays in evaluating functionals. With these preparations, we explain in Section~\ref{subsec:MatchingTraditional} how to compute $c_i^{(0)}(M)$, $c_{i,\text{heavy}}^{(1)}(M)$, and $c_{i,\text{mixed}}^{(1)}(M)$ using functional methods. The preparation sections~\ref{subsec:FunctionalReview} and~\ref{subsec:CDEReview} are basic reviews of the relevant subjects---readers familiar with them may wish to skip to Section~\ref{subsec:MatchingTraditional}.

\subsection{Basic review on functional methods}\label{subsec:FunctionalReview}

\subsubsection{Equivalence between functional methods and Feynman diagrams}\label{subsubsec:equivalence}

Feynman diagrams are a widely used technique in perturbative calculations of quantum field theory (QFT). The physical quantities the diagrams calculate for us, such as cross sections, decay rates, \textit{etc.}, can essentially be attributed to the calculation of the quantum correlation functions $\left\langle {\phi \left( {{x_1}} \right) \cdots \phi \left( {{x_n}} \right)} \right\rangle$ (and its LSZ reduction, \textit{i.e.} amplitudes). As is well known, Feynman diagram techniques originate from the functional integral formalism (a.k.a. path integral formalism) of the correlation functions,
\begin{equation}
\left\langle {\phi \left( {{x_1}} \right) \cdots \phi \left( {{x_n}} \right)} \right\rangle  = \frac{{\int {{\cal D}\phi {e^{iS\left[ \phi  \right]}}\phi \left( {{x_1}} \right) \cdots \phi \left( {{x_n}} \right)} }}{{\int {{\cal D}\phi {e^{iS\left[ \phi  \right]}}} }} . \label{eqn:PathIntegral}
\end{equation}
Therefore, one expects that there exists a functional method that is completely equivalent to diagrammatic techniques. Such a functional method is encoded in the well known ``one-particle-irreducible (1PI) effective action'', $\Gamma\left[\phi\right]$, which is a generating functional of all 1PI correlation functions:\footnote{There is a small caveat of this formula at $n=2$, where instead of generating the 1PI correlation function, the right hand side gives the inverse of the full two-point correlation function. This is due to the special role that the tree-level kinetic term plays---there is no 1PI two-point function for the tree-level kinetic term. Apart from the tree-level, Eq.~\eqref{eqn:1PIfunctional} holds for $n=2$ as well, namely that
\begin{equation}
\left\langle {\phi \left( {{x_1}} \right)\phi \left( {{x_2}} \right)} \right\rangle _{{\rm{1PI}}}^{{\rm{loop}}} = i\frac{{{\delta ^2}{\Gamma ^{{\rm{loop}}}}\left[ \phi  \right]}}{{\delta \phi \left( {{x_1}} \right)\delta \phi \left( {{x_2}} \right)}} . \label{eqn:1PIfunctional2}
\end{equation}
}
\begin{equation}
{\left\langle {\phi \left( {{x_1}} \right) \cdots \phi \left( {{x_n}} \right)} \right\rangle _{{\rm{1PI}}}} = i\frac{{{\delta ^n}\Gamma \left[ \phi  \right]}}{{\delta \phi \left( {{x_1}} \right) \cdots \delta \phi \left( {{x_n}} \right)}} . \label{eqn:1PIfunctional}
\end{equation}
Since the full correlation function $\left\langle {\phi \left( {{x_1}} \right) \cdots \phi \left( {{x_n}} \right)} \right\rangle$ is built from 1PI correlation functions $\left\langle {\phi \left( {{x_1}} \right) \cdots \phi \left( {{x_n}} \right)} \right\rangle _{{\rm{1PI}}}$, we see that any quantity calculated with Feynman diagrams is equivalently obtained through computing $\Gamma\left[\phi\right]$ and its functional derivatives. Therefore, diagrammatic methods and functional methods are just two sides of the same coin; the only matter is which one is simpler to use.

\subsubsection{Functional methods to one-loop order}\label{subsubsec:functional1loop}

Computing the 1PI generating functional $\Gamma\left[\phi\right]$ is generically not an easy task. Fortunately, however, it is simple enough up to one-loop order. For a weakly interacting QFT characterized by the Lagrangian
\begin{equation}
S\left[ \phi  \right] = \int {{d^4}x{\cal L}\left( \phi  \right)} ,
\end{equation}
the 1PI effective action to one-loop order is in general given by\footnote{In computing the 1PI effective action there is an intermediate step where external sources \(J_{\ph}\) are linearly coupled to the fields. The path integral is then a functional of the sources, \(Z[J_{\ph}] = e^{i E[J_{\ph}]} = \int D\ph e^{i\int \scL[\ph] + J_{\ph}\ph}\). \(\G[\ph]\) is then obtained as a Legendre transform of \(E[J_{\ph}]\). To one-loop order, the end result \(\G^{(1)}[\ph] \propto \log \det [-\d^2S/\d\ph^2]\) is simple enough because there is no explicit dependence on \(J_{\ph}[\ph]\). At higher loop-order, this is no longer true and more care is needed. See, for example,~\cite{Peskin:1995}}
\begin{equation}
\Gamma \left[ \phi  \right] = S\left[ \phi  \right] + \frac{i}{2}\log \det \left( { - \frac{{{\delta ^2}S\left[ \phi  \right]}}{{\delta {\phi ^2}}}} \right) . \label{eqn:GammaGeneral}
\end{equation}
Here we have used $\phi$ to collectively denote all the fields in the given QFT. We see that the tree-level piece $\Gamma^{(0)}\left[\phi\right]$ is trivially just the action of the theory $S$, while the one-loop piece $\Gamma^{(1)}\left[\phi\right]$ is obtained by evaluating a single functional determinant. For simplicity of the expression we have put $i/2$ as the pre-factor of the functional determinant term above; however, it is understood that this pre-factor needs to be tailored for each individual field in $\phi$, according to its statistics.

In order to demonstrate our argument in Section~\ref{subsubsec:equivalence} concretely, as well as to gain some experience on the typical procedure of using functional methods, let us work out an explicit example of using Eq.~\eqref{eqn:GammaGeneral} to compute a physical quantity. To demonstrate in the simplest context, we consider the standard $\phi^4$ theory,
\begin{equation}
{\cal L}\left( \phi  \right) = \frac{1}{2}\phi \left( { - {\partial ^2} - {m^2}} \right)\phi  - \frac{1}{{4!}}\lambda {\phi ^4} ,
\end{equation}
and compute the physical pole mass $m_p^2$ in terms of the renormalized mass parameter $m^2$ and $\lambda$ defined in $\overline{\text{MS}}$ scheme. 

At tree-level, the pole mass $m_p^2$ is just the renormalized mass parameter $m_p^2=m^2$. We are interested in the one-loop order correction to this. This information is encoded in the one-loop two-point 1PI correlation function $\left\langle {\phi \left( {{x_1}} \right)\phi \left( {{x_2}} \right)} \right\rangle _{{\rm{1PI}}}^{{\text{1-loop}}}$. In a Feynman diagram calculation, this means we should look for all one-loop two-point 1PI diagrams (there is only one for this simple case). In the functional approach, this corresponds to extracting only the $\phi^2$ piece in $\Gamma^{(1)}\left[\phi\right]$, as implied by Eq.~\eqref{eqn:1PIfunctional} (see also the discussion around Eq.~\eqref{eqn:1PIfunctional2}).

To compute $\Gamma^{(1)}\left[\phi\right]$, we first take the second functional derivative of the action $S\left[\phi\right]$:
\begin{equation}
 - \frac{{{\delta ^2}S\left[ \phi  \right]}}{{\delta {\phi ^2}}} = {\partial ^2} + {m^2} + \frac{1}{2}\lambda {\phi ^2} .
\end{equation}
Then according to Eq.~\eqref{eqn:GammaGeneral}, we have
\begin{align}
{\Gamma ^{\left( 1 \right)}}\left[ \phi  \right] &= \frac{i}{2}\log \det \left( {{\partial ^2} + {m^2} + \frac{1}{2}\lambda {\phi ^2}} \right) = \frac{i}{2}{\rm{Tr}}\log \left( {{\partial ^2} + {m^2} + \frac{1}{2}\lambda {\phi ^2}} \right) \nonumber \\
 &\supset \frac{i}{2}{\rm{Tr}}\log \left( {1 - \frac{1}{{ - {\partial ^2} - {m^2}}}\frac{1}{2}\lambda {\phi ^2}} \right) \supset \frac{i}{2}{\rm{Tr}}\left( { - \frac{1}{{ - {\partial ^2} - {m^2}}}\frac{1}{2}\lambda {\phi ^2}} \right) . \label{eqn:logexpand}
\end{align}
In the second line above, we have expanded the logarithm and kept only the $\phi^2$ piece. The last step is to evaluate this functional trace:
\begin{align}
{\Gamma ^{\left( 1 \right)}}\left[ \phi  \right] &=  - \frac{i\lambda}{4}\int d^4x\int \frac{d^4k}{\left( 2\pi  \right)^4} \braket{k| \frac{1}{-\pd^2 - m^2} | x} \braket{x|\ph^2|k} \nonumber \\
 &=  - \frac{{i\lambda }}{4}\int d^4x\, \phi ^2 \left( x \right)\int \frac{{{d^4}k}}{{{{\left( {2\pi } \right)}^4}}}\frac{1}{k^2 - m^2} \nonumber \\
 &= \int {{d^4}x\frac{1}{{{{\left( {4\pi } \right)}^2}}}\frac{\lambda }{4}{m^2}\left( {\log \frac{{{\mu ^2}}}{{{m^2}}} + 1} \right){\phi ^2}} . \label{eqn:phi4loop}
\end{align}
The first line of the above is just the definition of the functional trace, with a further insertion of unity, $1 = \int {d^4x\left| x \right\rangle \left\langle x \right|} $. Here $\ket{k}$ denotes the eigenstate of the derivative operator $\partial$ in the functional space, \textit{i.e.} $\bra{k} i\partial_\mu  = k_{\m}\bra{k}$, while $\ket{x}$ denotes the eigenstate of local operators, \textit{e.g.} $\bra{x} \ph^2 = \phi^2(x)\bra{x}$. They have the inner product $ \braket{x|k} = {e^{ - ikx}}$. We have made use of these to get the second line. The third line of Eq.~\eqref{eqn:phi4loop} is obtained by evaluating the $k$-integral (the ``loop'' integral) in $\overline{\text{MS}}$ scheme.

Now using Eq.~\eqref{eqn:1PIfunctional}, we take functional derivative of ${\Gamma ^{\left( 1 \right)}}\left[\phi\right]$ to get the two-point 1PI function $-i\Sigma(p^2)$:
\begin{equation}
- i\Sigma \left( {{p^2}} \right) = \int {{d^4}x{e^{ip\left( {{x_1} - {x_2}} \right)}}i\frac{{{\delta ^2}{\Gamma ^{\left( 1 \right)}}\left[ \phi  \right]}}{{\delta \phi \left( {{x_1}} \right)\delta \phi \left( {{x_2}} \right)}}}  = \frac{i}{{{{\left( {4\pi } \right)}^2}}}\frac{\lambda }{2}{m^2}\left( {\log \frac{{{\mu ^2}}}{{{m^2}}} + 1} \right) .
\end{equation}
This is the exact same result as one would get by Feynman diagram calculation. The pole mass follows from $\Sigma(p^2)$ as
\begin{equation}
m_p^2 = {m^2} + \Sigma \left( {{p^2} = {m^2}} \right) = {m^2} - \frac{1}{{{{\left( {4\pi } \right)}^2}}}\frac{\lambda }{2}{m^2}\left( {\log \frac{{{\mu ^2}}}{{{m^2}}} + 1} \right) .
\end{equation}

\subsubsection{Utility of functional methods compared to Feynman diagrams}\label{subsubsec:comparison}

With the help of the demonstrating example above, let us now make a detailed comparison between functional methods and diagrammatic methods. Overall, we see that the functional procedure is the same as the Feynman diagram calculation. Both start with identifying the relevant pieces for the correlation functions under concern, and then proceed with evaluating these pieces, during which the same loop integral shows up, and hence the same techniques for dealing with it (Feynman parameters, dimensional regularization, $\overline{\text{MS}}$ scheme, \textit{etc}.) get utilized. This is not surprising at all: since Feynman diagrams originate from the functional formalism, they are essentially just a diagrammatic representation of the functional method. However, despite computing the same quantities, the two differ in appearance, which results in a few advantages in using functional methods:
\begin{itemize}
  \item To compute a specific correlation function with Feynman diagrams, one needs to manually enumerate all diagrams interfering with each other, \textit{i.e.} having the same external legs but different internal structures. This enumeration can be exhausting if one deals with a theory of complicated interactions. What makes things worse is that these interfering diagrams sometimes have tricky minus signs relative to each other, \textit{e.g.} when there are identical external fermion legs. So it is easy to make a mistake in this step by either missing a diagram or messing up the interference sign.

      On the other hand, in the functional approach this enumeration step corresponds to expanding the logarithm of the functional determinant (\textit{e.g.} the second line of Eq.~\eqref{eqn:logexpand}) and collecting the same terms ($\phi^2$ terms in our demonstrating example). In a theory with complicated interactions, the functional derivative matrix is big, so expanding the logarithm is also tedious. But the point is a Taylor expansion is more systematic and mindless compared with a diagram enumeration. In addition, the correct interference sign will come out of the Taylor expansion automatically.

  \item In the diagrammatic approach, once the diagrams are enumerated, one composes various Feynman rules to obtain a ``diagram expression''. These include the rules on vertices, propagators, external legs, loop momentum integrals, symmetry factors, and so on. These rules come from breaking down the Taylor expanded functional expression (Eq.~\eqref{eqn:PathIntegral}) into components, such that we can have a diagrammatic representation of the calculation. We then compose these components back together to get the diagram expression. With functional methods, this composition step is avoided. The ``diagram expression'' comes out as a whole from the definition of the functional trace, \textit{e.g.} see the second line of Eq.~\ref{eqn:phi4loop}. Therefore, with functional methods, one need not remember or figure out any Feynman rules from a given Lagrangian.

  \item In Feynman diagram calculations, different correlation functions are computed by different sets of diagrams separately. On the other hand, $\Gamma\left[\phi\right]$ is a generating functional for \textit{all} 1PI correlation functions. Thus, functional methods provides us with a framework for dealing with different correlation functions together. This is especially important for gauge theories. Consider the SM as an example, where the six different correlation functions, $\left\langle {{H^{\dag 2}}{H^2}} \right\rangle$, $\left\langle {W{H^{\dag 2}}{H^2}} \right\rangle$, $\left\langle {{W^2}{H^{\dag 2}}{H^2}} \right\rangle$, $\left\langle {B{H^{\dag 2}}{H^2}} \right\rangle$, $\left\langle {{B^2}{H^{\dag 2}}{H^2}} \right\rangle$, and $\left\langle {WB{H^{\dag 2}}{H^2}} \right\rangle$ all originate from the gauge-invariant effective operator $\mathcal{O}_{HD}=\left|H^\dag DH\right|^2$. This situation might be easily recognized in the functional expression of $\Gamma\left[\phi\right]$, but would be very hard to see through individual sets of Feynman diagrams.
\end{itemize}

Let us be clear that we believe Feynman diagrams to be incredibly useful, especially for conceptual understanding. Moreover, when the desired correlation function is simple enough, Feynman diagrams frequently provide a quicker route to the answer. However, when the underlying theory is complicated or one is interested in many correlation functions, functional methods can organize and ease the evaluation in a way that is very difficult to see or do with diagrams.

\subsection{What is a ``covariant derivative expansion''?}\label{subsec:CDEReview}

Let us explain what is meant by a ``covariant derivative expansion'', as lately there has been some confusion about this term in the community.

In a gauge theory, the Lagrangian is built up by fields and the covariant derivative \(D_{\m} = \pd_{\m} -i g A_{\m}\). 
The essence of a ``covariant derivative expansion'' is really as straightforward as its name suggests:
\begin{quote}
For an expression that involves the covariant derivative $D_\mu$, if one expands the expression in a power series of $D$ while keeping $D_\mu$ intact (as opposed to splitting it into the partial derivative $\partial_\mu$ and the gauge fields $A_\mu$), then this expansion is called a ``covariant derivative expansion'' (CDE).
\end{quote}
For example, consider the gauge-promoted propagator $(-D^2-m^2)^{-1}$. The expansion
\begin{equation}
\frac{1}{{ - {D^2} - {m^2}}} = \frac{1}{{ - {m^2}}} + \frac{{{D^2}}}{{{m^4}}} +  \cdots , \label{eqn:DmCDE}
\end{equation}
qualifies as being called a CDE. Here is a practical example of using Eq.~\eqref{eqn:DmCDE}. Suppose that the tree-level 1PI effective action contains a piece
\begin{equation}
{\Gamma ^{\left( 0 \right)}}\left[ H \right] \supset {H^\dag }\frac{1}{{ - {D^2} - {m^2}}}H ,
\end{equation}
then one can use Eq.~\eqref{eqn:DmCDE} to Taylor expand the above into gauge-invariant local operators:
\begin{equation}
{\Gamma ^{\left( 0 \right)}}\left[ H \right] \supset {H^\dag }\frac{1}{{ - {D^2} - {m^2}}}H = \frac{1}{{ - {m^2}}}{H^\dag }H + \frac{1}{{{m^4}}}{H^\dag }{D^2}H +  \cdots .
\end{equation}

In contrast to a CDE, sometimes in manipulating an expression, one might first split $D_\mu$ into \(\pd_{\m} - i g A_{\m}\) and then expand out the two pieces differently. For example, instead of using Eq.~\eqref{eqn:DmCDE}, one could also expand the gauge-promoted propagator as
\begin{align}
\frac{1}{{ - {D^2} - {m^2}}} &= \frac{1}{{ - {\partial ^2} - {m^2} + ig{A^\mu }{\partial _\mu } + ig{\partial _\mu }{A^\mu } + {g^2}{A^2}}} \nonumber \\
 &= \frac{1}{{ - {\partial ^2} - {m^2}}} - \frac{1}{{ - {\partial ^2} - {m^2}}}\left( {ig{A^\mu }{\partial _\mu } + ig{\partial _\mu }{A^\mu } + {g^2}{A^2}} \right)\frac{1}{{ - {\partial ^2} - {m^2}}} +  \cdots . \label{eqn:DmPDE}
\end{align}
This expansion is clearly not a CDE.

One may wonder why someone would ever use Eq.~\eqref{eqn:DmPDE} instead of Eq.~\eqref{eqn:DmCDE}. To answer this question, let us consider calculating the 1PI effective action for a model with two Higgs doublets, $H$ and $\phi$:
\begin{equation}
{\cal L} \supset {\phi ^\dag }\left( { - {D^2} - {m^2}} \right)\phi - \lambda \left( {{\phi ^\dag }H} \right)\left( {{H^\dag }\phi } \right) .
\end{equation}
From Eq.~\eqref{eqn:GammaGeneral}, we know that the one-loop piece of the effective action would contain a piece like following
\begin{align}
{\Gamma ^{\left( 1 \right)}}\left[ H \right] &\supset i\log \det \left( { - \frac{{{\delta ^2}S\left[ {H,\phi } \right]}}{{\delta {\phi ^\dag }\delta \phi }}} \right) = i\log \det \left( {{D^2} + {m^2} + \lambda H{H^\dag }} \right) \nonumber \\
 &\supset -i {\rm{Tr}}\left( {\frac{1}{{ - {D^2} - {m^2}}}\lambda H{H^\dag }} \right) . \label{eqn:Dtrace}
\end{align}
To evaluate this functional trace, one would wish to follow the procedure in Eq.~\eqref{eqn:phi4loop}. However, there is a problem --- $\left|k\right\rangle$ is not an eigenstate of $D$:
\begin{equation}
\left\langle k \right.\left| {\frac{1}{{ - {D^2} - {m^2}}}} \right. \ne \frac{1}{{{k^2} - {m^2}}}\left\langle k \right| .
\end{equation}
The most straightforward work around to this problem is to expand the gauge-promoted propagator as in Eq.~\eqref{eqn:DmPDE}. This breaks down the functional trace of Eq.~\eqref{eqn:Dtrace} into many pieces, each of which then can be evaluated following the procedure in Eq.~\eqref{eqn:phi4loop}. In contrast to a CDE, we refer to this way of evaluating functional traces (or functional determinants) as a ``partial derivative expansion'' (PDE).

We want to emphasize that the PDE method works totally fine to evaluate any functional determinant. It is even more straightforward to use compared with a CDE. Its only drawback is that it is cumbersome. One can see that many terms are generated once we plug Eq.~\eqref{eqn:DmPDE} into Eq.~\eqref{eqn:Dtrace}, each of which would require multiple functional state insertions to evaluate.\footnote{These terms are actually in one-to-one correspondence with the relevant correlation functions that one would have to compute separately in a Feynman diagram method.} On the other hand, since the functional trace we are computing in Eq.~\eqref{eqn:Dtrace} is a gauge singlet, we know in the end we must arrive at a gauge-invariant expression, by recombining the $\partial_\mu$'s and the $A_\mu$'s. Therefore, this splitting of $D_\mu$ in a PDE is obviously a detour. A CDE method throughout which $D_\mu$ is kept intact is obviously preferable.

\subsubsection{Universal results for $\log\det [D^2  + m^2 + U(x)]$}\label{subsubsec:traditionalCDE}

While it is always straightforward to apply the PDE method to evaluate any particular functional determinant or trace, it is often not immediately clear whether or not a CDE method is available. In the next section and Appendix~\ref{app:Trace} we show how a CDE can be done on a wide class of functional traces. However, here we wish to include a few comments on the special (but, frequently encountered) case of the elliptic operator
\begin{equation}
D^2 + m^2 + U\left( x \right). \label{eqn:Elliptic}
\end{equation}
The functional determinant of this operator can always be evaluated using a CDE method. In~\cite{Henning:2014wua}, we showed how to do so, building on techniques introduced in~\cite{Gaillard:1985,Cheyette:1987}. Due to the advantage of CDE, we managed to obtain a universal result for \(\log \det [D^2 + m^2 + U(x)]\) up to dimension-six level, which we reproduce here for later reference:
\begin{align}
\log\det &\left[ {{D^2} + {m^2} + U\left( x \right)} \right] = \int {{d^4}x\frac{{ - i}}{{{{(4\pi )}^2}}}{\rm{tr}}} \Bigg\{ \nonumber \\
 & \hspace{5mm} {{m^4}\left[ { - \frac{1}{2}\left( {\ln \frac{{{m^2}}}{{{\mu ^2}}} - \frac{3}{2}} \right)} \right]} \nonumber \\
 &+ {m^2}\left[ { - \left( {\ln \frac{{{m^2}}}{{{\mu ^2}}} - 1} \right)U} \right] \nonumber \\
 &+ {m^0}\left[ { - \frac{1}{2}\ln \frac{{{m^2}}}{{{\mu ^2}}}{U^2} - \frac{1}{{12}}\left( {\ln \frac{{{m^2}}}{{{\mu ^2}}} - 1} \right){G'}_{\mu \nu }^2} \right] \nonumber \\
 &+ \frac{1}{{{m^2}}}\left[ { - \frac{1}{6}{U^3} + \frac{1}{{12}}{{(DU)}^2} - \frac{1}{{12}}U{G'}_{\mu \nu }^2 + \frac{1}{{60}}{{({D^\mu }{{G'}_{\mu \nu }})}^2} - \frac{1}{{90}}{G'}_\mu^\nu {G'}_\nu^\rho {G'}_\rho^\mu } \right] \nonumber \\
 &+ \frac{1}{{{m^4}}}\left[ {\frac{1}{{24}}{U^4} - \frac{1}{{12}}U{{(DU)}^2} + \frac{1}{{120}}{{({D^2}U)}^2} + \frac{1}{{60}}({D^\mu }U)({D^\nu }U){{G'}_{\mu \nu }}} \right. \nonumber \\
 & \hspace{10mm} \left. { + \frac{1}{{40}}{U^2}{G'}_{\mu \nu }^2 + \frac{1}{{60}}{{(U{{G'}_{\mu \nu }})}^2}} \right] \nonumber \\
 &+ \frac{1}{{{m^6}}}\left[ { - \frac{1}{{60}}{U^5} + \frac{1}{{20}}{U^2}{{(DU)}^2} + \frac{1}{{30}}{{(UDU)}^2}} \right] \nonumber \\
 & { + \frac{1}{{{m^8}}}\left[ {\frac{1}{{120}}{U^6}} \right]} \Bigg\} , \label{eqn:ellipticUniversal}
\end{align}
where ${{G'}_{\mu \nu }} \equiv \left[ {{D_\mu },{D_\nu }} \right]$. The above universal result assumes a single degenerate mass scale $m$. A generalization to accommodate multiple mass scales is given in~\cite{Drozd:2015rsp}.

The reason the above is so useful is because of the ubiquity of such determinants in one-loop calculations: the argument of \(\log \det (- \d^2S/\d\Ph^2)\) picks out quadratic terms in the action, which are of the form \(D^2 + m^2 + U(x)\). Note that this is also true for fermions; this is explained in~\cite{Henning:2014wua}, but we reproduce it here for future reference. While taking the functional derivative with respect to fermionic fields, one typically obtains an operator of the following form
\begin{equation}
 - i\slashed D + m + F\left( x \right) , \label{eqn:Fermionic}
\end{equation}
where $\slashed D \equiv \gamma^\mu D_\mu$. As the determinant is invariant under flipping the sign of the \(\g^\mu\) matrices, we may write
\begin{align}
\log \det \left[ { - i\slashed D + m + F\left( x \right)} \right] &= \log \det \left[ {i\slashed D + m + F\left( x \right)} \right] \nonumber \\
 &= \frac{1}{2}\log \det \left\{ {\left[ { - i\slashed D + m + F\left( x \right)} \right]\left[ {i\slashed D + m + F\left( x \right)} \right]} \right\} .
\end{align}
The product gives
\begin{align}
\left( { - i\slashed D + m + F} \right)\left( {i\slashed D + m + F} \right) &= {{\slashed D}^2} + {m^2} - i\slashed DF + \{m,F\} + {F^2} \nonumber \\
 &= {D^2} + {m^2} - \frac{i}{2}{\sigma ^{\mu \nu }}{{G'}_{\mu \nu }} - i\slashed DF + \{m,F\} + {F^2} ,
\end{align}
where, by definition, \(\slashed{D}F \equiv [\slashed{D},F]\), and we have used ${{\slashed D}^2} = {D^2} - \frac{i}{2}{\sigma ^{\mu \nu }}{{G'}_{\mu \nu }}$, with ${\sigma ^{\mu \nu }} \equiv \frac{i}{2}\left[ {{\gamma ^\mu },{\gamma ^\nu }} \right]$ defined as usual. We see that this is now clearly in the form of Eq.~\eqref{eqn:Elliptic} with
\begin{equation}
U_\text{ferm} \left( x \right) =  - \frac{i}{2}{\sigma ^{\mu \nu }}{{G'}_{\mu \nu }} - i\slashed DF + \{m,F\} + {F^2} .
\label{eqn:Uferm}
\end{equation}

\subsection{``Traditional'' matching with functional methods}\label{subsec:MatchingTraditional}

Matching a UV theory onto an EFT is done by equating the 1LPI effective actions of the two theories at the matching scale, Eq.~\eqref{eq:match_1LPI_intro}. In this subsection, we give explicit formulas for carrying out the general procedure of~\cite{Georgi:1991ch,Georgi:1992xg} through one-loop order using functional methods. We refer to this procedure as ``traditional'' matching, and show how it allows us to compute $c_i^{(0)}(M)$, $c_{i,\text{heavy}}^{(1)}(M)$, and $c_{i,\text{mixed}}^{(1)}(M)$. As we will see, the contributions to $c_{i,\text{heavy}}^{(1)}$ and $c_{i,\text{mixed}}^{(1)}$ are entangled in this approach; in the next section, we develop a new procedure to isolate these components.

Following the notations in Section~\ref{subsec:MatchingReview}, we consider a UV theory and an EFT
\begin{align}
{{\cal L}_{{\rm{UV}}}}\left( {\phi ,\Phi } \right) &= {{\cal L}_\phi }\left( \phi  \right) + {{\cal L}_\Phi }\left( \phi, \Phi \right) , \\
{{\cal L}_{{\rm{EFT}}}}\left( \phi  \right) &= {{\cal L}_\phi }\left( \phi  \right) + \sum\limits_i {{c_i}\left( \mu  \right){{\cal O}_i}\left( \phi  \right)} ,
\end{align}
with the corresponding actions \(S_{\text{UV}} = \int d^4x \, \scL_{\text{UV}}\) and \(S_{\text{EFT}} = \int d^4x\, \scL_{\text{EFT}}\). The matching criterion is requiring the 1LPI effective actions computed from the EFT and the UV theory agree at RG scale $\mu=M$:
\begin{equation}
{\Gamma _{{\rm{L,EFT}}}}\left[ {\phi} \right](\m = M) = {\Gamma _{{\rm{L,UV}}}}\left[ {\phi } \right](\m = M) . \label{eqn:MatchingCriterion}
\end{equation}
Since matching can be done order-by-order in perturbation theory, Eq.~\eqref{eqn:MatchingCriterion} implies that
\begin{subequations}
\label{eqn:MatchingCriterionTreeLoop}
\begin{align}
\Gamma_{{\rm{L,EFT}}}^{\left( 0 \right)}\left[ {\phi } \right](\m = M) &= \Gamma _{{\rm{L,UV}}}^{\left( 0 \right)}\left[ {\phi } \right](\m = M) , \label{eqn:MatchingCriterionTree} \\
\Gamma_{{\rm{L,EFT}}}^{\left( 1 \right)}\left[ {\phi } \right](\m = M) &= \Gamma _{{\rm{L,UV}}}^{\left( 1 \right)}\left[ {\phi } \right](\m = M) . \label{eqn:MatchingCriterionLoop}
\end{align}
\end{subequations}
Our task then is to solve Eqs.~\eqref{eqn:MatchingCriterionTree} and~\eqref{eqn:MatchingCriterionLoop}.

Let us first look at the UV side. In Section~\ref{subsubsec:functional1loop}, we explained how to compute the 1PI effective action $\Gamma$ to one-loop order. For the UV theory, this quantity is a functional of both the light fields $\phi$ and the heavy field $\Phi$: \({\Gamma _{{\rm{UV}}}}\left[ {\phi ,\Phi } \right]\). According to Eq.~\eqref{eqn:GammaGeneral}, the tree-level piece and the one-loop level piece are respectively
\begin{align}
\Gamma _{{\rm{UV}}}^{\left( 0 \right)}\left[ {\phi ,\Phi } \right] &= {S_{{\rm{UV}}}}\left[ {\phi ,\Phi } \right] , \\
\Gamma _{{\rm{UV}}}^{\left( 1 \right)}\left[ {\phi ,\Phi } \right] &= \frac{i}{2}\log \det \left( { - \frac{{{\delta ^2}{S_{{\rm{UV}}}}\left[ {\phi ,\Phi } \right]}}{{\delta {{\left( {\phi ,\Phi } \right)}^2}}}} \right) .
\end{align}
The only subtlety on the UV side is that the 1LPI effective action $\Gamma_\text{L,UV}$, which is a functional of the light fields $\phi$ only, is not the same as the 1PI effective action $\Gamma_\text{UV}$, which depends on both $\phi$ and $\Phi$. However, it is very easy to get $\Gamma_\text{L,UV}$ from $\Gamma_\text{UV}$ --- simply plug in the solution of the equations of motion of $\Phi_c\left[\phi\right]$:\footnote{Here, Eq.~\eqref{eqn:1LPI1PI} is a bit sloppy, to quickly convey the idea. The exact expression for the one-loop piece, as we will see in the next few equations, is to plug in $\Phi_c\left[\phi\right]$ after taking the functional derivative but before evaluating the functional determinant.}
\begin{equation}
{\Gamma _{{\rm{L,UV}}}}\left[ \phi  \right] = {\Gamma _{{\rm{UV}}}}\left[ {\phi ,{\Phi _c}\left[ \phi  \right]} \right] . \label{eqn:1LPI1PI}
\end{equation}
We therefore obtain
\begin{subequations}
\label{eqn:1LPI01UV}
\begin{align}
\Gamma _{{\rm{L,UV}}}^{\left( 0 \right)}\left[ \phi  \right] &= {S_{{\rm{UV}}}}\left[ {\phi ,{\Phi _c}\left[ \phi  \right]} \right] = \int {{d^4}x\left[ {{{\cal L}_\phi }\left( \phi  \right) + {{\cal L}_\Phi }\left( {\phi ,{\Phi _c}\left[ \phi  \right]} \right)} \right]} , \label{eqn:1LPI0UV} \\
\Gamma _{{\rm{L,UV}}}^{\left( 1 \right)}\left[ \phi  \right] &= \frac{i}{2}\log \det \left( {{{\left. { - \frac{{{\delta ^2}{S_{{\rm{UV}}}}\left[ {\phi ,\Phi } \right]}}{{\delta {{\left( {\phi ,\Phi } \right)}^2}}}} \right|}_{\Phi  = {\Phi _c}\left[ \phi  \right]}}} \right) . \label{eqn:1LPI1UV}
\end{align}
\end{subequations}

Now let us move on to the EFT side. For the EFT, the 1LPI effective action \textit{is} the 1PI effective action, ${\Gamma _{{\rm{L,EFT}}}}\left[ \phi  \right] = {\Gamma _{{\rm{EFT}}}}\left[ \phi  \right]$, as there is no heavy field. However, the loop order counting is a bit subtle for the EFT side, as the Lagrangian contains one-loop sized Wilson coefficients,\footnote{The loop order counting is defined in terms of the couplings of the UV theory.} \textit{i.e.} ${S_{{\rm{EFT}}}}\left[ \phi  \right] = S_{{\rm{EFT}}}^{\left( 0 \right)}\left[ \phi  \right] + S_{{\rm{EFT}}}^{\left( 1 \right)}\left[ \phi  \right]$, with
\begin{align}
S_{{\rm{EFT}}}^{\left( 0 \right)}\left[ \phi  \right] &= \int {{d^4}x\left[ {{{\cal L}_\phi }\left( \phi  \right) + \sum\limits_i {c_i^{\left( 0 \right)}\left( M \right){{\cal O}_i}\left( \phi  \right)} } \right]} , \label{eqn:SEFT0} \\
S_{{\rm{EFT}}}^{\left( 1 \right)}\left[ \phi  \right] &= \int {{d^4}x\left\{ {\sum\limits_i {\left[ {c_{i,\text{heavy}}^{\left( 1 \right)}\left( M \right) + c_{i,\text{mixed}}^{\left( 1 \right)}\left( M \right)} \right]{{\cal O}_i}\left( \phi  \right)} } \right\}} . \label{eqn:SEFT1}
\end{align}
After rearranging pieces according to the UV loop-order counting, we get
\begin{subequations}
\label{eqn:1LPI01EFT}
\begin{align}
\Gamma _{{\rm{L,EFT}}}^{\left( 0 \right)}\left[ \phi  \right] &= S_{{\rm{EFT}}}^{\left( 0 \right)}\left[ \phi  \right] = \int {{d^4}x\left[ {{{\cal L}_\phi }\left( \phi  \right) + \sum\limits_i {c_i^{\left( 0 \right)}\left( M \right){{\cal O}_i}\left( \phi  \right)} } \right]} , \label{eqn:1LPI0EFT} \\
\Gamma _{{\rm{L,EFT}}}^{\left( 1 \right)}\left[ \phi  \right] &= S_{{\rm{EFT}}}^{\left( 1 \right)}\left[ \phi  \right] + \frac{i}{2}\log \det \left( { - \frac{{{\delta ^2}S_{{\rm{EFT}}}^{\left( 0 \right)}\left[ \phi  \right]}}{{\delta {\phi ^2}}}} \right) \nonumber \\
 &= \int {{d^4}x\left\{ {\sum\limits_i {\left[ {c_{i,\text{heavy}}^{\left( 1 \right)}\left( M \right) + c_{i,\text{mixed}}^{\left( 1 \right)}\left( M \right)} \right]{{\cal O}_i}\left( \phi  \right)} } \right\}}  + \frac{i}{2}\log \det \left( { - \frac{{{\delta ^2}S_{{\rm{EFT}}}^{\left( 0 \right)}\left[ \phi  \right]}}{{\delta {\phi ^2}}}} \right) . \label{eqn:1LPI1EFT}
\end{align}
\end{subequations}

Using Eqs.~\eqref{eqn:1LPI01UV} and~\eqref{eqn:1LPI01EFT} in Eq.~\eqref{eqn:MatchingCriterionTreeLoop}, we arrive at the matching results:
\begin{align}
 &\sum\limits_i {c_i^{\left( 0 \right)}\left( M \right){{\cal O}_i}\left( \phi  \right)}  = {{\cal L}_\Phi }\left( {\phi ,{\Phi _c}\left[ \phi  \right]} \right) , \label{eqn:c0Functional} \\
 &\int {{d^4}x\left\{ {\sum\limits_i {\left[ {c_{i{\rm{,heavy}}}^{\left( 1 \right)}\left( M \right) + c_{i{\rm{,mixed}}}^{\left( 1 \right)}\left( M \right)} \right]{{\cal O}_i}\left( \phi  \right)} } \right\}} \nonumber \\
 & \qquad = \frac{i}{2}\log \det \left( {{{\left. { - \frac{{{\delta ^2}{S_{{\rm{UV}}}}\left[ {\phi ,\Phi } \right]}}{{\delta {{\left( {\phi ,\Phi } \right)}^2}}}} \right|}_{\Phi  = {\Phi _c}\left[ \phi  \right]}}} \right) - \frac{i}{2}\log \det \left( { - \frac{{{\delta ^2}S_{{\rm{EFT}}}^{\left( 0 \right)}\left[ \phi  \right]}}{{\delta {\phi ^2}}}} \right) . \label{eqn:c1Functional}
\end{align}
The tree-level matching result Eq.~\eqref{eqn:c0Functional} is very easy to calculate for any given UV theory. The one-loop result Eq.~\eqref{eqn:c1Functional} looks a bit complicated, but it is actually very intuitive to understand: After determining $c_i^{(0)}(M)$ (and hence $S_\text{EFT}^{(0)}[\phi]$) by tree-level matching, if one were to calculate the one-loop 1LPI using $S_\text{EFT}^{(0)}[\phi]$, it does not fully agree with the 1LPI calculated using the UV action $S_\text{UV}$. The mismatch between the two dictates the need of ${c_{i,\text{heavy}}^{(1)}(M)}$ and ${c_{i,\text{mixed}}^{(1)}(M)}$.

%% file: sec_MatchingDirect.tex
In the previous section we showed how to perform the matching analysis through one-loop order with functional methods. However, the one-loop matching result in Eq.~\eqref{eqn:c1Functional} is somewhat unsatisfactory because the results for ${c_{i,\text{heavy}}^{(1)}(M)}$ and ${c_{i,\text{mixed}}^{(1)}(M)}$ are entangled; namely, one has to compute them together. Ideally, we could obtain a disentangled result for each separate piece of \(c_i^{(1)}(M)\). Upon isolating ${c_{i,\text{heavy}}^{(1)}(M)}$ and ${c_{i,\text{mixed}}^{(1)}(M)}$, it is additionally desirable that the functional determinants involved can be evaluated with a CDE.

This section is devoted to the above tasks. In Sec~\ref{subsec:DirectGeneral} we first derive the disentangled results for ${c_{i,\text{heavy}}^{(1)}(M)}$ and ${c_{i,\text{mixed}}^{(1)}(M)}$. Then we show how to systematically evaluate the functional determinants involved by a new CDE technique. This new CDE technique is capable of evaluating functional determinants of a much wider class of operators beyond the elliptic operators that~\cite{Gaillard:1985,Cheyette:1987,Henning:2014wua,Drozd:2015rsp} are limited to. We show how to use these techniques in practice by considering two examples: A toy model with a heavy and a light singlet scalar in Section~\ref{subsec:ToyScalar}, and a more phenomenological example in Section~\ref{subsec:TripletScalar} --- the Standard Model with a heavy electroweak triplet scalar.

\subsection{General formalism for a direct computation of $c_{i,\text{mixed}}^{(1)}(M)$}\label{subsec:DirectGeneral}

\subsubsection{Integrating out the heavy field $\Phi$}\label{subsubsec:IntegrateOut}

In order to resolve ${c_{i,\text{heavy}}^{(1)}(M)}$ and ${c_{i,\text{mixed}}^{(1)}(M)}$ in the one-loop matching result Eq.~\eqref{eqn:c1Functional}, let us take a closer look at the first functional determinant:
\begin{equation}
\log\det \left( { - {{\left. {\frac{{{\delta ^2}{S_{{\rm{UV}}}}\left[ {\phi ,\Phi } \right]}}{{\delta {{\left( {\phi ,\Phi } \right)}^2}}}} \right|}_{\Phi  = {\Phi _c}\left[ \phi  \right]}}} \right) . \label{eqn:mixeddet}
\end{equation}
The problem with this determinant is that it involves functional derivatives with respect to both $\phi$ and $\Phi$. This makes the boundary between contributions to ${c_{i,\text{heavy}}^{(1)}(M)}$ and ${c_{i,\text{mixed}}^{(1)}(M)}$ unclear. However, we can cast the above determinant into a resolved form where there are no ``mixed'' functional derivatives:
\begin{equation}
\log \det \left( { - {{\left. {\frac{{{\delta ^2}{S_{{\rm{UV}}}}\left[ {\phi ,\Phi } \right]}}{{\delta {{\left( {\phi ,\Phi } \right)}^2}}}} \right|}_{{\Phi _c}}}} \right) = \log\det \left( { - {{\left. {\frac{{{\delta ^2}{S_{{\rm{UV}}}}\left[ {\phi ,\Phi } \right]}}{{\delta {\Phi ^2}}}} \right|}_{{\Phi _c}}}} \right) + \log\det \left( { - \frac{{{\delta ^2}{S_{{\rm{UV}}}}\left[ {\phi ,{\Phi _c}\left[ \phi  \right]} \right]}}{{\delta {\phi ^2}}}} \right) . \label{eqn:determinantsresolved}
\end{equation}
A derivation of this formula is given in Appendix~\ref{app:DeriveResolve}. Note the difference in the functional dependence of \(S_{\text{UV}}\) in the two determinants above. As in Eq.~\eqref{eqn:mixeddet}, in the first we vary the UV action with respect to \(\Ph\) and then evaluate on the classical solution \(\Ph_c[\ph]\). In the second determinant we first plug in the classical solution \(\Ph_c[\ph]\) (which is a non-local functional of \(\ph\)) and then vary with respect to \(\ph\).

Eq.~\eqref{eqn:determinantsresolved} looks a bit abstract, but it actually has a very intuitive physical understanding: integrating out the heavy field $\Phi$. To see this, let us recall from Eq.~\eqref{eqn:1LPI01UV} that the UV 1LPI effective action ${\Gamma _{{\rm{L,UV}}}}\left[ \phi  \right]$ is given by
\begin{equation}
{\Gamma _{{\rm{L,UV}}}}\left[ \phi  \right] = {S_{{\rm{UV}}}}\left[ {\phi ,{\Phi _c}\left[ \phi  \right]} \right] + \frac{i}{2}\log \det \left( {{{\left. { - \frac{{{\delta ^2}{S_{{\rm{UV}}}}\left[ {\phi ,\Phi } \right]}}{{\delta {{\left( {\phi ,\Phi } \right)}^2}}}} \right|}_{{\Phi _c}}}} \right) .
\end{equation}
Using Eq.~\eqref{eqn:determinantsresolved} in this expression, we get
\begin{equation}
{\Gamma _{{\rm{L,UV}}}}\left[ \phi  \right] = {S_{{\rm{UV}}}}\left[ {\phi ,{\Phi _c}\left[ \phi  \right]} \right] + \frac{i}{2}\log\det \left( { - \frac{{{\delta ^2}{S_{{\rm{UV}}}}\left[ {\phi ,{\Phi _c}\left[ \phi  \right]} \right]}}{{\delta {\phi ^2}}}} \right) + \frac{i}{2}\log\det \left( { - {{\left. {\frac{{{\delta ^2}{S_{{\rm{UV}}}}\left[ {\phi ,\Phi } \right]}}{{\delta {\Phi ^2}}}} \right|}_{{\Phi _c}}}} \right) . \label{eqn:1LPIUVresolved}
\end{equation}
To recognize what this result implies, let us imagine integrating out the heavy field $\Phi$ in the path integral,
\begin{equation}
{e^{i{S_{{\rm{eff}}}}\left[ \phi  \right]}} = \int {D\Phi {e^{i{S_{{\rm{UV}}}}\left[ {\phi ,\Phi } \right]}}} . \label{eqn:Seffdef}
\end{equation}
The above serves as a definition of \(S_{\text{eff}}[\ph]\). It is an inherently non-local object.
To one-loop order, $S_\text{eff}[\phi]$ is obtained by a saddle-point approximation of the path integral,
\begin{equation}
{S_{{\rm{eff}}}}\left[ \phi  \right] = {S_{{\rm{UV}}}}\left[ {\phi ,{\Phi _c}\left[ \phi  \right]} \right] + \frac{i}{2}\log\det \left( { - {{\left. {\frac{{{\delta ^2}{S_{{\rm{UV}}}}\left[ {\phi ,\Phi } \right]}}{{\delta {\Phi ^2}}}} \right|}_{{\Phi _c}}}} \right) .
\end{equation}
If we use this $S_\text{eff}[\phi]$ to compute the corresponding 1PI effective action $\Gamma_\text{eff}[\phi]$, following the general prescription in Eq.~\eqref{eqn:GammaGeneral}, we obtain exactly the same expression as in Eq.~\eqref{eqn:1LPIUVresolved}, namely that
\begin{equation}
{\Gamma _{{\rm{L,UV}}}}\left[ \phi  \right] = {\Gamma _{{\rm{eff}}}}\left[ \phi  \right] . \label{eqn:1LPIequivalence}
\end{equation}
Physically, Eq.~\eqref{eqn:1LPIequivalence} means that the theory $S_\text{eff}[\phi]$ as defined in Eq.~\eqref{eqn:Seffdef} is equivalent to the UV theory $S_\text{UV}[\phi,\Phi]$ with regard to the physics of the light fields $\phi$. This is a well-known statement (\textit{e.g.} \cite{Georgi:1991ch,Georgi:1992xg}). In the path integral formalism, Eq.~\eqref{eqn:PathIntegral}, this statement is almost trivially true by the definition of $S_\text{eff}[\phi]$:
\begin{align}
{\left\langle {\phi \left( {{x_1}} \right) \cdots \phi \left( {{x_n}} \right)} \right\rangle _{{\rm{UV}}}} &= \frac{{\int {{\cal D}\phi {\cal D}\Phi {e^{i{S_{{\rm{UV}}}}\left[ {\phi ,\Phi } \right]}}\phi \left( {{x_1}} \right) \cdots \phi \left( {{x_n}} \right)} }}{{\int {{\cal D}\phi {\cal D}\Phi {e^{i{S_{{\rm{UV}}}}\left[ {\phi ,\Phi } \right]}}} }} \nonumber \\
 &= \frac{{\int {{\cal D}\phi {e^{i{S_{{\rm{eff}}}}\left[ \phi  \right]}}\phi \left( {{x_1}} \right) \cdots \phi \left( {{x_n}} \right)} }}{{\int {{\cal D}\phi {e^{i{S_{{\rm{eff}}}}\left[ \phi  \right]}}} }} = {\left\langle {\phi \left( {{x_1}} \right) \cdots \phi \left( {{x_n}} \right)} \right\rangle _{{\rm{eff}}}} .
\end{align}
With the help of this integrating out picture, it is hopefully easier to understand and remember Eq.~\eqref{eqn:determinantsresolved}.

\subsubsection{Resolved matching results}\label{subsubsec:Disentangled}

With the functional derivatives with respect to $\phi$ and $\Phi$ disentangled, it is easy to separately identify the pieces ${c_{i,\text{heavy}}^{(1)}(M)}$ and ${c_{i,\text{mixed}}^{(1)}(M)}$. Combining Eqs.~\eqref{eqn:c1Functional} and~\eqref{eqn:determinantsresolved}, we arrive at our resolved matching results:
\begin{align}
\int {{d^4}x\sum\limits_i {c_{i{\rm{,heavy}}}^{\left( 1 \right)}\left( M \right){{\cal O}_i}\left( \phi  \right)} } &= \frac{i}{2}\log\det \left( { - {{\left. {\frac{{{\delta ^2}{S_{{\rm{UV}}}}\left[ {\phi ,\Phi } \right]}}{{\delta {\Phi ^2}}}} \right|}_{{\Phi _c}}}} \right) , \label{eqn:c1heavydirect} \\
\int {{d^4}x\sum\limits_i {c_{i{\rm{,mixed}}}^{\left( 1 \right)}\left( M \right){{\cal O}_i}\left( \phi  \right)} } &= \frac{i}{2}\log\det \left( { - \frac{{{\delta ^2}{S_{{\rm{UV}}}}\left[ {\phi ,{\Phi _c}\left[ \phi  \right]} \right]}}{{\delta {\phi ^2}}}} \right) - \frac{i}{2}\log \det \left( { - \frac{{{\delta ^2}S_{{\rm{EFT}}}^{\left( 0 \right)}\left[ \phi  \right]}}{{\delta {\phi ^2}}}} \right) . \label{eqn:c1mixeddirect}
\end{align}

We see that our matching result for ${c_{i,\text{heavy}}^{(1)}(M)}$ has reproduced Eq.~\eqref{eqn:c1heavy}. The functional derivative matrix in Eq.~\eqref{eqn:c1heavydirect} is always of the form \(D^2 + M^2 + U(x)\), so the universal results of~\cite{Henning:2014wua,Drozd:2015rsp} are immediately available. As the piece ${c_{i,\text{heavy}}^{(1)}(M)}$ has been intensively studied in~\cite{Henning:2014wua,Drozd:2015rsp}, our focus in this paper will be on the other one, ${c_{i,\text{mixed}}^{(1)}(M)}$.

One might wonder why Eq.~\eqref{eqn:c1mixeddirect} does not vanish, since the tree-level matching result (Eq.~\eqref{eqn:c0Functional}) states that $\mathcal{L}_\text{EFT}^{(0)}(\phi) = \mathcal{L}_\text{UV}\left(\phi, \Phi_c[\phi] \right)$. The answer is that these two Lagrangians are actually different. $\mathcal{L}_\text{UV}\left( \phi, \Phi_c[\phi] \right)$ is an inherently \textit{non-local} object, because of the non-local nature of $\Phi_c[ \phi]$. On the other hand, $\mathcal{L}_\text{EFT}^{(0)}(\phi)$ is by definition a sum of \textit{local} operators. Therefore, the true meaning of the tree-level matching result in Eq.~\eqref{eqn:c0Functional} is that $\mathcal{L}_{{\rm{EFT}}}^{(0)}(\phi)$ should be a local expansion of the non-local $\mathcal{L}_\text{UV}\left(\phi, \Phi_c[\phi] \right)$.

In fact, it is exactly this local vs. non-local difference between $\mathcal{L}_{{\rm{EFT}}}^{(0)}(\phi)$ and $\mathcal{L}_\text{UV}\left(\phi, \Phi_c[\phi] \right)$ that leads to nonzero ${c_{i,\text{mixed}}^{(1)}(M)}$. Using the local $\mathcal{L}_{{\rm{EFT}}}^{(0)}(\phi)$ in the functional determinant will give a result that mismatches with that obtained using the non-local $\mathcal{L}_\text{UV}\left(\phi, \Phi_c[\phi] \right)$ \cite{Witten:1975bh,Witten:1976kx}, and hence dictates the need for $c_{i,\text{mixed}}^{(1)}$.\footnote{Mathematically, this mismatch is due to the illegitimate expansion of propagator inside the momentum integral under dimensional regularization:
\begin{equation}
\int {\frac{{{d^4}p}}{{{{\left( {2\pi } \right)}^4}}}\frac{1}{{ - {M^2}}}\left( {1 + \frac{{{p^2}}}{{{M^2}}} + \frac{{{p^4}}}{{{M^4}}} +  \cdots } \right)}  \ne \int {\frac{{{d^4}p}}{{{{\left( {2\pi } \right)}^4}}}\frac{1}{{{p^2} - {M^2}}}} . \label{eqn:dimregmismatch}
\end{equation}
In dimensional regularization, the result of the right hand side (the UV theory) generically has a fractional power of $M^2$. However, the left hand side (the EFT) clearly can only yield integer powers of $M^2$, as $M^2$, $M^4$, \textit{etc.} participate only as overall factors of the integrals. This mismatch can be avoided by using other regularization scheme, such as a sharp cutoff of the momentum integral. In that case $c_{i,\text{mixed}}^{(1)}=0$. This is consistent with our understanding that $c_{i,\text{mixed}}^{(1)}$ is not physical by itself. More details on using a different regularization scheme are explained in Appendix~\ref{app:Wilsonian}.
}

Let us see a concrete example of this local vs. non-local difference. Consider a UV theory with a heavy complex scalar field $\Phi$:
\begin{equation}
{{\cal L}_{{\rm{UV}}}}\left( {\phi ,\Phi } \right) \supset {\Phi ^\dag }\left[ { - {D^2} - {M^2} - U\left( \phi  \right)} \right]\Phi  + \left[ {{\Phi ^\dag }B\left( \phi  \right) + c.c.} \right] .
\end{equation}
The solution to the equation of motion of $\Phi$ is
\begin{equation}
{\Phi _c}\left[ \phi  \right] =  - \frac{1}{{ - {D^2} - {M^2} - U\left( \phi  \right)}}B\left( \phi  \right) .
\end{equation}
We see that $\Phi_c[\phi]$ is NOT a local operator that only depends on the spacetime coordinate $x$, because the covariant derivative $D$ shows up in the denominator. Plugging this $\Phi_c[\phi]$ back into the UV Lagrangian, we get
\begin{equation}
{{\cal L}_{{\rm{UV}}}}\left( {\phi ,{\Phi _c}\left[ \phi  \right]} \right) =  - {B^\dag }\left( \phi  \right)\frac{1}{{ - {D^2} - {M^2} - U\left( \phi  \right)}}B\left( \phi  \right) , \label{eqn:Lnonlocal}
\end{equation}
which is also non-local. On the other hand, when matching the above UV theory with an EFT at tree-level, we solve the Wilson coefficients according to the matching equation
\begin{equation}
{\cal L}_{{\rm{EFT}}}^{\left( 0 \right)}\left( \phi  \right) = \sum\limits_i {c_i^{\left( 0 \right)}\left( M \right){{\cal O}_i}\left( \phi  \right)}  = {{\cal L}_{{\rm{UV}}}}\left( {\phi ,{\Phi _c}\left[ \phi  \right]} \right) .
\end{equation}
This solving step amounts to an inverse mass expansion of Eq.~\eqref{eqn:Lnonlocal} into a sum of local operators:
\begin{equation}
{\cal L}_{{\rm{EFT}}}^{\left( 0 \right)}\left( \phi  \right) =  - {B^\dag }\frac{1}{{{M^2}}}B - {B^\dag }\frac{1}{{{M^2}}}\left( { - {D^2} - U} \right)\frac{1}{{{M^2}}}B +  \cdots . \label{eqn:Llocal}
\end{equation}
We see that ${\cal L}_{{\rm{EFT}}}^{\left( 0 \right)}\left( \phi  \right)$ is local.\footnote{It is worth noting that the non-local character of $\mathcal{L}_\text{UV}\left(\phi, \Phi_c[\phi] \right)$ is due to the derivatives being in the denominator. As long as one does not expand out these derivatives, this character is not changed. For example, while using Eq.~\eqref{eqn:Lnonlocal} in Eq.~\eqref{eqn:c1mixeddirect}, it is legitimate to do the expansion
\begin{equation}
{{\cal L}_{{\rm{UV}}}}\left( {\phi ,{\Phi _c}\left[ \phi  \right]} \right) = {B^\dag }\left( \phi  \right)\frac{1}{{ - {D^2} - {M^2}}}B\left( \phi  \right) + {B^\dag }\left( \phi  \right)\frac{1}{{ - {D^2} - {M^2}}}U\left( \phi  \right)\frac{1}{{ - {D^2} - {M^2}}}B\left( \phi  \right) +  \cdots . \nonumber
\end{equation}
}

Now we understand that the two functional determinants in Eq.~\eqref{eqn:c1mixeddirect} are very close to each other---they only differ by this local vs. non-local difference. This actually provides us a way of using this formula more efficiently. In particular, the second term in Eq.~\eqref{eqn:c1mixeddirect} is always \textit{contained} in the first term. As we are only interested in their difference, instead of computing both terms separately and subtracting them in the end, one can compute only the first term and then \textit{drop} the pieces corresponding to the second term. This procedure is kind of similar with the usual loop diagram calculation in renormalized perturbation theory under $\overline{\text{MS}}$ scheme, where in principle one has contributions from both the renormalized coupling and the counterterm, but in practice one just computes the renormalized coupling part and then \textit{drops} the $1/\epsilon-\gamma+\log(4\pi)$.\footnote{This analogy is not purely mathematical, as the subtraction in Eq.~\eqref{eqn:c1mixeddirect} does have the physical meaning of removing IR divergences.}

So, in actual calculations, we will only compute the first term in Eq.~\eqref{eqn:c1mixeddirect}, coming from the non-local Lagrangian $\mathcal{L}_\text{UV}\left(\phi,\Phi_c[\phi]\right)$, and then we drop its ``local counterpart'' coming from the local Lagrangian $\mathcal{L}_\text{EFT}^{(0)}[\phi]$ that was matched with $\mathcal{L}_\text{UV}\left(\phi,\Phi_c[\phi]\right)$. With this strategy, we abbreviate Eq.~\eqref{eqn:c1mixeddirect} as
\begin{equation}
\int {{d^4}x\sum\limits_i {c_{i{\rm{,mixed}}}^{\left( 1 \right)}\left( M \right){{\cal O}_i}\left( \phi  \right)} }  = \frac{i}{2}\log \det {\left( { - \frac{{{\delta ^2}{S_{{\rm{UV}}}}\left[ {\phi ,{\Phi _c}\left[ \phi  \right]} \right]}}{{\delta {\phi ^2}}}} \right)_{\rm{d}}} , \label{eqn:c1mixeddirectabbr}
\end{equation}
where the subscript ``d'' is short for ``drop'' which reminds us to drop the ``local counterpart''.

To identify which pieces are the ``local counterparts'' to drop, we make the crucial observation that going from $\mathcal{L}_\text{UV}\left(\phi, \Phi_c[\phi] \right)$ (\textit{e.g.} Eq.~\eqref{eqn:Lnonlocal}) to $\mathcal{L}_\text{EFT}^{(0)}(\phi)$ (\textit{e.g.} Eq.~\eqref{eqn:Llocal}), one just \textit{truncates} the inverse mass expansion of the heavy propagator:
\begin{equation}
\frac{1}{{ - {D^2} - {M^2}}} =  - \frac{1}{{{M^2}}} - \frac{{ - {D^2}}}{{{M^4}}} +  \cdots  = {\left( {\frac{1}{{ - {D^2} - {M^2}}}} \right)_{{\text{truncated}}}} + {\left( {\frac{1}{{ - {D^2} - {M^2}}}} \right)_{{\text{rest}}}} . \label{eqn:DMsplit}
\end{equation}
For example, if the expansion is truncated at \(-1/M^2\), then we rewrite the propagator as
\begin{equation}
\frac{1}{-D^2-M^2} = -\frac{1}{M^2} + \frac{1}{M^2}\frac{-D^2}{-D^2-M^2}, \nonumber
\end{equation}
and identify the first and second pieces with \((-D^2-M^2)^{-1}_{\text{truncated}}\) and \((-D^2-M^2)^{-1}_{\text{rest}}\), respectively.

If one splits the heavy propagator into the ``truncated piece'' and the ``rest piece'', then term where \textit{all} heavy propagators in Eq.~\eqref{eqn:c1mixeddirectabbr} take the truncated piece must also come out of the second term in Eq.~\eqref{eqn:c1mixeddirect}. Hence, these are the ``local counterparts'' to drop. So the main task is just to identify the ``truncated piece'' ${\left( {\frac{1}{{ - {D^2} - {M^2}}}} \right)_{{\rm{truncated}}}}$ by examining how we went from $\mathcal{L}_\text{UV}\left(\phi, \Phi_c[\phi] \right)$ to $\mathcal{L}_\text{EFT}^{(0)}(\phi)$.

\subsubsection{Trace evaluation}\label{subsubsec:NewCDE}

Eq.~\eqref{eqn:c1mixeddirectabbr} is our master formula in this paper, with which one can calculate the piece ${c_{i,\text{mixed}}^{(1)}(M)}$ directly. To evaluate the functional determinant in Eq.~\eqref{eqn:c1mixeddirectabbr} one can, of course, use the PDE method described in Section~\ref{subsec:CDEReview}. However, it would be desirable to have a CDE method, which makes the evaluation more efficient.

As explained in Section~\ref{subsubsec:traditionalCDE}, the functional determinant of the elliptic operator can be evaluated by a CDE method. The method is intensively discussed in~\cite{Gaillard:1985,Cheyette:1987,Henning:2014wua,Drozd:2015rsp}, with a universal result of degenerate mass case represented in Eq.~\eqref{eqn:ellipticUniversal}. Unfortunately, the functional derivative matrix in Eq.~\eqref{eqn:c1mixeddirectabbr} is typically not an elliptic operator, due to the non-local nature of $\mathcal{L}_\text{UV}\left(\phi, \Phi_c[\phi] \right)$. Instead, its typical form is (as we shall see later in the examples)
\begin{align}
 - \frac{{{\delta ^2}{S_{{\rm{UV}}}}\left[ {\phi ,{\Phi _c}\left[ \phi  \right]} \right]}}{{\delta {\phi ^2}}} &= {D^2} + {m^2} + \sum\limits_n {{A_{n1}}\left( x \right)\left[ {\prod\limits_{i = 2}^n {\frac{1}{{ - {D^2} - {M^2}}}{A_{ni}}\left( x \right)} } \right]} \nonumber \\
 &= {D^2} + {m^2} + \left\{ {{A_{11}}\left( x \right) + {A_{21}}\left( x \right)\frac{1}{{ - {D^2} - {M^2}}}{A_{22}}\left( x \right)} \right. \nonumber \\
 & \qquad \qquad \left. { + {A_{31}}\left( x \right)\frac{1}{{ - {D^2} - {M^2}}}{A_{32}}\left( x \right)\frac{1}{{ - {D^2} - {M^2}}}{A_{33}}\left( x \right) +  \cdots } \right\} , \label{eqn:EllipticGen}
\end{align}
where $m^2$ is an IR scale, typically the mass of the light field $\phi$.\footnote{If the light field $\phi$ is massless, $m^2$ can be viewed as an IR regulator introduced for the calculation, and can be taken to zero in the end, as the end result is always free of IR divergence.} Therefore, we cannot make use of the universal result in Eq.~\eqref{eqn:ellipticUniversal}. Nevertheless, this kind of functional determinants can still be evaluated by a CDE method. The first step is to expand out the logarithm and convert it into a sum of functional traces
\begin{align}
 &\log \det \left( { - \frac{{{\delta ^2}{S_{{\rm{UV}}}}\left[ {\phi ,{\Phi _c}\left[ \phi  \right]} \right]}}{{\delta {\phi ^2}}}} \right) = {\rm{Tr}}\log \left( { - \frac{{{\delta ^2}{S_{{\rm{UV}}}}\left[ {\phi ,{\Phi _c}\left[ \phi  \right]} \right]}}{{\delta {\phi ^2}}}} \right) \nonumber \\
 &= {\rm{Tr}}\log \left[ {{D^2} + {m^2} + {A_{11}}\left( x \right) + {A_{21}}\left( x \right)\frac{1}{{ - {D^2} - {M^2}}}{A_{22}}\left( x \right) +  \cdots } \right] \nonumber \\
 &\supset {\rm{Tr}}\log \left[ {1 - \frac{1}{{ - {D^2} - {m^2}}}{A_{11}}\left( x \right) - \frac{1}{{ - {D^2} - {m^2}}}{A_{21}}\left( x \right)\frac{1}{{ - {D^2} - {M^2}}}{A_{22}}\left( x \right) +  \cdots } \right] \nonumber \\
 &=  - \sum\limits_{n = 1}^\infty  {\frac{1}{n}{\rm{Tr}}\left\{ {{{\left[ {\frac{1}{{ - {D^2} - {m^2}}}{A_{11}}\left( x \right) - \frac{1}{{ - {D^2} - {m^2}}}{A_{21}}\left( x \right)\frac{1}{{ - {D^2} - {M^2}}}{A_{22}}\left( x \right) +  \cdots } \right]}^n}} \right\}} . \label{eqn:TraceSeries}
\end{align}
The next step is to evaluate each of the functional traces in the last line above. These functional traces can be all captured by the following general form:
\begin{equation}
{\rm{Tr}}\left[ {\frac{{{{\left( { - {D^2}} \right)}^{{k_1}}}}}{{ - {D^2} - m_1^2}}{A_1}\left( x \right)\frac{{{{\left( { - {D^2}} \right)}^{{k_2}}}}}{{ - {D^2} - m_2^2}}{A_2}\left( x \right) \cdots \frac{{{{\left( { - {D^2}} \right)}^{{k_3}}}}}{{ - {D^2} - m_n^2}}{A_n}\left( x \right)} \right] , \label{eqn:TraceGeneral}
\end{equation}
where $k_i$ are integers. In Appendix~\ref{app:Trace}, we present a systematic CDE method, which allows one to evaluate any functional trace of this form.\footnote{Note that in the third line of deriving Eq.~\eqref{eqn:TraceSeries}, we have dropped a term $\Tr\log\left(D^2+m^2\right)$. This piece is not of our interest for the current scope, as it will always get dropped as the ``local counterpart'' later anyway, and hence does not contribute to $c_{i,\text{mixed}}^{(1)}$. However, if one actually wants to evaluate this piece, one can use the universal formula of Eq.~\eqref{eqn:ellipticUniversal} with \(U=0\). This is also discussed in Appendix~\ref{app:Trace}.}

\subsection{Demonstrating example: a toy scalar model}\label{subsec:ToyScalar}

In order to demonstrate the methodology described in Section~\ref{subsec:DirectGeneral} in the simplest context, let us consider a toy model with a light real scalar field $\phi$ and a heavy real scalar field $\Phi$, both of which are gauge singlets. The UV Lagrangian is
\begin{equation}
{{\cal L}_{{\text{UV}}}}\left( {\phi ,\Phi } \right) = \frac{1}{2}\Phi \left( { - {\partial ^2} - {M^2}} \right)\Phi  - \frac{\lambda }{{3!}}\Phi {\phi ^3} + \frac{1}{2}\phi \left( { - {\partial ^2} - {m^2}} \right)\phi  - \frac{\kappa }{{4!}}{\phi ^4} .
\end{equation}
For clarity of discussion, we ignore possible self-interactions of the heavy field \(\Ph\). We would like to match this theory one an EFT of \(\ph\) alone, $\mathcal{L}_\text{EFT}\left(\phi\right)$.

Let us focus on the dimension-four operator $\mathcal{O}_4\equiv\phi^4$ and the dimension-six operator $\mathcal{O}_6\equiv\phi^6$ with the Wilson coefficients $c_4$, $c_6$ normalized as \({{\cal L}_{{\text{EFT}}}}\left( \phi  \right) \supset {c_4}{\phi^4} + {c_6}{\phi^6}\). Specifically, we want to compute the piece $c_{4,\text{mixed}}^{(1)}$ and $c_{6,\text{mixed}}^{(1)}$, the contribution to which should involve the UV coupling $\lambda$.

Following the procedure described in Section~\ref{subsec:DirectGeneral}, we first compute the non-local Lagrangian ${{\cal L}_{{\rm{UV}}}}\left( {\phi ,{\Phi _c}\left[ \phi  \right]} \right)$:
\begin{align}
{\Phi _c}\left[ \phi  \right] &= \frac{1}{{ - {\partial ^2} - {M^2}}}\frac{\lambda }{6}{\phi ^3} , \\
{{\cal L}_{{\text{UV}}}}\left( {\phi ,{\Phi _c}\left[ \phi  \right]} \right) &= \frac{1}{2}\phi \left( { - {\partial ^2} - {m^2}} \right)\phi  - \frac{\kappa }{{4!}}{\phi ^4} - \frac{{{\lambda ^2}}}{{72}}{\phi ^3}\frac{1}{{ - {\partial ^2} - {M^2}}}{\phi ^3} . \label{eqn:toyLnonlocal}
\end{align}
Then we take its second variation with respect to the light field $\phi$
\begin{equation}
- \frac{{{\delta ^2}{S_{{\rm{UV}}}}\left[ {\phi ,{\Phi _c}\left[ \phi  \right]} \right]}}{{\delta {\phi ^2}}} = {\partial ^2} + {m^2} + \frac{\kappa }{2}{\phi ^2} + \frac{{{\lambda ^2}}}{{36}}\left[ {9{\phi ^2}\frac{1}{{ - {\partial ^2} - {M^2}}}{\phi ^2} + 6\phi {{\left( {\frac{1}{{ - {\partial ^2} - {M^2}}}{\phi ^3}} \right)}_x}} \right] .
\end{equation}
Note that here we have used a subscript ``$x$'' to mark a ``local'' term, \textit{i.e.} the state $\left|x\right\rangle$ in the functional space is an eigenstate of the expression inside the parentheses:
\begin{align}
{\left( {\frac{1}{{ - {\partial ^2} - {M^2}}}{\phi ^3}} \right)_x}\left| x \right\rangle  &= \left| x \right\rangle \left( {\frac{1}{{ - {\partial ^2} - {M^2}}}{\phi ^3}} \right) \nonumber \\
 &= \left| x \right\rangle \left[ -{\frac{1}{{{M^2}}}\phi ^3 + \frac{1}{{{M^4}}}{\partial ^2}\left( {{\phi ^3}} \right) -\frac{1}{M^6} \pd^4 \left(\ph^3\right) +  \cdots } \right] . \label{eqn:xfunctional}
\end{align}
The practical understanding of the notation \((\dots)_x\) is that derivatives act only within the parentheses. Now using Eq.~\eqref{eqn:c1mixeddirectabbr}, we get
\begin{align}
&S_{{\rm{EFT,mixed}}}^{\left( 1 \right)}\left[ \phi  \right] = \frac{i}{2}\log \det {\left( { - \frac{{{\delta ^2}{S_{{\rm{UV}}}}\left[ {\phi ,{\Phi _c}\left[ \phi  \right]} \right]}}{{\delta {\phi ^2}}}} \right)_{\rm{d}}} \nonumber \\
&\supset \frac{i}{2}\Tr\log {\left\{ {1 - \frac{1}{{ - {\partial ^2} - {m^2}}}\left\{ {\frac{\kappa }{2}{\phi ^2} + \frac{{{\lambda ^2}}}{{36}}\left[ {9{\phi ^2}\frac{1}{{ - {\partial ^2} - {M^2}}}{\phi ^2} + 6\phi {{\left( {\frac{1}{{ - {\partial ^2} - {M^2}}}{\phi ^3}} \right)}_x}} \right]} \right\}} \right\}_{\rm{d}}} \nonumber \\
&\supset  - \frac{i}{{144}}{\rm{Tr}}{\left\{ \begin{array}{l}
2{\lambda ^2}\frac{1}{{ - {\partial ^2} - {m^2}}}\left[ {9{\phi ^2}\frac{1}{{ - {\partial ^2} - {M^2}}}{\phi ^2} + 6\phi {{\left( {\frac{1}{{ - {\partial ^2} - {M^2}}}{\phi ^3}} \right)}_x}} \right]\\
 + {\lambda ^2}\kappa \frac{1}{{ - {\partial ^2} - {m^2}}}{\phi ^2}\frac{1}{{ - {\partial ^2} - {m^2}}}\left[ {9{\phi ^2}\frac{1}{{ - {\partial ^2} - {M^2}}}{\phi ^2} + 6\phi {{\left( {\frac{1}{{ - {\partial ^2} - {M^2}}}{\phi ^3}} \right)}_x}} \right]
\end{array} \right\}_{\rm{d}}} , \label{eqn:toysEFTUV}
\end{align}
where we have expanded the log and kept only the pieces that contain $\phi^4$, $\phi^6$ and involve $\lambda$.\footnote{The pieces that do not involve UV couplings will always be cancelled by the ``local counterparts''.}

Our next step is to identify and drop the ``local counterpart''. We see from Eq.~\eqref{eqn:toyLnonlocal} that if one were to compute ${\cal L}_{{\rm{EFT}}}^{\left( 0 \right)}\left( \phi  \right)$ up to mass dimension six, one would have arrived at
\begin{equation}
{\cal L}_{{\rm{EFT}}}^{\left( 0 \right)}\left( \phi  \right) \supset \frac{{{\lambda ^2}}}{{72{M^2}}}{\phi ^6} ,
\end{equation}
which amounts to taking
\begin{equation}
{\left( {\frac{1}{{ - {\partial ^2} - {M^2}}}} \right)_{{\text{truncated}}}} =  - \frac{1}{{{M^2}}} . \label{eqn:toySPtruncated}
\end{equation}
Therefore, we should split the heavy propagators in Eq.~\eqref{eqn:toysEFTUV} as
\begin{equation}
\frac{1}{{ - {\partial ^2} - {M^2}}} = {\left( {\frac{1}{{ - {\partial ^2} - {M^2}}}} \right)_{{\text{truncated}}}} + \frac{1}{{{M^2}}}\frac{{ - {\partial ^2}}}{{ - {\partial ^2} - {M^2}}} , \label{eqn:toySPsplit}
\end{equation}
and then drop the term with \textit{all} heavy propagators taking the first piece in the above. For example, for the first term in Eq.~\eqref{eqn:toysEFTUV}, the procedure is as following
\begin{align}
&\Tr{\left\{ {\frac{1}{{ - {\partial ^2} - {m^2}}}9{\phi ^2}\frac{1}{{ - {\partial ^2} - {M^2}}}{\phi ^2}} \right\}_{\rm{d}}} \nonumber \\
&= \Tr{\left\{ {\frac{1}{{ - {\partial ^2} - {m^2}}}9{\phi ^2}\left[ {{{\left( {\frac{1}{{ - {\partial ^2} - {M^2}}}} \right)}_{{\rm{truncated}}}} + \frac{1}{{{M^2}}}\frac{{ - {\partial ^2}}}{{ - {\partial ^2} - {M^2}}}} \right]{\phi ^2}} \right\}_{\rm{d}}} \nonumber \\
&= \Tr\left\{ {\frac{1}{{ - {\partial ^2} - {m^2}}}9{\phi ^2}\left[ {\frac{1}{{{M^2}}}\frac{{ - {\partial ^2}}}{{ - {\partial ^2} - {M^2}}}} \right]{\phi ^2}} \right\} ,
\end{align}
where in the second line we have split the heavy propagator according to Eqs.~\eqref{eqn:toySPtruncated} and~\eqref{eqn:toySPsplit}, and in the third line we have dropped its truncated piece. We reiterate that we need to drop the term with \textit{all} (instead of \textit{any}) heavy propagators taking the truncated piece.

After dropping all the ``local counterparts'', we get
\begin{align}
S_{{\rm{EFT,mixed}}}^{\left( 1 \right)}\left[ \phi  \right] &=  - \frac{i}{{144}}{\rm{Tr}}\left\{ \begin{array}{l}
\frac{{2{\lambda ^2}}}{{{M^2}}}\frac{1}{{ - {\partial ^2} - {m^2}}}\left[ {9{\phi ^2}\frac{{ - {\partial ^2}}}{{ - {\partial ^2} - {M^2}}}{\phi ^2} + 6\phi {{\left( {\frac{{ - {\partial ^2}}}{{ - {\partial ^2} - {M^2}}}{\phi ^3}} \right)}_x}} \right]\\
 + \frac{{{\lambda ^2}\kappa }}{{{M^2}}}\frac{1}{{ - {\partial ^2} - {m^2}}}{\phi ^2}\frac{1}{{ - {\partial ^2} - {m^2}}}\left[ {9{\phi ^2}\frac{{ - {\partial ^2}}}{{ - {\partial ^2} - {M^2}}}{\phi ^2} + 6\phi {{\left( {\frac{{ - {\partial ^2}}}{{ - {\partial ^2} - {M^2}}}{\phi ^3}} \right)}_x}} \right]
\end{array} \right\} \nonumber \\
 &\supset  - \frac{i}{{16}}{\rm{Tr}}\left\{ \begin{array}{l}
\frac{{2{\lambda ^2}}}{{{M^2}}}\frac{1}{{ - {\partial ^2} - {m^2}}}{\phi ^2}\frac{{ - {\partial ^2}}}{{ - {\partial ^2} - {M^2}}}{\phi ^2}\\
 + \frac{{{\lambda ^2}\kappa }}{{{M^2}}}\frac{1}{{ - {\partial ^2} - {m^2}}}{\phi ^2}\frac{1}{{ - {\partial ^2} - {m^2}}}{\phi ^2}\frac{{ - {\partial ^2}}}{{ - {\partial ^2} - {M^2}}}{\phi ^2}
\end{array} \right\} , \label{eqn:tracetoy}
\end{align}
where in the second line we have dropped the ``$x$'' pieces, because they start with two derivatives on $\phi^3$ (the ``\(x\)'' piece is evaluated similar to Eq.~\eqref{eqn:xfunctional}), which would not contribute to $\mathcal{O}_4=\phi^4$ or $\mathcal{O}_6=\phi^6$ in $S_{{\rm{EFT,mixed}}}^{\left( 1 \right)}\left[ \phi  \right]$.

The rest of the calculation is just to evaluate the functional traces in Eq.~\eqref{eqn:tracetoy}, which can be done by a CDE method, as we explain in Appendix~\ref{app:Trace}. The end results are
\begin{align}
{\rm{Tr}}\left( {\frac{1}{{ - {\partial ^2} - {m^2}}}{\phi ^2}\frac{{ - {\partial ^2}}}{{ - {\partial ^2} - {M^2}}}{\phi ^2}} \right) &\supset \frac{i}{{{{\left( {4\pi } \right)}^2}}}\int {{d^4}x{M^2}{\mkern 1mu} {\phi ^4}} \\
{\rm{Tr}}\left( {\frac{1}{{ - {\partial ^2} - {m^2}}}{\phi ^2}\frac{1}{{ - {\partial ^2} - {m^2}}}{\phi ^2}\frac{{ - {\partial ^2}}}{{ - {\partial ^2} - {M^2}}}{\phi ^2}} \right) &\supset \frac{i}{{{{\left( {4\pi } \right)}^2}}}\int {{d^4}x{\mkern 1mu} {\phi ^6}} . \label{eqn:tracetoyresult}
\end{align}
Therefore we get the one-loop mixed piece of the Wilson coefficient as
\begin{align}
S_{{\rm{EFT,mixed}}}^{\left( 1 \right)}\left[ \phi  \right] &\supset \int {{d^4}x{\mkern 1mu} \frac{1}{{{{\left( {4\pi } \right)}^2}}}\left( {\frac{{{\lambda ^2}}}{8}{\phi ^4} + \frac{{{\lambda ^2}\kappa }}{{16{M^2}}}{\phi ^6}} \right)} , \\
c_{{\rm{4,mixed}}}^{\left( 1 \right)}\left( {\mu  = M} \right) &= \frac{1}{{{{\left( {4\pi } \right)}^2}}}\frac{{{\lambda ^2}}}{8} , \\
c_{{\rm{6,mixed}}}^{\left( 1 \right)}\left( {\mu  = M} \right) &= \frac{1}{{{{\left( {4\pi } \right)}^2}}}\frac{{{\lambda ^2}\kappa }}{{16{M^2}}} .
\end{align}

\subsection{Example: the Standard Model with a heavy electroweak triplet scalar}\label{subsec:TripletScalar}

Now let us work out a model that is of some phenomenological relevance---the Standard Model with a heavy electroweak triplet scalar $\Phi^a$. The UV Lagrangian is
\begin{equation}
{{\cal L}_{{\rm{UV}}}} = {{\cal L}_{{\rm{SM}}}}\left( \phi  \right) + \frac{1}{2}{\Phi ^a}\left( { - {D^2} - {M^2}} \right){\Phi ^a} + 2\kappa {H^\dag }{t^a}H{\Phi ^a} - \eta {\left| H \right|^2}{\Phi ^a}{\Phi ^a} ,
\end{equation}
where $t^a=\sigma^a/2$ is the $SU(2)_W$ generator in the fundamental representation, and we have used ``$\phi$'' to collectively denote all the SM fields. As in the previous example, we ignore the self-interactions of the heavy field \(\Ph^a\) for simplicity. The Higgs sector of the SM is 
\begin{equation}
{{\cal L}_{{\rm{SM}}}}\left( \phi  \right) \supset {H^\dag }\left( { - {D^2} - {m^2}} \right)H - \frac{\lambda }{4}{\left| H \right|^4} ,
\end{equation}
where $m^2$ is just an IR regulator. We will take $m^2 \to 0$ at the end of the calculation. We are interested in the dimension-six operators ${{\cal O}_H} \equiv \frac{1}{2}{\left( {{D_\mu }{{\left| H \right|}^2}} \right)^2}$, ${{\cal O}_{HD}} \equiv {\left| {{H^\dag }DH} \right|^2}$, and ${{\cal O}_R} \equiv {\left| H \right|^2}{\left| {{D_\mu }H} \right|^2}$, with their Wilson coefficients $c_H^{}$, $c_{HD}^{}$, and $c_R^{}$ normalized as
\begin{equation}
{{\cal L}_{{\rm{EFT}}}}\left( \phi  \right) \supset {c_H}{{\cal O}_H} + {c_{HD}}{{\cal O}_{HD}} + {c_R}{{\cal O}_R} .
\end{equation}
These effective operators all have four powers of the Higgs field and two powers of the covariant derivative. We are specifically interested in the one-loop mixed pieces $c_{H{\rm{,mixed}}}^{\left( 1 \right)}$, $c_{HD{\rm{,mixed}}}^{\left( 1 \right)}$, and $c_{R{\rm{,mixed}}}^{\left( 1 \right)}$.\footnote{The \(c_{i,\text{heavy}}^{(1)}(M)\) can be obtained with the universal formula~\eqref{eqn:ellipticUniversal}; this example is shown explicitly in~\cite{Henning:2014wua}.}

Again following the procedure described in Section~\ref{subsec:DirectGeneral}, we first compute the non-local Lagrangian ${{\cal L}_{{\rm{UV}}}}\left( {\phi ,\Phi _c^a\left[ \phi  \right]} \right)$
\begin{align}
\Phi _c^a\left[ \phi  \right] &=  - \frac{1}{{ - {D^2} - {M^2} - 2\eta {{\left| H \right|}^2}}}2\kappa {H^\dag }{t^a}H , \\
{{\cal L}_{{\rm{UV}}}}\left( {\phi ,\Phi _c^a\left[ \phi  \right]} \right) &= {{\cal L}_{{\rm{SM}}}}\left( \phi  \right) - 2{\kappa ^2}{H^\dag }{t^a}H\frac{1}{{ - {D^2} - {M^2} - 2\eta {{\left| H \right|}^2}}}{H^\dag }{t^a}H \nonumber \\
 &= \left[ \begin{array}{l}
{{\cal L}_{{\rm{SM}}}}\left( \phi  \right) - 2{\kappa ^2}{H^\dag }{t^a}H\frac{1}{{ - {D^2} - {M^2}}}{H^\dag }{t^a}H\\
 - 4{\kappa ^2}\eta {H^\dag }{t^a}H\frac{1}{{ - {D^2} - {M^2}}}{\left| H \right|^2}\frac{1}{{ - {D^2} - {M^2}}}{H^\dag }{t^a}H
\end{array} \right] + {\cal O}\left( {{H^8}} \right) . \label{eqn:TSnlL}
\end{align}
Here we have thrown away the terms with more than six powers in $H$, because upon second variation, such terms would give effective operators with more than four powers in $H$, which cannot contribute to $\mathcal{O}_{H}$, $\mathcal{O}_{HD}$, or $\mathcal{O}_R$. It is easy to work out the tree-level Wilson coefficients from here. They are given by expanding ${{\cal L}_{{\rm{UV}}}}\left( {\phi ,\Phi _c^a\left[ \phi  \right]} \right)$ into local operators:
\begin{align}
{{\cal L}_{{\rm{UV}}}}\left( {\phi ,\Phi _c^a\left[ \phi  \right]} \right) &\supset  - 2{\kappa ^2}{H^\dag }{t^a}H\frac{1}{{ - {D^2} - {M^2}}}{H^\dag }{t^a}H \nonumber \\
 &\supset  - 2{\kappa ^2}{H^\dag }{t^a}H\left( { - \frac{1}{{{M^2}}} - \frac{{ - {D^2}}}{{{M^4}}}} \right){H^\dag }{t^a}H \nonumber \\
 &\supset \frac{{2{\kappa ^2}}}{{{M^4}}}{H^\dag }{t^a}H\left( { - {D^2}} \right){H^\dag }{t^a}H = \frac{{2{\kappa ^2}}}{{{M^4}}}{\left[ {D\left( {{H^\dag }{t^a}H} \right)} \right]^2} \nonumber \\
 &= \frac{{2{\kappa ^2}}}{{{M^4}}}{\left[ {\left( {D{H^\dag }} \right){t^a}H + {H^\dag }{t^a}\left( {DH} \right)} \right]^2} \nonumber \\
 &= \frac{{2{\kappa ^2}}}{{{M^4}}}\left[ {\frac{1}{4}{{\left( {{D_\mu }{{\left| H \right|}^2}} \right)}^2} - {{\left| {{H^\dag }{D_\mu }H} \right|}^2} + {{\left| H \right|}^2}{{\left| {{D_\mu }H} \right|}^2}} \right] ,
\end{align}
from which we can identify
\begin{equation}
c_H^{\left( 0 \right)} = \frac{{2{\kappa ^2}}}{{{M^4}}}\frac{1}{2}\;\;\;{\mkern 1mu} ,\;\;\;{\mkern 1mu} c_{HD}^{\left( 0 \right)} =  - \frac{{2{\kappa ^2}}}{{{M^4}}}\;\;\;{\mkern 1mu} ,\;\;\;{\mkern 1mu} c_R^{\left( 0 \right)} = \frac{{2{\kappa ^2}}}{{{M^4}}} . \label{eqn:TSctree}
\end{equation}

To continue with computing the one-loop mixed pieces, the next step is to take the second variation of the action $S_{{\rm{UV}}}\left[ \phi ,\Phi _c^a\left[ \phi  \right] \right]$ with respect to \textit{all} the light fields $\phi$ in SM, which includes the Higgs doublet, the gauge bosons, as well as the Fermions. This will give us a big functional matrix, and we need to evaluate its determinant. This evaluation is a big task to do. However, we know in the end, we will only extract the terms that contain $\mathcal{O}_{H}$, $\mathcal{O}_{HD}$, and $\mathcal{O}_{R}$ with coefficient involving the UV couplings, \textit{i.e.} $\kappa$ or $\eta$. This will greatly simplify the task. Furthermore, if one ignores the pieces in $c_{H{\rm{,mixed}}}^{\left( 1 \right)}$, $c_{HD{\rm{,mixed}}}^{\left( 1 \right)}$, and $c_{R{\rm{,mixed}}}^{\left( 1 \right)}$ that involves the SM gauge couplings, then taking only the sub-matrix from the Higgs field variation is sufficient. For purpose of simplicity, we will only evaluate the functional determinant of this sub-matrix, and compute the terms in Wilson coefficients that do not involve the SM gauge couplings. From Eq.~\eqref{eqn:TSnlL}, we get this sub-matrix as
\begin{align}
{\delta ^2}{\mathcal{L}_{{\rm{UV}}}}\left[ {\phi ,\Phi _c^a\left[ \phi  \right]} \right] &= \frac{1}{2}\left( {\begin{array}{*{20}{c}}
{\delta {H^\dag }}&{\delta {H^T}}
\end{array}} \right)\left[ \begin{array}{l}
 - {D^2} - {m^2} - \frac{\lambda }{2}\left( {\begin{array}{*{20}{c}}
{{a_1}}&{{b_1}}\\
{b_1^*}&{a_1^T}
\end{array}} \right)\\
 - 4{\kappa ^2}\left( {\begin{array}{*{20}{c}}
{{a_2}}&{{b_2}}\\
{b_2^*}&{a_2^T}
\end{array}} \right) - 8{\kappa ^2}\eta \left( {\begin{array}{*{20}{c}}
{{a_3}}&{{b_3}}\\
{b_3^*}&{a_3^T}
\end{array}} \right)
\end{array} \right]\left( {\begin{array}{*{20}{c}}
{\delta H}\\
{\delta {H^*}}
\end{array}} \right) , \nonumber
\end{align}
so that
\begin{align}
 - \frac{{{\delta ^2}{S_{{\rm{UV}}}}\left[ {\phi ,\Phi _c^a\left[ \phi  \right]} \right]}}{{\delta {\phi ^2}}} &\supset {D^2} + {m^2} + \frac{\lambda }{2}\left( {\begin{array}{*{20}{c}}
{{a_1}}&{{b_1}}\\
{b_1^*}&{a_1^T}
\end{array}} \right) + 4{\kappa ^2}\left( {\begin{array}{*{20}{c}}
{{a_2}}&{{b_2}}\\
{b_2^*}&{a_2^T}
\end{array}} \right) + 8{\kappa ^2}\eta \left( {\begin{array}{*{20}{c}}
{{a_3}}&{{b_3}}\\
{b_3^*}&{a_3^T}
\end{array}} \right) . \nonumber
\end{align}
where $a_1$, $b_1$, $a_2$, $b_2$, $a_3$, and $b_3$ are all functionals as well as $2\times2$ matrices in the $SU(2)_W$ space. Their detailed expressions are
\begin{align*}
{a_1} &= {\left| H \right|^2} + H{H^\dag } , \\ 
{b_1} &= H{H^T} , \\ 
{a_2} &= {t^a}{\left( {\frac{1}{{ - {D^2} - {M^2}}}{H^\dag }{t^a}H} \right)_x} + {t^a}H\frac{1}{{ - {D^2} - {M^2}}}{H^\dag }{t^a} , \\
{b_2} &= {t^a}H\frac{1}{{ - {D^2} - {M^2}}}{H^T}{t^{a*}} ,\\
{a_3} &= \left[ \begin{array}{l}
{t^a}{\left( {\frac{1}{{ - {D^2} - {M^2}}}{{\left| H \right|}^2}\frac{1}{{ - {D^2} - {M^2}}}{H^\dag }{t^a}H} \right)_x} + {t^a}H\frac{1}{{ - {D^2} - {M^2}}}{\left| H \right|^2}\frac{1}{{ - {D^2} - {M^2}}}{H^\dag }{t^a} \\
 + {t^a}H\frac{1}{{ - {D^2} - {M^2}}}{H^\dag }{\left( {\frac{1}{{ - {D^2} - {M^2}}}{H^\dag }{t^a}H} \right)_x} + {\left( {\frac{1}{{ - {D^2} - {M^2}}}{H^\dag }{t^a}H} \right)_x}H\frac{1}{{ - {D^2} - {M^2}}}{H^\dag }{t^a} \\
 + \frac{1}{2}{\left( {\frac{1}{{ - {D^2} - {M^2}}}{H^\dag }{t^a}H} \right)_x}{\left( {\frac{1}{{ - {D^2} - {M^2}}}{H^\dag }{t^a}H} \right)_x}
\end{array} \right] , \\ 
{b_3} &= {t^a}H\frac{1}{{ - {D^2} - {M^2}}}{\left| H \right|^2}\frac{1}{{ - {D^2} - {M^2}}}{H^T}{t^{a*}} + 2{t^a}H\frac{1}{{ - {D^2} - {M^2}}}{H^T}{\left( {\frac{1}{{ - {D^2} - {M^2}}}{H^\dag }{t^a}H} \right)_x} . 
\end{align*}
Now using Eq.~\eqref{eqn:c1mixeddirectabbr}, we get
\begin{align}
S_{{\rm{EFT,mixed}}}^{\left( 1 \right)}\left[ \phi  \right] &= \frac{i}{2}\log \det {\left[ {{D^2} + {m^2} + \frac{\lambda }{2}\left( {\begin{array}{*{20}{c}}
{{a_1}}&{{b_1}}\\
{b_1^*}&{a_1^T}
\end{array}} \right) + 4{\kappa ^2}\left( {\begin{array}{*{20}{c}}
{{a_2}}&{{b_2}}\\
{b_2^*}&{a_2^T}
\end{array}} \right) + 8{\kappa ^2}\eta \left( {\begin{array}{*{20}{c}}
{{a_3}}&{{b_3}}\\
{b_3^*}&{a_3^T}
\end{array}} \right)} \right]_{\rm{d}}} \nonumber \\
 &\supset \frac{i}{2}{\rm{Tr}}\log {\left\{ {1 - \frac{1}{{ - {D^2} - {m^2}}}\left[ {\frac{\lambda }{2}\left( {\begin{array}{*{20}{c}}
{{a_1}}&{{b_1}}\\
{b_1^*}&{a_1^T}
\end{array}} \right) + 4{\kappa ^2}\left( {\begin{array}{*{20}{c}}
{{a_2}}&{{b_2}}\\
{b_2^*}&{a_2^T}
\end{array}} \right) + 8{\kappa ^2}\eta \left( {\begin{array}{*{20}{c}}
{{a_3}}&{{b_3}}\\
{b_3^*}&{a_3^T}
\end{array}} \right)} \right]} \right\}_{\rm{d}}} , \label{eqn:TSSeff}
\end{align}
Next we would like to expand out and truncate the log. We notice that ${a_1},{b_1},{a_2},{b_2}\sim{H^2}$ and ${a_3},{b_3}\sim{H^4}$, and we want to keep terms with four powers in $H$, and also involve $\kappa$ or $\eta$. This restricts us to three terms
\begin{equation}
S_{{\rm{EFT,mixed}}}^{\left( 1 \right)}\left[ \phi  \right] \supset \frac{i}{2}{\rm{Tr}}{\left[ \begin{array}{l}
 - 2\lambda {\kappa ^2}\frac{1}{{ - {D^2} - {m^2}}}\left( {\begin{array}{*{20}{c}}
{{a_1}}&{{b_1}}\\
{b_1^*}&{a_1^T}
\end{array}} \right)\frac{1}{{ - {D^2} - {m^2}}}\left( {\begin{array}{*{20}{c}}
{{a_2}}&{{b_2}}\\
{b_2^*}&{a_2^T}
\end{array}} \right)\\
 - 8{\kappa ^4}\frac{1}{{ - {D^2} - {m^2}}}\left( {\begin{array}{*{20}{c}}
{{a_2}}&{{b_2}}\\
{b_2^*}&{a_2^T}
\end{array}} \right)\frac{1}{{ - {D^2} - {m^2}}}\left( {\begin{array}{*{20}{c}}
{{a_2}}&{{b_2}}\\
{b_2^*}&{a_2^T}
\end{array}} \right)\\
 - \frac{{8{\kappa ^2}\eta }}{{ - {D^2} - {m^2}}}\left( {\begin{array}{*{20}{c}}
{{a_3}}&{{b_3}}\\
{b_3^*}&{a_3^T}
\end{array}} \right)
\end{array} \right]_{\rm{d}}} = {S_1} + {S_2} + {S_3} , 
\end{equation}
with $S_1$, $S_2$, and $S_3$ defined as
\begin{subequations}
\label{eqn:S123}
\begin{align}
{S_1} &\equiv \frac{i}{2}{\rm{Tr}}{\left[ { - 2\lambda {\kappa ^2}\frac{1}{{ - {D^2} - {m^2}}}\left( {\begin{array}{*{20}{c}}
{{a_1}}&{{b_1}}\\
{b_1^*}&{a_1^T}
\end{array}} \right)\frac{1}{{ - {D^2} - {m^2}}}\left( {\begin{array}{*{20}{c}}
{{a_2}}&{{b_2}}\\
{b_2^*}&{a_2^T}
\end{array}} \right)} \right]_{\rm{d}}} , \label{eqn:S1} \\
{S_2} &\equiv \frac{i}{2}{\rm{Tr}}{\left[ { - 8{\kappa ^4}\frac{1}{{ - {D^2} - {m^2}}}\left( {\begin{array}{*{20}{c}}
{{a_2}}&{{b_2}}\\
{b_2^*}&{a_2^T}
\end{array}} \right)\frac{1}{{ - {D^2} - {m^2}}}\left( {\begin{array}{*{20}{c}}
{{a_2}}&{{b_2}}\\
{b_2^*}&{a_2^T}
\end{array}} \right)} \right]_{\rm{d}}} , \label{eqn:S2} \\
{S_3} &\equiv \frac{i}{2}{\rm{Tr}}{\left[ { - \frac{{8{\kappa ^2}\eta }}{{ - {D^2} - {m^2}}}\left( {\begin{array}{*{20}{c}}
{{a_3}}&{{b_3}}\\
{b_3^*}&{a_3^T}
\end{array}} \right)} \right]_{\rm{d}}} . \label{eqn:S3}
\end{align}
\end{subequations}

Our next step is to identify and drop the ``local counterparts'', and then evaluate these functional traces using the CDE method described in Appendix~\ref{app:Trace}. To identify the ``local counterparts'', we notice that computing $c_{HD}^{(0)}$ from Eq.~\eqref{eqn:TSnlL} amounts to taking
\begin{equation}
{\left( {\frac{1}{{ - {D^2} - {M^2}}}} \right)_{{\rm{truncated}}}} =  - \frac{1}{{{M^2}}} - \frac{{ - {D^2}}}{{{M^4}}} .
\end{equation}
So we should split the heavy propagator as
\begin{equation}
\frac{1}{{ - {D^2} - {M^2}}} = {\left( {\frac{1}{{ - {D^2} - {M^2}}}} \right)_{{\rm{truncated}}}} + \frac{1}{{{M^4}}}\frac{{{D^4}}}{{ - {D^2} - {M^2}}} , \label{eqn:TSPsplit}
\end{equation}
and all heavy propagators taking the first term in this splitting should be the ``local counterpart'' to drop. We leave the details of evaluating $S_1$, $S_2$, and $S_3$ to Appendix~\ref{app:TripletScalar}. The end results are (see Eqs.~\eqref{eqn:S1app}-\eqref{eqn:S3app})
\begin{align}
{S_1} &\supset \frac{1}{{{{\left( {4\pi } \right)}^2}}}\frac{{\lambda {\kappa ^2}}}{{{M^4}}}\int {{d^4}x\left[ {\frac{{13}}{8}{{\left( {{D_\mu }{{\left| H \right|}^2}} \right)}^2} - \frac{3}{2}{{\left| {{H^\dag }{D_\mu }H} \right|}^2} + \frac{{25}}{4}{{\left| H \right|}^2}{{\left| {{D_\mu }H} \right|}^2}} \right]} , \\
{S_2} &\supset \frac{1}{{{{\left( {4\pi } \right)}^2}}}\frac{{{\kappa ^4}}}{{{M^6}}}\int {{d^4}x\left[ { - 2{{\left( {{D_\mu }{{\left| H \right|}^2}} \right)}^2} - {{\left| {{H^\dag }{D_\mu }H} \right|}^2} - \frac{{21}}{2}{{\left| H \right|}^2}{{\left| {{D_\mu }H} \right|}^2}} \right]} , \\
{S_3} &\supset \frac{1}{{{{\left( {4\pi } \right)}^2}}}\frac{{{\kappa ^2}\eta }}{{{M^4}}}\int {{d^4}x\left[ { - 7{{\left( {{D_\mu }{{\left| H \right|}^2}} \right)}^2} + 16{{\left| {{H^\dag }{D_\mu }H} \right|}^2} - 21{{\left| H \right|}^2}{{\left| {{D_\mu }H} \right|}^2}} \right]} .
\end{align}
Combining the above, we get
\begin{align}
c_{H{\rm{,mixed}}}^{\left( 1 \right)}\left( {\mu  = M} \right) &= \frac{1}{{{{\left( {4\pi } \right)}^2}}}\frac{{{\kappa ^2}}}{{{M^4}}}\left( {\frac{{13}}{4}\lambda  - 14\eta  - 4\frac{{{\kappa ^2}}}{{{M^2}}}} \right) , \\
c_{HD{\rm{,mixed}}}^{\left( 1 \right)}\left( {\mu  = M} \right) &= \frac{1}{{{{\left( {4\pi } \right)}^2}}}\frac{{{\kappa ^2}}}{{{M^4}}}\left( { - \frac{3}{2}\lambda  + 16\eta  - \frac{{{\kappa ^2}}}{{{M^2}}}} \right) , \\
c_{R{\rm{,mixed}}}^{\left( 1 \right)}\left( {\mu  = M} \right) &= \frac{1}{{{{\left( {4\pi } \right)}^2}}}\frac{{{\kappa ^2}}}{{{M^4}}}\left( {\frac{{25}}{4}\lambda  - 21\eta  - \frac{{21}}{2}\frac{{{\kappa ^2}}}{{{M^2}}}} \right) .
\end{align}
The result of $c_{HD,\text{mixed}}^{(1)}(\mu=M)$ agrees with that in the literature, \textit{e.g.} Eq.~(31) in~\cite{delAguila:2016zcb}.

In some literatures, the wavefunction correction effect is also included in the one-loop Wilson coefficients. This piece is also easily computed with our method. One just needs to compute the Wilson coefficient $c_K$ of the kinetic operator ${{\cal O}_K} \equiv {\left| {DH} \right|^2}$. 
To one-loop order, the only contribution to \(c_K\) comes from $c_{K{\rm{,mixed}}}^{\left( 1 \right)}$. To compute this contribution, we just go back to Eq.~\eqref{eqn:TSSeff} and keep the terms with two powers in $H$ while expanding and truncating the log. This gives
\begin{align}
S_{{\rm{EFT,mixed}}}^{\left( 1 \right)}\left[ \phi  \right] &\supset \frac{i}{2}{\rm{Tr}}\log {\left\{ {1 - \frac{1}{{ - {D^2} - {m^2}}}\left[ {\frac{\lambda }{2}\left( {\begin{array}{*{20}{c}}
{{a_1}}&{{b_1}}\\
{b_1^*}&{a_1^T}
\end{array}} \right) + 4{\kappa ^2}\left( {\begin{array}{*{20}{c}}
{{a_2}}&{{b_2}}\\
{b_2^*}&{a_2^T}
\end{array}} \right) + 8{\kappa ^2}\eta \left( {\begin{array}{*{20}{c}}
{{a_3}}&{{b_3}}\\
{b_3^*}&{a_3^T}
\end{array}} \right)} \right]} \right\}_{\rm{d}}} \nonumber \\
 &\supset  - 2i{\kappa ^2}{\rm{Tr}}{\left[ {\frac{1}{{ - {D^2} - {m^2}}}\left( {\begin{array}{*{20}{c}}
{{a_2}}&{{b_2}}\\
{b_2^*}&{a_2^T}
\end{array}} \right)} \right]_{\rm{d}}} \equiv {S_K} . \label{eqn:SK}
\end{align}

As before, one proceeds with first dropping the ``local counterparts'' and then evaluating the functional trace using CDE. We give the details of evaluating $S_K$ in Appendix~\ref{app:TripletScalar}. The end result is (see Eq.~\eqref{eqn:SKapp})
\begin{equation}
{S_K} \supset \frac{1}{{{{\left( {4\pi } \right)}^2}}}\frac{{{\kappa ^2}}}{{{M^2}}}\int {{d^4}x\left[ {\frac{3}{2}{{\left| {DH} \right|}^2}} \right]} ,
\end{equation}
which gives
\begin{equation}
{{\cal L}_{{\rm{EFT}}}}\left( \phi  \right) \supset \frac{1}{{{{\left( {4\pi } \right)}^2}}}\frac{3}{2}\frac{{{\kappa ^2}}}{{{M^2}}}{\left| {DH} \right|^2} .
\end{equation}
This result agrees with that in the literature, \textit{e.g.} Eq.~(32) in~\cite{delAguila:2016zcb}. As explained in~\cite{delAguila:2016zcb}, sometimes people rescale the field $H$ to absorb this correction of the kinetic term, in order to make the kinetic term canonical:
\begin{equation}
H \to \left[ {1 - \frac{1}{{{{\left( {4\pi } \right)}^2}}}\frac{3}{4}\frac{{{\kappa ^2}}}{{{M^2}}}} \right]H .
\end{equation}
Rescaling $H$ will also rescale $\mathcal{O}_{H}$, $\mathcal{O}_{HD}$, and $\mathcal{O}_{R}$; hence, the one-loop pieces of their Wilson coefficients are modified:
\begin{align*}
\left\{ {{{\cal O}_H},{{\cal O}_{HD}},{{\cal O}_R}} \right\} &\to \left[ {1 - \frac{1}{{{{\left( {4\pi } \right)}^2}}}3\frac{{{\kappa ^2}}}{{{M^2}}}} \right]\left\{ {{{\cal O}_H},{{\cal O}_{HD}},{{\cal O}_R}} \right\} , \\
\left\{ {c_{H{\rm{,mixed}}}^{\left( 1 \right)},c_{HD{\rm{,mixed}}}^{\left( 1 \right)},c_{R{\rm{,mixed}}}^{\left( 1 \right)}} \right\} &\to \left\{ {c_{H{\rm{,mixed}}}^{\left( 1 \right)},c_{HD{\rm{,mixed}}}^{\left( 1 \right)},c_{R{\rm{,mixed}}}^{\left( 1 \right)}} \right\} - \frac{1}{{{{\left( {4\pi } \right)}^2}}}3\frac{{{\kappa ^2}}}{{{M^2}}}\left\{ {c_H^{\left( 0 \right)},c_{HD}^{\left( 0 \right)},c_R^{\left( 0 \right)}} \right\} .
\end{align*}
Using the tree-level matching results in Eq.~\eqref{eqn:TSctree}, we get
\begin{align}
c_{H{\rm{,mixed}}}^{\left( 1 \right)}\left( {\mu  = M} \right) &\to \frac{1}{{{{\left( {4\pi } \right)}^2}}}\frac{{{\kappa ^2}}}{{{M^4}}}\left( {\frac{{13}}{4}\lambda  - 14\eta  - 7\frac{{{\kappa ^2}}}{{{M^2}}}} \right) , \\
c_{HD{\rm{,mixed}}}^{\left( 1 \right)}\left( {\mu  = M} \right) &\to \frac{1}{{{{\left( {4\pi } \right)}^2}}}\frac{{{\kappa ^2}}}{{{M^4}}}\left( { - \frac{3}{2}\lambda  + 16\eta  + 5\frac{{{\kappa ^2}}}{{{M^2}}}} \right) , \label{eqn:TScshifted} \\
c_{R{\rm{,mixed}}}^{\left( 1 \right)}\left( {\mu  = M} \right) &\to \frac{1}{{{{\left( {4\pi } \right)}^2}}}\frac{{{\kappa ^2}}}{{{M^4}}}\left( {\frac{{25}}{4}\lambda  - 21\eta  - \frac{{33}}{2}\frac{{{\kappa ^2}}}{{{M^2}}}} \right) .
\end{align}
Eq.~\eqref{eqn:TScshifted} agrees with~\cite{Khandker:2012zu}, upon neglecting the SM gauge coupling piece. We see that functional method has the strength to compute different Wilson coefficients at the same time.

%% file: sec_Yukawa.tex
To highlight aspects of the general discussions on matching from the previous sections, here we consider a simple example of matching a Yukawa theory with a heavy scalar $\Phi$ onto a low energy EFT containing the four-fermion interaction \(\scO_S = (\psb\ps)^2/2\). We will show how to do the matching in both the ``traditional'' procedure of Sec.~\ref{subsec:MatchingTraditional} as well as the ``direct'' approach of Sec.~\ref{sec:MatchingDirect}.

The UV Lagrangian is given by
\begin{equation}
\scL_\text{UV}(\Ph,\ps) = \frac{1}{2}\Ph \big(-\pd^2 -  M^2\big) \Ph + \psb\big(i\pds - m)\ps - \l \Ph \psb\ps ,
\label{eq:L_UV_yuk}
\end{equation}
with the hierarchy \(M \gg m\). For simplicity, in this discussion we will ignore the \(\Ph^4\) self-interaction of the heavy scalar. At the scale \(M\), we match it onto the EFT given by
\begin{align}
\scL_\text{EFT}(\ps) &= \psb\big(i\pds - m)\ps + \frac{c_S^{}}{2}\big(\psb\ps\big)^2 +\dots ,
\end{align}
where the dots indicate other higher-dimensional operators.

We give some detail to the explicit computations to provide a full-picture of the procedure. So as not to lose the physics in the math, let us summarize the computations:
\begin{itemize}
\item In the ``traditional'' approach, we first compute the 1LPI effective actions of the UV theory and the EFT and then equate them to obtain the Wilson coefficients. Subsection~\ref{sec:yukawa_1LPI} is devoted to the computation of the 1LPI effective actions.
\item In Subsection~\ref{sec:yukawa_match} we apply the matching condition, Eq.~\eqref{eqn:MatchingCriterion}, to determine the Wilson coefficients.
\item In Subsection~\ref{sec:yukawa_direct} we show the direct computation of the Wilson coefficient $c_{S,\text{mixed}}^{(1)}$ using the non-local action $S_\text{UV}\left(\psi,\Phi_c[\psi]\right)$, following the general procedure described in Section~\ref{sec:MatchingDirect}.
\end{itemize}
Finally, we mention that the lecture notes~\cite{Skiba:2010xn} analyze this Yukawa model in detail using diagrammatic methods, and serve as a complementary text to this section.

\subsection{Computing the 1LPI effective actions}\label{sec:yukawa_1LPI}

In this subsection we compute the 1LPI effective actions \(\G_{\text{L,UV}}\) and \(\G_{\text{L,EFT}}\) at tree and one-loop level. In obtaining the effective actions we will expand in powers of \(m^2/M^2\) and keep only the leading order in this expansion. For loop calculations we work in the \(\overline{\text{MS}}\) renormalization scheme.

\subsubsection{Tree-level}

As discussed around Eq.~\eqref{eqn:1LPI0UV}, the tree-level 1LPI effective action of the UV theory is simply obtained by solving \(\d S_{\text{UV}}/\d \Ph = 0\) to obtain
\begin{equation}
{\Phi _c}\left[ \psi  \right] = \frac{1}{{ - {\partial ^2} - {M^2}}}\lambda \bar \psi \psi , \label{eq:Ph_c}
\end{equation}
and then plugging this back into the Lagrangian,
\begin{equation}
\Gamma _{{\rm{L,UV}}}^{\left( 0 \right)}\left[ \psi  \right] = {S_{{\rm{UV}}}}\left[ {\psi ,{\Phi _c}\left[ \psi  \right]} \right] = \int {{d^4}x\left[ {\bar \psi \left( {i\slashed \partial  - m} \right)\psi  - \frac{1}{2}{\lambda ^2}\bar \psi \psi \frac{1}{{ - {\partial ^2} - {M^2}}}\bar \psi \psi } \right]} . \label{eq:yuk_UV_tree}
\end{equation}
For the EFT, the 1LPI effective action is simply the tree-level piece of the EFT action (Eq.~\eqref{eqn:1LPI0EFT}):
\begin{equation}
\Gamma _{{\rm{L,EFT}}}^{\left( 0 \right)}\left[ \psi  \right] = S_{{\rm{EFT}}}^{\left( 0 \right)}\left[ \psi  \right] = \int {{d^4}x\left[ {\bar \psi \left( {i\slashed \partial  - m} \right)\psi  + \frac{1}{2}{c_S^{(0)}}{{\left( {\bar \psi \psi } \right)}^2}} \right]} . \label{eq:yuk_EFT_tree}
\end{equation}

\subsubsection{One-loop in the UV theory}

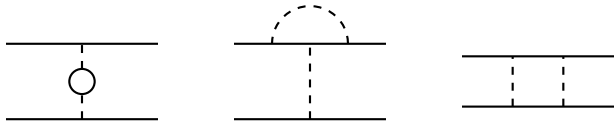
\begin{figure}
\begin{center}
\begin{tikzpicture}

\begin{scope}[xshift=-6cm]
\def\len{3};
\def\unit{\len/3};
\draw[-,thick] (0,0)--(2*\unit,0);
\draw[-,thick] (0,\unit)--(2*\unit,\unit);
\draw[-,dashed,thick] (\unit,0)--(\unit,\unit/3);
\draw[-,dashed,thick] (\unit,2*\unit/3)--(\unit,\unit);
\draw[thick] (\unit,\unit/2) circle (\unit/6);
\end{scope}

\begin{scope}[xshift=-3cm]
\def\len{3};
\def\unit{\len/3};
\draw[-,thick] (0,0)--(2*\unit,0);
\draw[-,thick] (0,\unit)--(2*\unit,\unit);
\draw[-,dashed,thick] (\unit,0)--(\unit,\unit);

\draw [thick,dashed,domain=0:180] plot ({\unit + 0.5*\unit * cos(\x)}, {\unit +  0.5*\unit *  sin(\x)});
\end{scope}

\begin{scope}[xshift=0cm,yshift=.167cm]
\def\len{2};
\def\unit{\len/3};
\draw[-,thick] (0,0)--(\len,0);
\draw[-,thick] (0,\unit)--(\len,\unit);
\draw[-,dashed,thick] (\unit,0)--(\unit,\unit);
\draw[-,dashed,thick] (2*\unit,0)--(2*\unit,\unit);
\end{scope}

\end{tikzpicture}
\end{center}
\caption{One-loop diagrams contributing to the 1LPI effective action in the UV theory. Solid lines represent fermions, dashed lines scalars.}
\label{fig:yuk_full_diag}
\end{figure}

The one-loop 1LPI in the UV theory contains diagrams with both heavy and light fields, see Figure~\ref{fig:yuk_full_diag}. In the functional approach, following the discussion of Section~\ref{subsec:MatchingTraditional}, to obtain \(\G_{\text{L,UV}}^{(1)}[\ps]\) we take the second variation of ${{\cal L}_{{\rm{UV}}}}\left( {\psi ,\Phi } \right)$
\begin{align}\renewcommand\arraystretch{1.4}
{\delta ^2}{{\cal L}_{{\rm{UV}}}}\left( {\psi ,\Phi } \right) = \frac{1}{2}\left( {\begin{array}{*{20}{c}}
{\delta \Phi }&{\delta {\psi ^T}}&{\delta \bar \psi }
\end{array}} \right)\left( {\begin{array}{*{20}{c}}
A&{\bar \Gamma }&{ - {\Gamma ^T}}\\
{ - {{\bar \Gamma }^T}}&0&{ - {B^T}}\\
\Gamma &B&0
\end{array}} \right)\left( {\begin{array}{*{20}{c}}
{\delta \Phi }\\
{\delta \psi }\\
{\delta {{\bar \psi }^T}}
\end{array}} \right) , \label{eq:dL_UV_matrix}
\end{align}
where
\begin{equation}
A = -\pd^2 - M^2, \ \ B = i \pds - m -\l \Ph_c, \ \ \G = -\l \ps, \ \ \Gb = - \l \psb. \label{eq:def_ABG}
\end{equation}

This is the first time in this paper that we take functional derivatives with respect to fields of mixed statistics. So let us procede slowly for clarity. From Eq.~\eqref{eqn:1LPI1UV}, we know that the one-loop 1LPI effective action $\Gamma_\text{L,UV}^{(1)}$ should be proportional to the determinant of the above matrix. However, what should be the proper pre-factor? We know that the ``$i/2$'' in Eq.~\eqref{eqn:1LPI1UV} needs to be appropriately tailored for each field. But in Eq.~\eqref{eq:dL_UV_matrix} there are both the scalar field $\Phi$ and the fermionic fields $\left(\psi, \bar\psi^T\right)$.

If the functional derivative matrix were block diagonal, \textit{i.e.} $\Gamma=\bar\Gamma=0$, it would be straightforward to obtain the appropriate expression of Eq.~\eqref{eqn:1LPI1UV}. In this case, the scalar block and the fermionic block would decouple, and we would obtain
\begin{equation}
\Gamma _{{\rm{L,UV}}}^{\left( 1 \right)}\left[ \psi  \right] = \frac{i}{2}\log \det \left( { - A} \right) - \frac{i}{2}\log \det \left[ { - \left( {\begin{array}{*{20}{c}}
0&{ - {B^T}}\\
B&0
\end{array}} \right)} \right] . \label{eqn:blockdiagonal}
\end{equation}
Note that we have $-i/2$ here for the second term, instead of the usual $-i$. This is because in the way of taking the functional derivative in Eq.~\eqref{eq:dL_UV_matrix}, we have paired the two degrees of freedom $\psi$ and $\bar\psi$ into one, bigger vector:
\begin{equation}
\chi  = \left( {\begin{array}{*{20}{c}}
{\delta \psi }\\
{\delta {{\bar \psi }^T}}
\end{array}} \right) ,
\end{equation}
which means we are reinterpreting the path integral from two integrals into one integral:
\begin{equation}
\int {D\bar \psi D\psi \exp \left[ {i\delta \bar \psi \left( {\frac{{{\delta ^2}S}}{{\delta \bar \psi \delta \psi }}} \right)\delta \psi } \right]}  = \int {D\chi \exp \left[ {i\frac{1}{2}\delta {\chi ^T}\left( {\frac{{{\delta ^2}S}}{{\delta {\chi ^2}}}} \right)\delta \chi } \right]} ,
\end{equation}
and hence the factor $1/2$.

Eq.~\eqref{eqn:blockdiagonal} would have been nice, but the reality in Eq.~\eqref{eq:dL_UV_matrix} is that $\Gamma\ne0$. However, we can first make the functional matrix in Eq.~\eqref{eq:dL_UV_matrix} block diagonal by changing basis, which amounts to a Gaussian elimination operation on the matrix:\footnote{This is equivalent to completing the square on the fermionic fields then performing a linear shift. This shift matrix diagnolizes~\eqref{eq:dL_UV_matrix}.}
\begin{equation}
\left( {\begin{array}{*{20}{c}}
A&{\bar \Gamma }&{ - {\Gamma ^T}}\\
{ - {{\bar \Gamma }^T}}&0&{ - {B^T}}\\
\Gamma &B&0
\end{array}} \right) \to \left( {\begin{array}{*{20}{c}}
{A - \bar \Gamma {B^{ - 1}}\Gamma }&0&{ - {\Gamma ^T}}\\
{ - {{\bar \Gamma }^T}}&0&{ - {B^T}}\\
\Gamma &B&0
\end{array}} \right) \to \left( {\begin{array}{*{20}{c}}
{A - \bar \Gamma {B^{ - 1}}\Gamma  + {\Gamma ^T}{B^{T - 1}}{{\bar \Gamma }^T}}&0&0\\
{ - {{\bar \Gamma }^T}}&0&{ - {B^T}}\\
\Gamma &B&0
\end{array}} \right) . \label{eqn:diagonalized}
\end{equation}
Now we can make use of Eq.~\eqref{eqn:blockdiagonal}. This is the general way we proceed with for functional determinants of mixed statistics.\footnote{This prescription can be more elegantly summarized into a definition of a new determinant, called ``superdeterminant'' (\textit{e.g.}~\cite{Neufeld:1998js}), where one stipulates different statistics for different blocks of the ``supermatrix''. With this new definition, our 1LPI effective action can be written as~\cite{Neufeld:1998js}:
\[
\Gamma _{{\rm{L,UV}}}^{\left( 1 \right)}\left[ \psi  \right] \propto \log {\rm{Sdet}}\left( {\begin{array}{*{20}{c}}
A&{\bar \Gamma }&{ - {\Gamma ^T}}\\
{ - {{\bar \Gamma }^T}}&0&{ - {B^T}}\\
\Gamma &B&0
\end{array}} \right) = \text{STr}\log \left( {\begin{array}{*{20}{c}}
A&{\bar \Gamma }&{ - {\Gamma ^T}}\\
{ - {{\bar \Gamma }^T}}&0&{ - {B^T}}\\
\Gamma &B&0
\end{array}} \right), 
\]
where ``Sdet'' is the superdeterminant, ``STr'' is the supertrace. The functional trace can be written in a form \(\text{STr} \log(\widetilde{D}^2 + Y)\) where  \(\widetilde{D}_{\m} = D_{\m} - iX_{\m}\) with \(X_{\m}\) and \(Y\) supermatrices. It would be interesting (and likely straightforward) to generalize the CDE procedure for evaluating the determinant of elliptic operators to evaluating superdeterminants of ``super elliptic operators''.}

Using Eq.~\eqref{eqn:blockdiagonal} on the matrix in Eq.~\eqref{eqn:diagonalized}, we get
\begin{align}
\Gamma _{{\rm{L,UV}}}^{\left( 1 \right)}\left[ \psi  \right] &= \frac{i}{2}\log \det \left( { - A + \bar \Gamma {B^{ - 1}}\Gamma  - {\Gamma ^T}{B^{T - 1}}{{\bar \Gamma }^T}} \right) - \frac{i}{2}\log \det \left[ { - \left( {\begin{array}{*{20}{c}}
0&{ - {B^T}}\\
B&0
\end{array}} \right)} \right] \nonumber \\
& = \frac{i}{2} \underbrace{\log \det[-A]}_{(i)} - i \underbrace{\log \det[-B]}_{(ii)} + \frac{i}{2}\underbrace{\log \det[1 - A^{-1}\Gb B^{-1}\G + A^{-1}\G^T B^{-1\, T} \Gb^T]}_{(iii)}.
\label{eq:three_terms}
\end{align}

\begin{figure}
\begin{center}
\begin{tikzpicture}

\begin{scope}[xshift=-6cm]
\def\len{3};
\def\unit{\len/3};
\draw[-,thick,white] (0,0)--(2*\unit,0);
\draw[-,dashed,thick] (\unit,0)--(\unit,\unit/3);
\draw[-,dashed,thick] (\unit,2*\unit/3)--(\unit,\unit);
\draw[thick] (\unit,\unit/2) circle (\unit/6);

\draw[-,thick] (\unit - 0.1, 0.1)--(\unit + 0.1,-0.1);
\draw[-,thick] (\unit + 0.1, 0.1)--(\unit - 0.1,-0.1);
\draw[-,thick] (\unit - 0.1,\unit + 0.1)--(\unit + 0.1,\unit-0.1);
\draw[-,thick] (\unit + 0.1,\unit + 0.1)--(\unit - 0.1,\unit-0.1);
\end{scope}

\begin{scope}[xshift=-3cm]
\def\len{3};
\def\unit{\len/3};
\draw[-,thick] (0,\unit)--(2*\unit,\unit);
\draw[-,dashed,thick] (\unit,0)--(\unit,\unit);

\draw [thick,dashed,domain=0:180] plot ({\unit + 0.5*\unit * cos(\x)}, {\unit +  0.5*\unit *  sin(\x)});

\draw[-,thick] (\unit - 0.1, 0.1)--(\unit + 0.1,-0.1);
\draw[-,thick] (\unit + 0.1, 0.1)--(\unit - 0.1,-0.1);
\end{scope}

\begin{scope}[xshift=0cm,yshift=.167cm]
\def\len{2};
\def\unit{\len/3};
\draw[-,thick] (0,0)--(\len,0);
\draw[-,thick] (0,\unit)--(\len,\unit);
\draw[-,dashed,thick] (\unit,0)--(\unit,\unit);
\draw[-,dashed,thick] (2*\unit,0)--(2*\unit,\unit);
\end{scope}

\end{tikzpicture}
\end{center}
\caption{How the diagrams of Figure~\ref{fig:yuk_full_diag} appear in the functional evaluation of \(\G_{\text{L,UV}}^{(1)}[\ps]\). Here, external fermion lines correspond to \(\ps, \bar\psi\), internal fermions to \(\delta\psi, \delta\bar\psi\), dashed lines to \(\delta\Phi\), while the cross denotes an insertion of \(\Ph_c\).}
\label{fig:yuk_UV_func}
\end{figure}
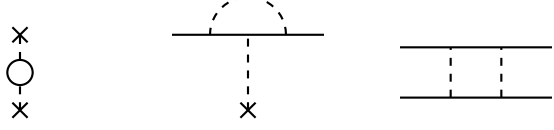

Let us briefly examine these three terms. The first term, \(\log\det[-A] = \Tr \log[\pd^2 + M^2]\), is trivial and only affects the normalization of the path integral. The last two terms contain non-trivial contributions to the effective action. Figure~\ref{fig:yuk_UV_func} shows how the Feynman diagrams in the full theory are reinterpreted in the functional calculation. The first diagram of Figure~\ref{fig:yuk_UV_func} lies within term (ii), while the diagrams involving both heavy and light propagators lie in term (iii). 

\vspace{3mm}
\noindent\textbf{Evaluating term (ii)}
\vspace{3mm}

\begin{equation}
(ii): -i \log\det[-B] = -i \Tr \log[-i\pds + m + \l \Ph_c(x)]
\end{equation}

The evaluation of this determinant is fairly simple because it can be converted into the functional determinant of an elliptic operator of the sort in subsection~\ref{subsubsec:traditionalCDE}. This allows us to apply the universal results in Eq.~\eqref{eqn:ellipticUniversal}.

From Eq.~\eqref{eqn:Uferm}, we have \(U_{\text{ferm}} = -i \pds F + 2mF + F^2\) with \(F = \l \Ph_c \approx -\l^2 \psb\ps/M^2\), which we then plug into the universal formula. \((\psb\ps)^2\) is clearly only contained in the \(\text{tr}\, U_{\text{ferm}}\) and \(\text{tr}\, U_{\text{ferm}}^2\) terms of Eq.~\eqref{eqn:ellipticUniversal}. It is apparent that to leading order in $m^2/M^2$, these terms vanish. Hence,
\begin{align}
\G_{\text{L,UV},(ii)}^{(1)} = 0 . \label{eq:term_ii}
\end{align}

\vspace{3mm}
\noindent \textbf{Evaluating term (iii)}
\vspace{3mm}

To compute term (iii) we expand the logarithm and then evaluate the functional traces:
\begin{equation}
\frac{i}{2}\log \det[1 - A^{-1}\Gb B^{-1}\G + A^{-1}\G^T B^{-1\, T}\Gb^T] = -\frac{i}{2}\sum_{n=1}^{\infty} \frac{1}{n} \Tr\big[ (A^{-1}\Gb B^{-1}\G - A^{-1}\G^T B^{-1\, T}\Gb^T)^n \big] .
\end{equation}
As \(\G \propto \ps\), keeping the first two terms is sufficient for our purposes:
\begin{align}
\text{(iiia):} &\quad -i \Tr\big[A^{-1}\Gb B^{-1} \G\big], \\
\text{(iiib):} &\quad -\frac{i}{2} \Tr\big[(A^{-1}\Gb B^{-1} \G)^2 - A^{-1}\Gb B^{-1} \G A^{-1}\G^T B^{-1\, T}\Gb^T\big], \label{eq:iiib}
\end{align}
where we have used transposition and cyclic properties of the trace (keeping appropriate track of signs when exchanging fermions) and \(A^{-1\, T} = A^{-1}\).

\vspace{3mm}
\noindent \textbf{Term (iiia)}
\vspace{3mm}

We can evaluate the trace in term (iiia) by using the techniques of Appendix~\ref{app:Trace}. To bring it to traces of the sort in Eq.~\eqref{eqn:TraceGeneral_app}, we expand the fermion propagator,
\begin{equation}
\frac{1}{i\pds - m - \l \Ph_c} = \frac{1}{i\pds - m} + \frac{1}{i\pds - m}\l\Ph_c \frac{1}{i\pds -m} + \dots, \nonumber
\end{equation}
so that the trace contains
\begin{equation}
\text{Tr}\left[ \frac{1}{-\pd^2 -M^2} \Gb \frac{i\pds + m}{-\pd^2 - m^2} \G +  \frac{1}{-\pd^2 -M^2} \Gb \frac{i\pds + m}{-\pd^2 - m^2} \l \Ph_c \frac{i\pds + m}{-\pd^2 - m^2} \right] . \label{eqn:iiia_exp_prop}
\end{equation}

We evaluate these traces following the general prescription of Eq.~\eqref{eqn:gen_prescription}, doing the derivative expansion, and then evaluating the momentum integrals. For the first trace in the above equation, this procedure gives:
\begin{align}
\int d^4x & \frac{d^4p}{(2\pi)^4} \frac{1}{(i\pd - p)^2 - M^2}\Gb \frac{(i\pds - \slashed{p}) + m}{(i\pd - p)^2 - m^2} \G \nonumber \\
& \supset \int d^4x \frac{d^4p}{(2\pi)^4} \frac{1}{p^2- M^2}\Gb \frac{1}{p^2-m^2} \left( 1 + \frac{2i p \cdot \pd}{p^2-m^2} \right)\big(i\pds - \slashed{p} + m\big)\G \nonumber \\
&= \int d^4x \frac{d^4p}{(2\pi)^4} \left( \frac{m \Gb\G + \Gb i\pds \G}{(p^2-M^2)(p^2-m^2)} - \frac{2 p_{\m}p_{\n} \big( \Gb i\g^{\m}\pd^{\n}\G\big)}{(p^2-M^2)(p^2-m^2)^2} \right) \nonumber \\
&= \frac{i}{(4\pi)^2} \int d^4x \left(-\log \frac{M^2}{\m^2} + 1\right)  m\Gb\G + \frac{1}{2}\left(-\log \frac{M^2}{\m^2} + \frac{1}{2}\right) \cdot \Gb i\pds \G. \nonumber
\end{align}
In the derivative expansion of the second line, we kept only linear order in derivatives and dropped total derivatives. In evaluating the momentum integrals in the last line we kept only the leading order terms in \(m^2/M^2\).

The evaluation of the second trace in Eq.~\eqref{eqn:iiia_exp_prop} proceeds similarly:
\begin{align}
\int d^4x & \frac{d^4p}{(2\pi)^4} \frac{1}{(i\pd - p)^2 - M^2}\Gb \frac{(i\pds - \slashed{p}) + m}{(i\pd - p)^2 - m^2} \l \Ph_c \G \frac{(i\pds - \slashed{p}) + m}{(i\pd - p)^2 - m^2} \nonumber \\
&\supset \int d^4x  \frac{d^4p}{(2\pi)^4} \frac{(p^2+m^2)\l\Ph_c \Gb\G}{(p^2-M^2)(p^2-m^2)^2} \nonumber \\
& = \frac{i}{(4\pi)^2} \int d^4x \left(-\log \frac{M^2}{\m^2} + 1\right) \l \Ph_c \Gb\G. \nonumber
\end{align}
Again, we have kept only the leading term in $m^2/M^2$ expansion (note that this allows us in the second line to drop the term proportional to \(m^2\), which saves some work).

Plugging in \(\Ph_c \approx -\l \psb\ps/M^2\), \(\G = -\l \ps\), \(\Gb = - \l\psb\), we find in the end
\begin{align}
\G_{\text{L,UV},(iiia)}^{(1)} &\supset \int d^4x \Bigg\{ \frac{1}{(4\pi)^2} \frac{\l^2}{2}\left( - \log \frac{M^2}{\m^2} + \frac{1}{2} \right) \left( \bar\psi i\slashed\partial \psi \right) + \frac{\l^2}{(4\pi)^2} \left( - \log \frac{M^2}{\mu^2} + 1\right) \left( m \bar\psi \psi \right) \nonumber \\
& + \frac{2}{(4\pi)^2} \frac{\l^4}{M^2} \left(\log \frac{M^2}{\mu^2} - 1\right) \frac{1}{2}\left(\bar\psi \psi\right)^2 \Bigg\} . \label{eq:term_iiia}
\end{align}

\vspace{3mm}
\noindent \textbf{Term (iiib)}
\vspace{3mm}

Evaluating (iiib) proceeds very similarly to the previous evaluation of \(\Tr (A^{-1}\Gb B^{-1}\G)\). However, since the present term is already order \((\Gb\G)^2 \propto (\psb\ps)^2\), we can take \(1/(i\pds - m - \l\Ph_c) \approx 1/(i\pds -m)\) for the calculation. For the first term of Eq.~\eqref{eq:iiib},
\begin{align}
-\frac{i}{2} \Tr \big[ (A^{-1}\Gb B^{-1}\G)^2 \big] &\approx -\frac{i}{2} \int d^4x\, \frac{d^4p}{(2\pi)^4}\, \frac{1}{p^2-M^2} \frac{\Gb(-\slashed{p} + m)\G}{p^2 - m^2} \frac{1}{p^2 - M^2} \frac{\Gb(-\slashed{p} + m)\G}{p^2 - m^2}  \nonumber \\
&=-\frac{i}{2}\int d^4x\,\frac{d^4p}{(2\pi)^4}\, \frac{ m^2 (\Gb\G)^2+ p_{\m}p_{\n}(\Gb\g^{\m}\G)(\Gb\g^{\n}\G) }{(p^2-M^2)^2(p^2-m^2)^2}. \label{eq:iiib_p1}
\end{align}
In the first line, we followed the prescription of Eq.~\eqref{eqn:gen_prescription} and then took the zeroth order in the derivative expansion since the above is already proportional to \((\Gb\G)^2 \propto (\psb\ps)^2\). Similarly for the second term of Eq.~\eqref{eq:iiib}
\begin{align}
&\frac{i}{2} \Tr \big[ A^{-1}\Gb B^{-1}\G A^{-1}\G^TB^{-1\, T} \Gb^T \big] \nonumber \\
&\approx \frac{i}{2} \int d^4x\, \frac{d^4p}{(2\pi)^4}\, \frac{1}{p^2-M^2} \frac{\Gb(-\slashed{p} + m)\G}{p^2 - m^2} \frac{1}{p^2 - M^2} \frac{\G^T(\slashed{p}^T + m)\Gb^T}{p^2 - m^2}  \nonumber \\
&=-\frac{i}{2}\int d^4x\,\frac{d^4p}{(2\pi)^4}\, \frac{m^2 (\Gb\G)^2- p_{\m}p_{\n}(\Gb\g^{\m}\G)(\Gb\g^{\n}\G) }{(p^2-M^2)^2(p^2-m^2)^2}, \label{eq:iiib_p2}
\end{align}
where we have used \((i\pd_{\m})^T = - i \pd_{\m}\) in the first line and \(\G^T\Gb^T = -\Gb \G\), \(\G^T \g^{\m\, T}\Gb^T = -\Gb \g^{\m}\G\) in the second line. Adding Eqs.~\eqref{eq:iiib_p1} and~\eqref{eq:iiib_p2} we see the operator \((\psb \g^{\m} \ps)^2\) cancels---as expected, since the heavy field \(\Ph\) is a scalar---and we have
\begin{equation}
\text{(iiib)} =  -i \int d^4x\, \frac{d^4p}{(2\pi)^4}\, \frac{m^2 (\Gb\G)^2}{(p^2-M^2)^2(p^2-m^2)^2}. \nonumber
\end{equation}
This term vanishes at leading order of $m^2/M^2$, hence
\begin{equation}
\G_{\text{L,UV},(iiib)}^{(1)} = 0 . \label{eq:term_iiib}
\end{equation}

\vspace{3mm}
\noindent \textbf{Combining terms (i)-(iii)}
\vspace{3mm}

Gathering the results from Eqs.~\eqref{eq:term_ii},~\eqref{eq:term_iiia}, and~\eqref{eq:term_iiib} we have
\begin{align}
\G_{\text{L,UV}}^{(1)}[\ps] &= \int d^4x\, \Bigg\{\psb i\pds \ps \cdot \frac{1}{(4\pi)^2} \frac{\l^2}{2}\left[ - \log \frac{M^2}{\m^2} + \frac{1}{2} \right] + m \psb \ps \cdot \frac{\l^2}{(4\pi)^2} \left[ - \log \frac{M^2}{\m^2} + 1 \right] \nonumber \\
&+\frac{1}{2}(\psb \ps)^2 \cdot \frac{2}{(4\pi)^2} \frac{\l^4}{M^2}\left[ \log \frac{M^2}{\mu^2} -1 \right]\Bigg\} . \label{eq:yuk_G_UV}
\end{align}

\subsubsection{One-loop in the EFT}

\begin{figure}
\begin{center}
\begin{tikzpicture}

\begin{scope}[xshift=-6cm]
\def\len{3};
\def\unit{\len/3};
\draw[-,thick] (.3,0) to [out=30, in=180] (\unit,.2) to [out=0,in=150] (2*\unit-.3,0);

\draw[-,thick] (.3,\unit) to [out=-30, in=180] (\unit,\unit -.2 ) to [out=0,in=-150] (2*\unit-.3,\unit);

\draw[-,thick] (\unit,.25) to [out=135,in=-135] (\unit,.75);
\draw[-,thick] (\unit,.25) to [out=45,in=-45] (\unit,.75);

\draw[-,thick,white] (0,0)--(2*\unit,0);

\end{scope}

\begin{scope}[xshift=-3cm]
\def\len{3};
\def\unit{\len/3};
\draw[-,thick,white] (0,0)--(2*\unit,0);

\draw[-,thick] (.3,0) to [out=30, in=180] (\unit,.2) to [out=0,in=150] (2*\unit-.3,0);

\draw [thick,domain=110:430] plot ({\unit + 0.25*\unit * cos(\x)}, {.55+  0.25*\unit *  sin(\x)});

\draw[-,thick] (0,\unit)--(.9,.78);
\draw[-,thick] (2*\unit,\unit)--(1.1,.78);

\end{scope}

\begin{scope}[xshift=0cm,yshift=.167cm,scale=.667]
\def\len{3};
\def\unit{\len/3};
\draw[-,thick] (0,0)--(\unit,\unit/2-.05);
\draw[-,thick] (\unit,\unit/2-.05) to [out=-70, in=180] (1.5*\unit,0) to [out=0,in=-110] (2*\unit,\unit/2-.05);
\draw[-,thick] (\len,0)--(2*\unit,\unit/2-.05);

\draw[-,thick] (0,\unit)--(\unit,\unit/2+.05);
\draw[-,thick] (\unit,\unit/2+.05) to [out=70, in=180] (1.5*\unit,\unit) to [out=0,in=110] (2*\unit,\unit/2+.05);
\draw[-,thick] (\len,\unit)--(2*\unit,\unit/2+.05);

\end{scope}

\end{tikzpicture}
\end{center}
\caption{How the diagrams of Figure~\ref{fig:yuk_full_diag} appear in the functional evaluation of \(\G_{\text{L,EFT}}^{(1)}[\ps]\). In order to follow the flow of fermion number, the four-fermion vertices intentionally do not touch.}
\label{fig:yuk_EFT_func}
\end{figure}
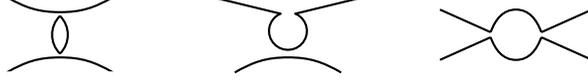

We now turn to computing \(\G_{\text{L,EFT}}^{(1)}[\ps]\). The diagrams relevant to determining \((\psb\ps)^2\) are shown in Figure~\ref{fig:yuk_EFT_func}---they are simply the shrinking of the scalar propagators in Figure~\ref{fig:yuk_full_diag}. Following the functional approach in Section~\ref{subsec:MatchingTraditional}, specifically Eq.~\eqref{eqn:1LPI1EFT}, we see that there are two pieces in \(\G_{\text{L,EFT}}^{(1)}[\ps]\). The first piece is simply
\begin{equation}
\Gamma _{{\rm{L,EFT}}}^{\left( 1 \right)}\left[ \psi  \right] \supset S_{{\rm{EFT}}}^{\left( 1 \right)}\left[ \psi  \right] = \int {{d^4}x\left[ {\frac{1}{2}c_S^{\left( 1 \right)}{{\left( {\bar \psi \psi } \right)}^2}} \right]} . \label{eq:yuk_G_EFT1}
\end{equation}
To compute the second piece, we take the second variation of ${\cal L}_\text{EFT}^{(0)}(\psi)$
\begin{equation}
{\delta ^2}{\cal L}_{{\rm{EFT}}}^{\left( 0 \right)}\left( \psi  \right) = \frac{1}{2}\left( {\begin{array}{*{20}{c}}
{\delta \bar \psi }&{\delta {\psi ^T}}
\end{array}} \right)\left[ {\left( {\begin{array}{*{20}{c}}
{i\slashed \partial  - m}&0\\
0&{ - {{\left( {i\slashed \partial  - m} \right)}^T}}
\end{array}} \right) - F} \right]\left( {\begin{array}{*{20}{c}}
{\delta \psi }\\
{\delta {{\bar \psi }^T}}
\end{array}} \right) ,
\end{equation}
where
\begin{equation}
F \equiv c_S^{(0)}\left( \begin{array}{cc} -(\psb\ps) - \ps \psb & \ps \ps^T \\ \psb^T\psb & (\psb\ps) - \psb^T\ps^T \end{array} \right).
\end{equation}
Then following Eq.~\eqref{eqn:1LPI1UV}, we obtain
\begin{equation}
\G_{\text{L,EFT}}^{(1)} \left[ \psi  \right] \supset -\frac{i}{2} \log\det \left[ \left( \begin{array}{cc} i\slashed \partial & 0 \\ 0 & -(i\slashed \partial)^T \end{array} \right) - m \s_3 - F \right], \nonumber 
\end{equation}
where \(\s_3\) is the Pauli matrix. As before (see the discussion around Eq.~\eqref{eqn:blockdiagonal}), the proper pre-factor is $-i/2$ instead of $-i$. This determinant can again be converted into the functional determinant of an elliptic operator, which allows us to immediately apply the universal results in Eq.~\eqref{eqn:ellipticUniversal}. Although it is straightforward, let us show how to do this conversion for purpose of pedagogy.

The trace is invariant under flipping the sign of \(\g^\mu\) matrices, so we may write
\begin{align}
\G_{\text{L,EFT}}^{(1)} \left[ \psi  \right] &= -\frac{i}{4}\left\{ \log\det \left[ \left( \begin{array}{cc} -i\slashed \partial & 0 \\ 0 & (i\slashed \partial)^T \end{array} \right) - m \s_3 - F \right] + \log\det \left[ \left( \begin{array}{cc} i\slashed \partial & 0 \\ 0 & -(i\slashed \partial)^T \end{array} \right) - m \s_3 - F \right] \right\} \nonumber \\
&= -\frac{i}{4} \log\det \left[ \left( \begin{array}{cc} \slashed \partial^2 & 0 \\ 0 & \slashed \partial^{T\, 2} \end{array} \right) + \left( \begin{array}{cc} m^2 & 0 \\ 0 & m^2 \end{array} \right) + U \right], \nonumber
\end{align}
with
\begin{equation}
U \equiv m \{\s_3,F\} + F^2 + \left[\left( \begin{array}{cc} i\slashed \partial & 0 \\ 0 & - (i\slashed \partial)^T \end{array} \right), F \right].
\end{equation}
Here, \(\slashed \partial^2 = \slashed \partial^{T\, 2} = \partial^2\).\footnote{More generally, for a covariant derivative \(D_\mu=\partial_\mu-igA_\mu\), the \(\slashed D^2\) would contain a piece proportional to \([D_{\m},D_{\n}]\), which would be added to the definition of \(U\) as in Eq.~\eqref{eqn:Uferm}.}

Now we are in a position to immediately apply the universal formula Eq.~\eqref{eqn:ellipticUniversal}. Up to dimension-six, we have
\begin{equation}
\G_{\text{L,EFT}}^{(1)}\left[ \psi  \right] \supset -\frac{1}{(4\pi)^2} \frac{1}{4} \left[ -\left( \log \frac{m^2}{\m^2} - 1\right) m^2\tr\, U - \left(\log \frac{m^2}{\m^2}\right) \frac{1}{2} \tr \, U^2 \right].
\end{equation}
Up to dimension-six operators we can ignore the \([\slashed \partial, F]\) and take \(U = m\{\s_3,F\} + F^2\). The traces are easy to evaluate,\footnote{For example, \(\{\s_3, F\} = 2c_S^{} \left( \begin{smallmatrix} - (\overline{\ps}\ps) - \ps \overline{\ps} & 0 \\ 0 & (\overline{\ps}\ps) - \overline{\ps}^T \ps^T \end{smallmatrix}\right) \) so that
\begin{align}
\tr \{\s_3,F\}^2 &= 4c_S^2 \tr \left(\begin{array}{cc} (\psb\ps)^2 + 3(\psb\ps) \ps \psb & 0 \\ 0 & (\psb\ps)^2 - 3(\psb\ps) \psb^T\ps^T \end{array} \right) \nonumber \\
&= 4c_S^2\big[(\psb\ps)^2(4 - 3) + (\psb\ps)^2(4-3)\big] = 8 c_S^2(\psb\ps)^2, \nonumber
\end{align}
where we used identities such as \(\tr \, \ps \psb = -(\psb\ps)\) and \( (\ps^T\psb^T) = -(\psb\ps)\).
}
\begin{align}
m^2 \tr\, U &= 8 m^2 c_S^{2} \cdot \frac{1}{2}(\psb\ps)^2 , \nonumber \\
\frac{1}{2} \tr\, U^2 &\supset \frac{m^2}{2} \tr \{\s_3,F\}^2 = 8m^2 c_{S}^2 \cdot \frac{1}{2} (\psb\ps)^2 , \nonumber
\end{align}
and we obtain
\begin{equation}
\G_{\text{L,EFT}}^{(1)}[\ps] \supset \frac{1}{2}(\psb\ps)^2 \cdot \frac{2}{(4\pi)^2}m^2 c_S^2 \left( 2 \log \frac{m^2}{\m^2} - 1 \right) \, .
\label{eq:yuk_G_EFT2}
\end{equation}
We see that this term vanishes at leading order of $m^2/M^2$. So the one-loop 1LPI is purely given by Eq.~\eqref{eq:yuk_G_EFT1}:
\begin{equation}
\G_{\text{L,EFT}}^{(1)}[\ps] = \int d^4x \left[ \frac{1}{2}c_S^{(1)}\left(\bar\psi\psi\right)^2 \right]. \label{eq:yuk_G_EFT}
\end{equation}

\subsection{Matching}\label{sec:yukawa_match}

To match the UV theory onto the EFT we use the matching condition (\textit{i.e.} Eq.~\eqref{eqn:MatchingCriterion}),
\begin{equation}
\G_{\text{L,EFT}}[\ps](c,\m = M) = \G_{\text{L,UV}}[\ps](\l,\m = M) , \nonumber
\end{equation}
to determine the Wilson coefficients. At tree-level this is very easy and the result is:
\begin{equation}
c_S^{(0)} = \frac{\l^2}{M^2} ,
\end{equation}

At one-loop we equalize \(\G_\text{L,UV}^{(1)}\) and \(\G_\text{L,EFT}^{(1)}\) given in Eqs.~\eqref{eq:yuk_G_UV} and~\eqref{eq:yuk_G_EFT}. Let us focus on the dimension-six operator \(\mathcal{O}_S=(\psb\ps)^2/2\). Using \(c_S^{(0)} = \l^2/M^2\), \(\G_\text{L,EFT}^{(1)}=\G_\text{L,UV}^{(1)}\) gives
\begin{equation}
c_S^{(1)} = - \frac{2}{(4\pi)^2} \frac{\l^4}{M^2} .
\end{equation}

In summary, we find that the Wilson coefficient up to one-loop order is
\begin{equation}
c_S^{} = \frac{\l^2}{M^2}\left(1  - \frac{\l^2}{8\pi^2} \right).
\end{equation}
This result is in agreement with the calculation in~\cite{Skiba:2010xn}. As is seen from Figure~\ref{fig:yuk_full_diag}, there is no one-loop contribution to $c_S^{}$ with only heavy fields as the propagators. So the components of our $c_S^{}$ should be $c_S^{(0)}=\frac{\l^2}{M^2}$, $c_{S,\text{heavy}}^{(1)}=0$, and $c_{S,\text{mixed}}^{(1)} = - \frac{2}{(4\pi)^2} \frac{\l^4}{M^2}$.

\subsection{Directly computing $c_{S,\text{mixed}}^{(1)}$}\label{sec:yukawa_direct}

In this subsection, we make use of the technique described in Section~\ref{subsec:DirectGeneral} to directly compute $c_{S,\text{mixed}}^{(1)}$. This is the same routine we did in Sections~\ref{subsec:ToyScalar} and~\ref{subsec:TripletScalar}. We first compute the non-local Lagrangian ${{\cal L}_{{\rm{UV}}}}\left( {\psi ,{\Phi _c}\left[ \psi  \right]} \right)$:
\begin{align}
{\Phi _c}\left[ \psi  \right] &= \frac{1}{{ - {\partial ^2} - {M^2}}}\lambda \bar \psi \psi , \nonumber \\
{{\cal L}_{{\rm{UV}}}}\left( {\psi ,{\Phi _c}\left[ \psi  \right]} \right) &= \bar \psi \left( {i\slashed \partial  - m} \right)\psi  - \frac{1}{2}{\lambda ^2}\bar \psi \psi \frac{1}{{ - {\partial ^2} - {M^2}}}\bar \psi \psi , \nonumber
\end{align}
and then take its second functional variation:
\begin{equation}
{\delta ^2}{{\cal L}_{{\rm{UV}}}}\left( {\psi ,{\Phi _c}\left[ \psi  \right]} \right) = \frac{1}{2}\left( {\begin{array}{*{20}{c}}
{\delta {\psi ^T}}&{\delta \bar \psi }
\end{array}} \right)\left( {\begin{array}{*{20}{c}}
{{\lambda ^2}a}&{ - {{\left( {i\slashed \partial  - m + {\lambda ^2}b} \right)}^T}}\\
{i\slashed \partial  - m + {\lambda ^2}b}&{{\lambda ^2}c}
\end{array}} \right)\left( {\begin{array}{*{20}{c}}
{\delta \psi }\\
{\delta {{\bar \psi }^T}}
\end{array}} \right) ,
\end{equation}
where
\begin{align}
a &\equiv {{\bar \psi }^T}\frac{1}{{ - {\partial ^2} - {M^2}}}\bar \psi , \nonumber \\
b &\equiv  - {\left( {\frac{1}{{ - {\partial ^2} - {M^2}}}\bar \psi \psi } \right)_x} - \psi \frac{1}{{ - {\partial ^2} - {M^2}}}{\psi ^T} , \nonumber \\
c &\equiv \psi \frac{1}{{ - {\partial ^2} - {M^2}}}{\psi ^T} . \nonumber
\end{align}
We remind the reader that the notation \((\dots)_x\) means the quantity is an eigenstate of \(\ket{x}\), which means in practice that the derivatives only act within the parentheses---see the discussion around Eq.~\eqref{eqn:xfunctional}. According to our master result Eq.~\eqref{eqn:c1mixeddirectabbr}, we have
\begin{align}
S_{{\rm{EFT,mixed}}}^{\left( 1 \right)}\left[ \psi  \right] &=  - \frac{i}{2}\log \det {\left( {\begin{array}{*{20}{c}}
{{\lambda ^2}a}&{ - {{\left( {i\slashed \partial  - m + {\lambda ^2}b} \right)}^T}}\\
{i\slashed \partial  - m + {\lambda ^2}b}&{{\lambda ^2}c}
\end{array}} \right)_{\rm{d}}} \nonumber \\
 &\supset \frac{i}{2}{\lambda ^4}{\rm{Tr}}{\left\{ \begin{array}{l}
\frac{1}{{{M^4}}}\frac{1}{{i\slashed \partial  - m}}\bar \psi \psi \frac{1}{{i\slashed \partial  - m}}\bar \psi \psi  + \frac{2}{{{M^2}}}\frac{1}{{ - {\partial ^2} - {M^2}}}\bar \psi \frac{1}{{i\slashed \partial  - m}}\bar \psi \psi \frac{1}{{i\slashed \partial  - m}}\psi \\
 - \frac{1}{{ - {\partial ^2} - {M^2}}}\bar \psi \frac{1}{{i\slashed \partial  - m}}\psi \frac{1}{{ - {\partial ^2} - {M^2}}}\bar \psi \frac{1}{{i\slashed \partial  - m}}\psi \\
 + \frac{1}{{ - {\partial ^2} - {M^2}}}\bar \psi \frac{1}{{i\slashed \partial  - m}}\psi \frac{1}{{ - {\partial ^2} - {M^2}}}{\psi ^T}{\left( {\frac{1}{{i\slashed \partial  - m}}} \right)^T}{{\bar \psi }^T}
\end{array} \right\}_{\rm{d}}} \nonumber \\
 &= \frac{i}{2}\frac{{{\lambda ^4}}}{{{M^4}}}{\rm{Tr}}\left\{ \begin{array}{l}
2\frac{{ - {\partial ^2}}}{{ - {\partial ^2} - {M^2}}}\bar \psi \frac{1}{{i\slashed \partial  - m}}\bar \psi \psi \frac{1}{{i\slashed \partial  - m}}\psi \\
 + 2\frac{{ - {\partial ^2}}}{{ - {\partial ^2} - {M^2}}}\bar \psi \frac{1}{{i\slashed \partial  - m}}\psi \bar \psi \frac{1}{{i\slashed \partial  - m}}\psi \\
 - 2\frac{{ - {\partial ^2}}}{{ - {\partial ^2} - {M^2}}}\bar \psi \frac{1}{{i\slashed \partial  - m}}\psi {\psi ^T}{\left( {\frac{1}{{i\slashed \partial  - m}}} \right)^T}{{\bar \psi }^T}\\
 - \frac{{ - {\partial ^2}}}{{ - {\partial ^2} - {M^2}}}\bar \psi \frac{1}{{i\slashed \partial  - m}}\psi \frac{{ - {\partial ^2}}}{{ - {\partial ^2} - {M^2}}}\bar \psi \frac{1}{{i\slashed \partial  - m}}\psi \\
 + \frac{{ - {\partial ^2}}}{{ - {\partial ^2} - {M^2}}}\bar \psi \frac{1}{{i\slashed \partial  - m}}\psi \frac{{ - {\partial ^2}}}{{ - {\partial ^2} - {M^2}}}{\psi ^T}{\left( {\frac{1}{{i\slashed \partial  - m}}} \right)^T}{{\bar \psi }^T}
\end{array} \right\} .
\end{align}
After evaluating all of these functional traces using the CDE techniques of Appendix~\ref{app:Trace}, we get
\begin{equation}
S_{{\rm{EFT,mixed}}}^{\left( 1 \right)}\left[ \psi  \right] = \int {{d^4}x\left[ {\frac{1}{{{{\left( {4\pi } \right)}^2}}}\frac{{ - {\lambda ^4}}}{{{M^2}}}{{\left( {\bar \psi \psi } \right)}^2}} \right]} ,
\end{equation}
which gives
\begin{equation}
c_{S,\text{mixed}}^{\left( 1 \right)} = \frac{1}{{{{\left( {4\pi } \right)}^2}}}\frac{{ - 2{\lambda ^4}}}{{{M^2}}} .
\end{equation}
This agrees with our result in Section~\ref{sec:yukawa_match}.

%% file: sec_Running.tex
While all of the previous sections have focused on how to use functional methods to match a UV theory onto an EFT, in this section we will discuss a slightly different---but closely related---topic: how to do running analysis for an EFT with functional methods. RG evolution is very important for an EFT since, by definition, an EFT is intended for studying low-energy physics. After obtaining the EFT at the matching scale $M$, one uses it to study physics processes at a much lower scale \(v \ll M\). To do so, the Wilson coefficients $c_i(\mu=M)$ need to be evolved down to the scale \(\mu=v\) using the renormalization group equation (RGE).

Suppose we have an EFT
\begin{equation}
{{\cal L}_{{\rm{EFT}}}}\left( \phi  \right) = {{\cal L}_\phi }\left( \phi  \right) + \sum\limits_i {{c_i}\left( \mu  \right){{\cal O}_i}\left( \phi  \right)} ,
\end{equation}
where $\mathcal{L}_\phi$ denotes the renormalizable part, while $\mathcal{O}_i$ are higher dimensional operators. In this case, the Wilson coefficients $c_i$ have negative mass dimension, meaning they are suppressed by certain powers of $M$ viewed from low energy scale $v$. Then to the leading order of $v/M$, the one-loop RGE has the generic linear form
\begin{equation}
\m \frac{d}{d\m} c_i(\m) = \frac{1}{(4\pi)^2} \sum_j \g_{ij} c_j. \label{eqn:WilsonRGE}
\end{equation}
This equation is governed by the anomalous dimension matrix \(\g_{ij}\) which, by simple dimensional analysis, is only a function of the marginal couplings of \(\scL_{\ph}\). 
Solving this RGE, the relation between $c_i(\mu=v)$ and $c_i(\mu=M)$ is
\begin{equation}
c_i(v) = c_i(M) + \frac{1}{(4\pi)^2}\sum_j \g_{ij}c_j(M) \log \frac{v}{M}. \nonumber
\end{equation}

In this section, we will show how to obtain the RGE by functional methods, as well as how to use it to obtain $c_i(\mu=v)$. We first describe in Subsection~\ref{subsec:RunningGeneral} how to obtain one-loop RGE by functional methods for general theories. This applies broadly to QFT, and is not limited to EFT. We then give two explicit examples in Subsections~\ref{subsec:RunningToyScalar} and~\ref{subsec:RunningYukawa} of computing the RGE in the scalar EFT and the four-fermion EFT considered earlier in this work. Finally, in Subsection~\ref{subsec:RunningTripletScalar}, we show how to obtain the low-energy Wilson coefficients of the electroweak triplet scalar model by using the (known) anomalous dimension matrix of the SM EFT.

\subsection{One-loop running with functional methods}\label{subsec:RunningGeneral}

Suppose we are interested in the coefficient (coupling) $\lambda$ of an operator $\mathcal{O}_\lambda$ in the Lagrangian
\begin{equation}
{\cal L}\left( \phi  \right) \supset {{\cal O}_K}\left( \phi  \right) + \lambda {{\cal O}_\lambda }\left( \phi  \right) ,
\end{equation}
where $\phi$ collectively denotes all the fields as before and $\mathcal{O}_K$ collectively denotes all the kinetic terms, canonically normalized. The procedure of deriving the one-loop RGE of $\lambda$ with functional methods is very simple. One first computes the 1PI effective action $\Gamma[\phi]$ to one-loop order, from the result of which one can identify the coefficients of the kinetic terms $a_K$ and that of the operator of interest $a_\lambda$:
\begin{equation}
\Gamma \left[ \phi  \right] \supset \int {{d^4}x\left[ {{a_k}\left( \mu  \right){{\cal O}_K}\left( \phi  \right) + {a_\lambda }\left( \mu  \right){{\cal O}_\lambda }\left( \phi  \right)} \right]} .
\end{equation}
Next, one canonically normalizes all kinetic terms canonically by appropriate rescalings of the fields,
\begin{equation}
\Gamma \left[ \phi  \right] \to \left[ {{{\cal O}_K}\left( \phi  \right) + {{a'}_\lambda }\left( \mu  \right){{\cal O}_\lambda }\left( \phi  \right)} \right] .
\end{equation}
Then, the one-loop RGE of $\lambda$ is simply given by
\begin{equation}
\mu \frac{d}{d\mu}{{a'}_\lambda }\left( \mu  \right) = 0 . \label{eqn:RGEderive}
\end{equation}

\subsection{Toy scalar model}\label{subsec:RunningToyScalar}

Let us take the toy scalar model of Section~\ref{subsec:ToyScalar} as our first example. In this model, the EFT that we match onto is of the form
\begin{equation}
{{\cal L}_{{\rm{EFT}}}}\left( \phi  \right) = \frac{1}{2}\phi \left( { - {\partial ^2} - {m^2}} \right)\phi  - \frac{\kappa }{{4!}}{\phi ^4} + {c_6}{\phi ^6} .
\end{equation}
We want the RGE for the Wilson coefficient $c_6$. Following the prescription outlined above, we need to compute the one-loop 1PI effective action, 
\begin{equation}
{\Gamma _{{\rm{EFT}}}}\left[ \phi  \right] = {S_{{\rm{EFT}}}}\left[ \phi  \right] + \frac{i}{2}\log \det \left( { - \frac{{{\delta ^2}{S_{{\rm{EFT}}}}\left[ \phi  \right]}}{{\delta {\phi ^2}}}} \right) . \nonumber 
\end{equation}
The first term is trivial. For the second term, we take the functional derivative
and get
\begin{subequations}
\begin{align}
\frac{i}{2}\log \det \left( { - \frac{{{\delta ^2}{S_{{\rm{EFT}}}}\left[ \phi  \right]}}{{\delta {\phi ^2}}}} \right) &= \frac{i}{2}\log \det \left( {{\partial ^2} + {m^2} + \frac{\kappa }{2}{\phi ^2} - 30{c_6}{\phi ^4}} \right)  \label{eqn:scalar_det} \\
 &\supset \frac{i}{2}{\rm{Tr}}\log \left( {1 - \frac{1}{{ - {\partial ^2} - {m^2}}}\frac{\kappa }{2}{\phi ^2} + \frac{1}{{ - {\partial ^2} - {m^2}}}30{c_6}{\phi ^4}} \right) \label{eqn:scalar_det_exp}.
\end{align}
\end{subequations}
As the functional determinant is over an elliptic operator of the form \(\pd^2 + m^2 + U(x)\), we can use the universal results in Eq.~\eqref{eqn:ellipticUniversal} with \(U = \k \ph^2/2 - 30 c_6 \ph^4\). For this specific example, this is the fastest route to obtaining the RGE.\footnote{The relevant piece for the RGE in this case is the \(\ph^6\) piece contained in the \(\text{tr}\, U^2\) term of Eq.~\eqref{eqn:ellipticUniversal}.} To showcase methods that apply more generally, however, we factor the logarithm as in the second line above and evaluate the trace using the CDE techniques described in Appendix~\ref{app:Trace}.

Expanding the logarithm in Eq.~\eqref{eqn:scalar_det_exp} will generate a lot of terms. However, according to Sec.~\ref{subsec:RunningGeneral}, we only need two operators: the kinetic term ${{\cal O}_K} = \frac{1}{2}\phi \left( { - {\partial ^2}} \right)\phi$ and ${{\cal O}_6} = {\phi ^6}$. These correspond to keeping terms with two powers and six powers of $\phi$, respectively. For the terms with two powers of $\phi$, there is only one candidate:
\begin{equation}
{\rm{Tr}}\left( { - \frac{1}{{ - {\partial ^2} - {m^2}}}\frac{\kappa }{2}{\phi ^2}} \right) . \nonumber
\end{equation}
We have actually evaluated this functional trace in Section~\ref{subsubsec:functional1loop}, specifically in Eqs.~\eqref{eqn:logexpand} and~\eqref{eqn:phi4loop}. We see that the result does not contain the kinetic operator ${{\cal O}_K} = \frac{1}{2}\phi \left( { - {\partial ^2}} \right)\phi$. Therefore, we can drop this piece of functional trace.

Moving on to the terms with six powers of $\phi$, we find two candidates
\begin{align}
&{\rm{Tr}}\left( { - \frac{1}{3}\frac{1}{{ - {\partial ^2} - {m^2}}}\frac{\kappa }{2}{\phi ^2}\frac{1}{{ - {\partial ^2} - {m^2}}}\frac{\kappa }{2}{\phi ^2}\frac{1}{{ - {\partial ^2} - {m^2}}}\frac{\kappa }{2}{\phi ^2}} \right) , \nonumber \\
&{\rm{Tr}}\left( {\frac{1}{{ - {\partial ^2} - {m^2}}}\frac{\kappa }{2}{\phi ^2}\frac{1}{{ - {\partial ^2} - {m^2}}}30{c_6}{\phi ^4}} \right) . \nonumber
\end{align}
The first term in the above is finite; namely, it does not depend on the RG scale $\mu$ explicitly and therefore does not contribute to the RGE. So we can drop this piece. Then the only piece left is the second line above. This functional trace is easily evaluated to find 
\begin{equation}
{\rm{Tr}}\left( {\frac{1}{{ - {\partial ^2} - {m^2}}}\frac{\kappa }{2}{\phi ^2}\frac{1}{{ - {\partial ^2} - {m^2}}}30{c_6}{\phi ^4}} \right) \supset \int {{d^4}x\left\{ {\left[ {\frac{i}{{{{\left( {4\pi } \right)}^2}}}15\kappa {c_6}\log \frac{{{\mu ^2}}}{{{m^2}}}} \right]{\phi ^6}} \right\}} . \nonumber
\end{equation}

Gathering the pieces, the 1PI effective action up to one-loop order is
\begin{equation}
{\Gamma _{{\rm{EFT}}}}\left[ \phi  \right] \supset \int {{d^4}x\left\{ {\frac{1}{2}\phi \left( { - {\partial ^2} - {m^2}} \right)\phi  - \frac{\kappa }{{4!}}{\phi ^4} + \left[ {{c_6} + \frac{1}{{{{\left( {4\pi } \right)}^2}}}\frac{{ - 15\kappa {c_6}}}{2}\log \frac{{{\mu ^2}}}{{{m^2}}}} \right]{\phi ^6}} \right\}} .
\end{equation}
We derive the RGE of $c_6$ by requiring
\begin{equation}
\mu \frac{d}{{d\mu }}\left[ {{c_6} + \frac{1}{{{{\left( {4\pi } \right)}^2}}}\frac{{ - 15\kappa {c_6}}}{2}\log \frac{{{\mu ^2}}}{{{m^2}}}} \right] = 0 , \nonumber
\end{equation}
which gives
\begin{equation}
\mu \frac{d}{{d\mu }}{c_6} = \frac{1}{{{{\left( {4\pi } \right)}^2}}}15\kappa {c_6} .
\end{equation}
We identity the anomalous dimension matrix element $\gamma_{66}^{}=15\kappa$.

\subsection{Yukawa model}\label{subsec:RunningYukawa}

As a second example, we study the running of the Wilson coefficient $c_S$ in the EFT for the Yukawa model discussed in Section~\ref{sec:Yukawa}. To make the running more interesting, we supplement it by a light scalar $\phi$, which we assume to be degenerate with $\psi$ for simplicity. The EFT Lagrangian is
\begin{equation}
{{\cal L}_{{\rm{EFT}}}}\left( {\psi ,\phi } \right) = \bar \psi \left( {i\slashed \partial  - m} \right)\psi  + \frac{1}{2}{c_S}{\left( {\bar \psi \psi } \right)^2} + \frac{1}{2}\phi \left( { - {\partial ^2} - {m^2}} \right)\phi  - y\phi \bar \psi \psi .
\end{equation}
As before, we ignore other interactions for purpose of pedagogy.

To one-loop, the 1PI effective action is \(\G_{\text{EFT}} = S_{\text{EFT}} + \frac{i}{2} \log \det \big( -\d^2S_{\text{EFT}}/\d(\ps,\ph)^2 \big)\). To compute the loop contribution, we take the functional variation of the Lagrangian, 
\begin{equation}
{\delta ^2}{{\cal L}_{{\rm{EFT}}}}\left( {\psi ,\phi } \right) = \frac{1}{2}\left( {\begin{array}{*{20}{c}}
{\delta \phi }&{\delta {\psi ^T}}&{\delta \bar \psi }
\end{array}} \right)\left[ {\left( {\begin{array}{*{20}{c}}
{ - {\partial ^2} - {m^2}}&{ - y\bar \psi }&{y{\psi ^T}}\\
{y{{\bar \psi }^T}}&0&{ - {B^T}}\\
{ - y\psi }&B&0
\end{array}} \right) - F} \right]\left( {\begin{array}{*{20}{c}}
{\delta \phi }\\
{\delta \psi }\\
{\delta {{\bar \psi }^T}}
\end{array}} \right) ,
\end{equation}
where,
\begin{align}
B &\equiv i\slashed \partial  - m - y\phi , \\
F &\equiv {c_S}\left( {\begin{array}{*{20}{c}}
0&0&0\\
0&{{{\bar \psi }^T}\bar \psi }&{\bar \psi \psi  - {{\bar \psi }^T}{\psi ^T}}\\
0&{ - \bar \psi \psi  - \psi \bar \psi }&{\psi {\psi ^T}}
\end{array}} \right) .
\end{align}
We see that this functional derivative matrix is very similar to the one in Eq.~\eqref{eq:dL_UV_matrix}. We follow the same procedure described around Eq.~\eqref{eq:dL_UV_matrix} to evaluate its functional determinant, and keep only the fermion kinetic operator $\bar\psi i\slashed\partial \psi$ and the operator $\mathcal{O}_S\equiv\frac{1}{2}\left(\bar\psi\psi\right)^2$. These calculations are very similar to others contained in this work, so we omit the detailed steps. The end result for the 1PI effective action is
\begin{align}
{\Gamma _{{\rm{EFT}}}}\left[\psi, \phi\right] & \supset \int d^4x \Bigg\{ \psb i \pds \ps \left(1 - \frac{1}{(4\pi)^2} \frac{y^2}{2} \log \frac{m^2}{\m^2} \right) \nonumber \\
&+ \frac{1}{2}(\psb\ps)^2 \left[ c_S^{} + \frac{1}{(4\pi)^2}\left( 2y^2c_S^{}\log \frac{m^2}{\m^2} + 4 m^2 c_S^2 \log \frac{m^2}{\m^2} + \dots \right) \right] \Bigg\} , \label{eqn:YukawaRunningGEFT}
\end{align}
where the dots indicate one-loop finite terms, which cannot contribute to the RGE.

To get the RGE, we need to canonically normalize the kinetic term by rescaling $\psi$, 
\begin{equation}
\psi  \to \frac{1}{\sqrt{ 1 - \frac{1}{(4\pi)^2}\frac{y^2}{2}\log \frac{m^2}{\mu ^2}}} \ps \approx \left[ 1 + \frac{1}{(4\pi)^2}\frac{y^2}{4}\log \frac{m^2}{\mu ^2} \right]\psi ,
\end{equation}
which rescales $\mathcal{O}_S$ as
\begin{equation}
\frac{1}{2}{\left( {\bar \psi \psi } \right)^2} \to \left[ {1 + \frac{1}{{{{\left( {4\pi } \right)}^2}}}{y^2}\log \frac{{{m^2}}}{{{\mu ^2}}}} \right]\frac{1}{2}{\left( {\bar \psi \psi } \right)^2} ,
\end{equation}
and shifts the second line of Eq.~\eqref{eqn:YukawaRunningGEFT} into
\begin{equation}
\frac{1}{2}(\psb\ps)^2 \left[ c_S^{} + \frac{1}{(4\pi)^2}\left(3y^2c_S^{}\log \frac{m^2}{\m^2} + 4m^2 c_S^2 \log \frac{m^2}{\m^2} \right) \right] .
\end{equation}
The one-loop RGE of $c_S$ is found by requiring that the term in brackets vanish upon taking the derivative \(\frac{d}{d\log\m}\). This gives
\begin{equation}
\mu \frac{d}{{d\mu }}{c_S} = \frac{1}{{{{\left( {4\pi } \right)}^2}}}\left( {6{y^2}{c_S} + 8{m^2}c_S^2} \right) .
\end{equation}
Let us make a few brief physics comments. The second term is suppressed relative to the first by a factor \(m^2c_S \sim m^2/M^2 \ll 1\), and therefore is higher-order from the EFT point of view. 
From the leading order, \textit{i.e.} the first term, we get the anomalous dimension matrix element
\begin{equation}
{\gamma _{SS}} = 6{y^2} ,
\end{equation}
in agreement with the calculation of~\cite{Skiba:2010xn}.

\subsection{Electroweak triplet scalar model}\label{subsec:RunningTripletScalar}

The anomalous dimension matrix for dimension-six operators in the SM EFT has been intensively studied~\cite{Jenkins:2013zja,Jenkins:2013wua,Alonso:2013hga,Grojean:2013kd,Elias-Miro:2013gya,Elias-Miro:2013mua,Elias-Miro:2013eta}. In this section, we take the triplet scalar model discussed in Section~\ref{subsec:TripletScalar} and compute the Wilson coefficient $c_{HD}^{}(\mu=v)$ at a scale $v<M$ using the known \(\g_{ij}\).

As we have computed in Section~\ref{subsec:TripletScalar}, $c_{HD}^{}$ at the matching scale $M$ is\footnote{Here we take the kinetic term rescaled version of $c_{HD{\rm{,mixed}}}^{\left( 1 \right)}$.} (see  Eqs.~\eqref{eqn:TSctree} and~\eqref{eqn:TScshifted})
\begin{equation}
{c_{HD}^{}}\left( M \right) =  - \frac{{2{\kappa ^2}}}{{{M^4}}} + \frac{1}{{{{\left( {4\pi } \right)}^2}}}\frac{{{\kappa ^2}}}{{{M^4}}}\left( { - \frac{3}{2}\lambda + 16\eta + 5\frac{{{\kappa ^4}}}{{{M^2}}} } \right) . \label{eqn:cHDM}
\end{equation}
Upon integrating the RGE, the Wilson coefficient at \(\m = v\) at leading order is given by
\begin{equation}
{c_{HD}^{}}\left( v \right) = {c_{HD}^{}}\left( M \right) + \frac{1}{{{{\left( {4\pi } \right)}^2}}}\sum\limits_j {{\gamma _{ij}}c_j^{\left( 0 \right)}\left( M \right)} \log \frac{v}{M} . \label{eqn:cHDvgeneral}
\end{equation}
In the triplet scalar model, the relevant nonzero $c_j^{(0)}$ are given in Eq.~\eqref{eqn:TSctree}, which we reproduce here
\begin{equation}
c_H^{\left( 0 \right)} = \frac{{2{\kappa ^2}}}{{{M^4}}}\frac{1}{2}\;\;\;{\mkern 1mu} ,\;\;\;{\mkern 1mu} c_{HD}^{\left( 0 \right)} =  - \frac{{2{\kappa ^2}}}{{{M^4}}}\;\;\;{\mkern 1mu} ,\;\;\;{\mkern 1mu} c_R^{\left( 0 \right)} = \frac{{2{\kappa ^2}}}{{{M^4}}} .
\end{equation}
Therefore, to compute ${c_{HD}^{}}\left( v \right)$ using Eq.~\eqref{eqn:cHDvgeneral}, we only need to know three elements of the anomalous dimension matrix: $\gamma_{H\to HD}^{}$, $\gamma_{HD\to HD}^{}$, and $\gamma_{R\to HD}^{}$. These can be extracted from~\cite{Elias-Miro:2013eta} (Table 7) upon doing an operator basis transformation from $\left(\mathcal{O}_H, \mathcal{O}_T, \mathcal{O}_R\right)$ to $\left(\mathcal{O}_H, \mathcal{O}_{HD}, \mathcal{O}_R\right)$, with $\mathcal{O}_T$ defined as
\begin{equation}
{{\cal O}_T} \equiv \frac{1}{2}{\left[ {{H^\dag }\left( {{D_\mu }H} \right) - \left( {{D_\mu }{H^\dag }} \right)H} \right]^2} = {{\cal O}_H} - 2{{\cal O}_{HD}} .
\end{equation}
Performing this basis transformation, we obtain
\begin{align}
{\gamma_{H \to HD}^{}} &=  - 3g_1^2 , \\
{\gamma_{HD \to HD}^{}} &= 3\lambda  - \frac{3}{2}g_1^2 + \frac{9}{2}g_2^2 + 12y_t^2 , \\
{\gamma_{R \to HD}^{}} &= 3g_1^2 .
\end{align}
In the above, SM Yukawa couplings other than top Yukawa $y_t$ are neglected. Using these in Eq.~\eqref{eqn:cHDvgeneral}, we obtain
\begin{align}
{c_{HD}^{}}\left( v \right) &=  - \frac{{2{\kappa ^2}}}{{{M^4}}} + \frac{1}{{{{\left( {4\pi } \right)}^2}}}\frac{{{\kappa ^2}}}{{{M^4}}}\left( { - \frac{3}{2}\lambda  + 16\eta  + 5\frac{{{\kappa ^2}}}{{{M^2}}}} \right) \nonumber \\
 & \qquad - \frac{1}{{{{\left( {4\pi } \right)}^2}}}\frac{{2{\kappa ^2}}}{{{M^4}}}\left( {3\lambda  - 3g_1^2 + \frac{9}{2}g_2^2 + 12y_t^2} \right)\log \frac{v}{M} .
\end{align}
This result is in agreement with~\cite{Khandker:2012zu} upon neglecting the gauge coupling term in the one-loop finite piece, and any Yukawa couplings other than $y_t$.\footnote{The disagreement on the $y_t^2$ term is due to a typo in~\cite{Khandker:2012zu}.}

%% file: sec_Summary.tex
One-loop matching and running analyses are crucial steps in an EFT approach of studying low energy physics. In this paper, we discussed how to perform these analyses more efficiently using functional methods in a manifestly gauge-covariant fashion. We clarified a few basic concepts, developed some new computational techniques, and showed a number of explicit examples for demonstration. In this section, we provide a summary of the central results in this paper.

In the introduction section, we reviewed the standard matching condition between the UV theory and the EFT --- 1LPI diagrams to agree at the matching scale. Up to one-loop order, the Wilson coefficients at the matching scale can be decomposed into three pieces:
\begin{equation}
{c_i}\left( M \right) = c_i^{\left( 0 \right)}\left( M \right) + c_{i,{\rm{heavy}}}^{\left( 1 \right)}\left( M \right) + c_{i,{\rm{mixed}}}^{\left( 1 \right)}\left( M \right) .
\end{equation}
The tree-level piece $c_i^{(0)}(M)$ and the one-loop heavy piece $c_{i,\text{heavy}}^{(1)}(M)$ have been discussed intensively elsewhere~\cite{Henning:2014wua}. Our focus in this paper is hence on the one-loop mixed piece $c_{i,\text{mixed}}^{(1)}(M)$, resulting from the exchange of both heavy and light fields. 

Our main purpose in section~\ref{sec:MatchingFunctional} was to explain how to do a ``traditional'' matching analysis with functional methods. 
We derived in subsection~\ref{subsec:MatchingTraditional} the matching results
\begin{align}
 &\sum\limits_i {c_i^{\left( 0 \right)}\left( M \right){{\cal O}_i}\left( \phi  \right)}  = {{\cal L}_\Phi }\left( {\phi ,{\Phi _c}\left[ \phi  \right]} \right) , \\
 &\int {{d^4}x\left\{ {\sum\limits_i {\left[ {c_{i{\rm{,heavy}}}^{\left( 1 \right)}\left( M \right) + c_{i{\rm{,mixed}}}^{\left( 1 \right)}\left( M \right)} \right]{{\cal O}_i}\left( \phi  \right)} } \right\}} \nonumber \\
 & \qquad = \frac{i}{2}\log \det \left( {{{\left. { - \frac{{{\delta ^2}{S_{{\rm{UV}}}}\left[ {\phi ,\Phi } \right]}}{{\delta {{\left( {\phi ,\Phi } \right)}^2}}}} \right|}_{\Phi  = {\Phi _c}\left[ \phi  \right]}}} \right) - \frac{i}{2}\log \det \left( { - \frac{{{\delta ^2}S_{{\rm{EFT}}}^{\left( 0 \right)}\left[ \phi  \right]}}{{\delta {\phi ^2}}}} \right) ,
\end{align}
where the tree-level piece is isolated, but the two one-loop level pieces, $c_{i,\text{heavy}}^{(1)}(M)$ and $c_{i,\text{mixed}}^{(1)}(M)$ are entangled.

In section~\ref{sec:MatchingDirect}, we resolved these two pieces by using a non-local Lagrangian, and arrived at the isolated matching results
\begin{align}
\int {{d^4}x\sum\limits_i {c_{i{\rm{,heavy}}}^{\left( 1 \right)}\left( M \right){{\cal O}_i}\left( \phi  \right)} } &= \frac{i}{2}\log\det \left( { - {{\left. {\frac{{{\delta ^2}{S_{{\rm{UV}}}}\left[ {\phi ,\Phi } \right]}}{{\delta {\Phi ^2}}}} \right|}_{{\Phi _c}}}} \right) , \\
\int {{d^4}x\sum\limits_i {c_{i{\rm{,mixed}}}^{\left( 1 \right)}\left( M \right){{\cal O}_i}\left( \phi  \right)} } &= \frac{i}{2}\log\det \left( { - \frac{{{\delta ^2}{S_{{\rm{UV}}}}\left[ {\phi ,{\Phi _c}\left[ \phi  \right]} \right]}}{{\delta {\phi ^2}}}} \right) - \frac{i}{2}\log \det \left( { - \frac{{{\delta ^2}S_{{\rm{EFT}}}^{\left( 0 \right)}\left[ \phi  \right]}}{{\delta {\phi ^2}}}} \right) .
\end{align}
We also showed how to systematically evaluate the functional determinants involved by a new CDE technique, which is capable of evaluating a much wider class of traces than previous methods. This new CDE technique is detailed in Appendix~\ref{app:Trace}. Our resolved matching formula using non-local Lagrangian, as well as our new CDE technique of evaluating functional traces are demonstrated by two examples --- a toy scalar model in subsection~\ref{subsec:ToyScalar}, and the heavy triplet scalar extension of the SM in subsection~\ref{subsec:TripletScalar}. In the triplet scalar example, we computated three different Wilson coefficients at the same time, demonstrating one advantage of using functional methods; namely, the ability of functional methods to easily handle and organize information that is encoded in many different correlation functions.

To give a full picture of all aspects of matching, we focused on a toy Yukawa model in section~\ref{sec:Yukawa}, showing how to obtain the Wilson coefficients using both the ``traditional'' procedure of section~\ref{sec:MatchingFunctional} as well as the ``direct'' procedure of section~\ref{sec:MatchingDirect}.

In section~\ref{sec:Running} we discussed the process of renormalization group running in EFTs. We showed how functional methods easily allow us to derive the RG equation for the Wilson coefficients. Since the relevant functional can be evaluated with a CDE, this provides an improved
computational technique for obtaining the anomalous dimension matrix of an EFT.

%% file: app_Trace.tex
Evaluating functional traces is the major task in using functional methods at one-loop order. In this appendix, we explain how to perform this task efficiently with a covariant derivative expansion. Before getting started, however, we emphasize that a CDE is not the only way of evaluating functional traces. The task can be done by other methods, such as a partial derivative expansion (PDE).\footnote{The essence to the methods of CDE and PDE are described in Section~\ref{subsec:CDEReview}.} But, for obvious reasons, a CDE greatly simplifies the calculation compared with a PDE.

Let us consider a generic functional that depends on the position and momentum operators, ${f\left( {\widehat x,\widehat p} \right)}$.\footnote{The hats remind us that the arguments are operators whose representation depends on what representation we pick to trace over the functional. See~\cite{Henning:2014wua}, section 2.2.} To evaluate its trace over the functional space by using a derivative expansion procedure (either a CDE or PDE), the general initial steps are as follows:
\begin{subequations}
\label{eqn:gen_prescription}
\begin{align}
{\rm{Tr}}\left[ {f\left( {\widehat x,\widehat p} \right)} \right] &= \int dp\, \braket{p|\, f(\widehat{x},\widehat{p})\, |p}  \label{eqn:tracedef} \\
 &= \int dx\, dp\, \braket{p|x}\braket{x| \, f(\widehat{x},\widehat{p})\, |p} \label{eqn:unity} \\
 &= \int dx \, dp \, e^{ipx} f\left( x,i \pd_x \right)e^{ - ipx} \label{eqn:sandwich} \\
 &= \int {dx \, dp \, f\left( {x,i{\partial _x} - p} \right)} . \label{eqn:tracegeneral}
\end{align}
\end{subequations}
Let us explain these steps one by one. Eq.~\eqref{eqn:tracedef} is just the definition of the trace over the functional space. In Eq.~\eqref{eqn:unity}, we inserted unity: $1 = \int {dx\left| x \right\rangle \left\langle x \right|}$. To obtain Eq.~\eqref{eqn:sandwich}, we have used the identity
\begin{equation}
\left\langle x \right.\left| {f\left( {\widehat x,\widehat p} \right)} \right|\left. p \right\rangle  = f\left( {x,i{\partial _x}} \right)\left\langle {x}
 \mathrel{\left | {\vphantom {x p}} \right. \kern-\nulldelimiterspace} {p} \right\rangle  = f\left( {x,i{\partial _x}} \right){e^{ - ipx}} ,
\end{equation}
where the $i\partial_x$ inside $f$ is understood to act on the rest that follows, \textit{i.e.} $e^{-ipx}$ in this case. To get Eq.~\eqref{eqn:tracegeneral}, we have used the Baker-Campbell-Hausdorff formula
\begin{equation}
{e^{ipx}}i{\partial _x}{e^{ - ipx}} = i{\partial _x} + p ,
\end{equation}
and made a sign flip for later convenience: $p\to -p$. Note that this Baker-Campbell-Hausdorff shift of $i\partial_x$ holds for the case of a covariant derivative $D_\mu=\partial_\mu-igA_\mu(x)$ as well, \textit{i.e.} ${e^{ipx}}iD{e^{ - ipx}} = iD + p$. Upon arriving at Eq.~\eqref{eqn:tracegeneral}, one can expand out the derivative $i\partial_x$ (or $iD$ in the covariant derivative case), and evaluate the integral over $p$.

To see this more concretely, consider the frequently encountered functional\footnote{In this and similar expressions \(iD\) technically should be understood symbolically as \(i D = \widehat{p} + g A(\widehat{x})\), whose position representation is \(\bra{x} \big(\widehat{p} + gA(\widehat{x})\big) = \big(i \pd_x + gA(x)\big) \bra{x}\). A perhaps better notation would use \(i\widehat{D}\). As we will always pick a position representation in this work, there is no possibility of confusion and so we drop the hats for the rest of this appendix and in the main text.}
\begin{equation}
f\left( {\widehat x,\widehat p} \right) = \frac{1}{{ - {D^2} - {M^2}}}B\left( \widehat{x} \right) .
\end{equation}
Making use of the general prescription from Eq.~\eqref{eqn:tracedef} to Eq.~\eqref{eqn:tracegeneral}, we get
\begin{equation}
{\rm{Tr}}\left[ {\frac{1}{{ - {D^2} - {M^2}}}B\left( x \right)} \right] = \int {{d^4}x\int {\frac{{{d^4}p}}{{{{\left( {2\pi } \right)}^4}}}{\rm{tr}}\left[ {\frac{1}{{{{\left( {iD - p} \right)}^2} - {M^2}}}B\left( x \right)} \right]} } , \label{eqn:DMB}
\end{equation}
where we have followed the notation in~\cite{Henning:2014wua} of using ``$\text{Tr}$'' to denote a trace over both the functional space and any internal indices (gauge, spin, flavor, \textit{etc}), and ``$\text{tr}$'' to denote a trace over the internal indices only. We can now expand out the covariant derivatives relative to the leading term $p^2-M^2$ to obtain an expansion in powers of \(D\) and the ``free propagator'' \((p^2 - M^2)^{-1}\):
\begin{align}
\frac{1}{{{{\left( {iD - p} \right)}^2} - {M^2}}}B\left( x \right) &= \frac{1}{{{p^2} - {M^2} - 2i{p^\mu }{D_\mu } - {D^2}}}B\left( x \right) \nonumber \\
 &= \frac{1}{{\left( {{p^2} - {M^2}} \right)\left[ {1 - \frac{1}{{{p^2} - {M^2}}}\left( {2i{p^\mu }{D_\mu } + {D^2}} \right)} \right]}}B\left( x \right) \nonumber \\
 &= \sum\limits_{n = 0}^\infty  {{{\left[ {\frac{1}{{{p^2} - {M^2}}}\left( {2i{p^\mu }{D_\mu } + {D^2}} \right)} \right]}^n}\frac{1}{{{p^2} - {M^2}}}} B\left( x \right) . \label{eqn:DMBCDE}
\end{align}
Here the covariant derivatives act on the operator $B(x)$ to form local operators. Plugging this expression back into Eq.~\eqref{eqn:DMB} and performing the integral over $p$, one will arrive at a result with a spacetime integral over a local operator, which is clearly in the form of an effective action. We see that throughout this procedure, $D_\mu$ is kept intact. Therefore, by definition, this is a CDE procedure.

The prescription described above works well for evaluating any functional trace in the form of Eq.~\eqref{eqn:TraceGeneral}, which we reproduce here:
\begin{equation}
{\rm{Tr}}\left[ {\frac{{{{\left( { - {D^2}} \right)}^{{k_1}}}}}{{ - {D^2} - m_1^2}}{A_1}\left( x \right)\frac{{{{\left( { - {D^2}} \right)}^{{k_2}}}}}{{ - {D^2} - m_2^2}}{A_2}\left( x \right) \cdots \frac{{{{\left( { - {D^2}} \right)}^{{k_3}}}}}{{ - {D^2} - m_n^2}}{A_n}\left( x \right)} \right] , \label{eqn:TraceGeneral_app}
\end{equation}
Later in this appendix, we show a few explicit examples.

An important note to make here is that although the prescription we just described works on a wide class of functionals, it actually fails to be a CDE in the special case that $B(x)$ is a constant, \textit{i.e.} for evaluating the trace
\begin{equation}
{\rm{Tr}}\left( {\frac{1}{{ - {D^2} - {M^2}}}} \right) .
\end{equation}
When $B(x)$ is a constant, $\partial_\mu B(x)=0$, we have
\begin{equation}
\left( {2i{p^\mu }{D_\mu } + {D^2}} \right)B\left( x \right) = 2g{p^\mu }{A_\mu }\left( x \right) - ig{D^\mu }{A_\mu }\left( x \right) . \label{eqn:Aexplicit}
\end{equation}
We see that $D_\mu$ is broken into components with the gauge field $A_\mu$ showing up explicitly. There is no problem with plugging Eq.~\eqref{eqn:Aexplicit} back into Eq.~\eqref{eqn:DMB} and evaluating the trace. It works, but it is a PDE instead of a CDE---after evaluating the $p$ integrals, one would need to recombine the $D_\mu$ and $A_\mu$ into field strengths to form a gauge invariant expression.

There is a very nice trick, introduced in~\cite{Gaillard:1985}, to keep $D_\mu$ intact while evaluating this trace. This is to make a further insertion of ${e^{iD\frac{\partial }{{\partial p}}}}$ and ${e^{-iD\frac{\partial }{{\partial p}}}}$:
\begin{align}
{\rm{Tr}}\left( {\frac{1}{{ - {D^2} - {M^2}}}} \right) &= \int {{d^4}x\int {\frac{{{d^4}p}}{{{{\left( {2\pi } \right)}^4}}}{\rm{tr}}\left[ {\frac{1}{{{{\left( {iD - p} \right)}^2} - {M^2}}}} \right]} } \nonumber \\
 &= \int {{d^4}x\int {\frac{{{d^4}p}}{{{{\left( {2\pi } \right)}^4}}}\left\{ {{e^{iD\frac{\partial }{{\partial p}}}}{\rm{tr}}\left[ {\frac{1}{{{{\left( {iD - p} \right)}^2} - {M^2}}}} \right]{e^{ - iD\frac{\partial }{{\partial p}}}}} \right\}} }.
\end{align}
Using the Baker-Campbell-Hausdorff formula will then convert every $D_\mu$ into commutators with itself (for details, see~\cite{Gaillard:1985,Cheyette:1987,Henning:2014wua,Drozd:2015rsp}). The problem shown in Eq.~\eqref{eqn:Aexplicit} is gone after this conversion, because for constant $B(x)$
\begin{equation}
\left[ {{D_\mu },{D_\nu }} \right]B\left( x \right) = \left[ {{D_\mu },{D_\nu }} \right] .
\end{equation}
We see that the $D_\mu$'s do not get broken down. This is ``the CDE'' method of~\cite{Gaillard:1985,Cheyette:1987,Henning:2014wua}.

\subsection{Variations on a theme}
A natural generalization of the term in Eq.~\eqref{eqn:DMB} that is frequently encountered is
\begin{equation}
\frac{1}{-D^2-M^2-U(x)}B(x).
\label{eqn:DMUB}
\end{equation}
Such pieces arise when the propagators are functions of background fields, \textit{i.e.} \(\scL \supset \Ph\big(-D^2 -M^2 - U(x)\big) \Ph\).

To handle terms like Eq.~\eqref{eqn:DMUB} there are basically two options. One option is to first expand \((-D^2 - M^2 - U)^{-1}\) is a power series of \(U\),
\begin{equation}
\frac{1}{-D^2-M^2-U}B = \sum_{n=0}^{\infty}\left[\frac{1}{-D^2-M^2}U\right]^n\frac{1}{-D^2-M^2}B,
\label{eqn:DMUB_Uexp}
\end{equation}
which just produces a specific case of Eq.~\eqref{eqn:TraceGeneral_app}. From here, a covariant derivative expansion along the lines of Eq.~\eqref{eqn:DMBCDE} and the examples of the next subsection can be done. This is the approach taken for the the triplet scalar and Yukawa models of Sections~\ref{subsec:TripletScalar} and~\ref{sec:Yukawa}, respectively. The other option is to do the expansion of \(U\) together with the covariant derivative expansion. Analogous to Eq.~\eqref{eqn:DMBCDE}, we have
\begin{equation}
\frac{1}{(iD-p)^2-M^2-U}B = \sum_{n=0}^{\infty}\left[ \frac{1}{p^2 - M^2} \big(2i p^\mu D_\mu + D^2 + U \big)\right]^n \frac{1}{p^2 - M^2} B.
\label{eqn:DMUBCDE}
\end{equation}
Which option to take is just a matter of preference.

Let us make a, perhaps obvious, comment that applies to the expansions in Eqs.~\eqref{eqn:DMBCDE} and~\eqref{eqn:DMUBCDE}. In general \(D_{\m}\), \(M\), and \(U(x)\) are matrix valued. Therefore, if \([M,U] \ne 0\), then the ordering in these expansions is important as \((p^2 - M^2)^{-1}\) will not commute through \((2i p\cdot D + D^2 + U)\). If all the matrices commute, then Eq.~\eqref{eqn:DMBCDE} can more compactly be expressed by \(\sum_{n=0} \frac{1}{(p^2 - M^2)^{n+1}}(2i p \cdot D + D^2)^n B\) and similarly for Eq.~\eqref{eqn:DMUBCDE}.

For completeness, let us briefly address the case of fermion propagators, \([i\slashed{D} - M - F(x)]^{-1}\). We can either first expand in a power series of \(F\),
\begin{equation}
\frac{1}{i\slashed{D} - M - F(x)} = \frac{i\slashed{D} + M}{-D^2 - M^2} + \frac{i\slashed{D} + M}{-D^2 - M^2}F(x) \frac{i\slashed{D} + M}{-D^2 - M^2} + \dots,
\end{equation}
and then develop a CDE following the prescription of Eq.~\eqref{eqn:gen_prescription} and the steps done for the bosonic propagator. Alternatively, we may first write
\begin{equation}
\frac{1}{i\slashed{D} - M - F(x)} = \frac{i\slashed{D} + M + F(x)}{-D^2 - M^2 - U_{\text{ferm}}(x)},
\label{eqn:DMF}
\end{equation}
with \(U_{\text{ferm}}(x)\) given in Eq.~\eqref{eqn:Uferm}, and then develop a CDE.

\subsection{Example trace evaluations}\label{subapp:example_traces}
To show how to use the CDE technique explained above, we now consider three example traces of the form in Eq.~\eqref{eqn:TraceGeneral_app}. Each of these traces are encountered in the main text.

\subsubsection{Trace Evaluation Example 1: $T$}\label{subapp:example1}

Let us first consider an example case of Eq.~\eqref{eqn:TraceGeneral_app} of the form
\begin{equation}
T\left( {{A_1},{A_2}} \right) \equiv {\rm{Tr}}\left[ {\frac{1}{{ - {D^2} - {m^2}}}{A_1}\left( x \right)\frac{1}{{ - {D^2} - {M^2}}}{A_2}\left( x \right)} \right] .
\end{equation}
Following the prescription of Eq.~\eqref{eqn:gen_prescription} we get
\begin{equation}
T = \int {{d^4}x\int {\frac{{{d^4}p}}{{{{\left( {2\pi } \right)}^4}}}{\rm{tr}}\left[ {\frac{1}{{{{\left( {iD - p} \right)}^2} - {m^2}}}{A_1}\frac{1}{{{{\left( {iD - p} \right)}^2} - {M^2}}}{A_2}} \right]} } . \label{eqn:Tshifted}
\end{equation}
Now we do the ``CDE step'', \textit{i.e.} expand out the covariant derivatives as in Eq.~\eqref{eqn:DMBCDE}. For demonstration purposes, we only keep up to two powers of $D_\mu$, in which case we have
\begin{align}
\frac{1}{{{{\left( {iD - p} \right)}^2} - {m^2}}} &= \frac{1}{{{p^2} - {m^2}}} + \frac{1}{{{{\left( {{p^2} - {m^2}} \right)}^2}}}2i{p^\mu }{D_\mu } + \frac{{\left( {1 - 4/d} \right){p^2} - {m^2}}}{{{{\left( {{p^2} - {m^2}} \right)}^3}}}{D^2} , \label{eqn:CDE0m} \\
\frac{1}{{{{\left( {iD - p} \right)}^2} - {M^2}}} &= \frac{1}{{{p^2} - {M^2}}} + \frac{1}{{{{\left( {{p^2} - {M^2}} \right)}^2}}}2i{p^\mu }{D_\mu } + \frac{{\left( {1 - 4/d} \right){p^2} - {M^2}}}{{{{\left( {{p^2} - {M^2}} \right)}^3}}}{D^2} , \label{eqn:CDE0M}
\end{align}
As this is the key step in our method, we would like to make a few comments on it.
\begin{itemize}
  \item To determine up to which power of $D$ to truncate this CDE step, one simply sums over the mass dimensions of $A_1$, $A_2$, and compares it with the mass dimension of the effective operators under consideration. For example, suppose $A_1=A_2=\phi^2$, which sums up to $\phi^4$, and we want effective operators up to dimension-six. Then we should keep up to two powers of $D$. For all the examples discussed in this paper, we never need to do this CDE step beyond the second power of $D$.
  \item Because all calculations in this paper require at most two powers of $D$, we have made the symmetrization $p^\mu p^\nu \to \frac{1}{d}g^{\mu\nu}p^2$ in Eqs.~\eqref{eqn:CDE0m} and~\eqref{eqn:CDE0M}. Here $d=4-2\epsilon$, as usual in dimensional regularization. Terms proportional to $1-4/d$ can give a nonzero contribution if the momentum integral has a divergence.
\end{itemize}

Now using Eqs.~\eqref{eqn:CDE0m} and~\eqref{eqn:CDE0M} in Eq.~\eqref{eqn:Tshifted} and keeping only up to two powers of $D$, we get
\begin{align}
T &\supset \int {{d^4}x\int {\frac{{{d^4}p}}{{{{\left( {2\pi } \right)}^4}}}{\rm{tr}}\left\{ \begin{array}{l}
\left[ {\frac{1}{{{p^2} - {m^2}}} + \frac{1}{{{{\left( {{p^2} - {m^2}} \right)}^2}}}2ipD + \frac{{\left( {1 - 4/d} \right){p^2} - {m^2}}}{{{{\left( {{p^2} - {m^2}} \right)}^3}}}{D^2}} \right]{A_1}\\
 \times \left[ {\frac{1}{{{p^2} - {M^2}}} + \frac{1}{{{{\left( {{p^2} - {M^2}} \right)}^2}}}2ipD + \frac{{\left( {1 - 4/d} \right){p^2} - {M^2}}}{{{{\left( {{p^2} - {M^2}} \right)}^3}}}{D^2}} \right]{A_2}
\end{array} \right\}} } \nonumber \\
 &= \int {{d^4}x\int {\frac{{{d^4}p}}{{{{\left( {2\pi } \right)}^4}}}{\rm{tr}}\left\{ {\frac{1}{{{p^2} - {m^2}}}{A_1}\left[ {\frac{1}{{{p^2} - {M^2}}} + \frac{{\left( {1 - 4/d} \right){p^2} - {M^2}}}{{{{\left( {{p^2} - {M^2}} \right)}^3}}}{D^2}} \right]{A_2}} \right\}} } \nonumber \\
 &= \int {d^4}x\Bigg\{ \left[ {\int {\frac{{{d^4}p}}{{{{\left( {2\pi } \right)}^4}}}\frac{1}{{\left( {{p^2} - {M^2}} \right)\left( {{p^2} - {m^2}} \right)}}} } \right]{\rm{tr}}\left( {{A_1}{A_2}} \right) \nonumber \\
 & \qquad \qquad + \left[ {\int {\frac{{{d^4}p}}{{{{\left( {2\pi } \right)}^4}}}\frac{{\left( {1 - 4/d} \right){p^2} - {M^2}}}{{{{\left( {{p^2} - {M^2}} \right)}^3}\left( {{p^2} - {m^2}} \right)}}} } \right]{\rm{tr}}\left( {{A_1}{D^2}{A_2}} \right) \Bigg\} . \label{eqn:Ttruncated}
\end{align}
In going from the first to the second line we have used the fact that total derivatives vanish under the position integral, so that we can replace the entry from Eq.~\eqref{eqn:CDE0m} by \(1/(p^2 - m^2)\). The momentum integrals above can be worked out easily
\begin{align}
{I_1} &\equiv \int {\frac{{{d^4}p}}{{{{\left( {2\pi } \right)}^4}}}\frac{1}{{\left( {{p^2} - {M^2}} \right)\left( {{p^2} - {m^2}} \right)}}}  = \frac{i}{{{{\left( {4\pi } \right)}^2}}}\left( {\ln \frac{{{\mu ^2}}}{{{M^2}}} + \frac{{{m^2}}}{{{M^2} - {m^2}}}\ln \frac{{{m^2}}}{{{M^2}}} + 1} \right) , \\
{I_2} &\equiv \int {\frac{{{d^4}p}}{{{{\left( {2\pi } \right)}^4}}}\frac{{\left( {1 - 4/d} \right){p^2} - {M^2}}}{{{{\left( {{p^2} - {M^2}} \right)}^3}\left( {{p^2} - {m^2}} \right)}}}  = \frac{i}{{{{\left( {4\pi } \right)}^2}}}\frac{{ - {M^2}}}{{{{\left( {{M^2} - {m^2}} \right)}^2}}}\left( {\frac{{{m^2}}}{{{M^2} - {m^2}}}\ln \frac{{{m^2}}}{{{M^2}}} + \frac{{{M^2} + {m^2}}}{{2{M^2}}}} \right) ,
\end{align}
where we have used dimensional regularization in the $\overline{\text{MS}}$ renormalization scheme. In practical calculations, we only keep operators in the EFT up to some give mass dimension. This requires us to expand \(I_{1,2}\) in \(m^2/M^2\) and truncate at some order consistent with the operator dimensions kept in the EFT. If we take zeroth order in $m^2/M^2$, we get
\begin{align}
{I_1} &= \frac{i}{{{{\left( {4\pi } \right)}^2}}}\left( {\ln \frac{{{\mu ^2}}}{{{M^2}}} + 1} \right) , \label{eqn:I1} \\
{I_2} &= \frac{i}{{{{\left( {4\pi } \right)}^2}}}\frac{1}{{{M^2}}}\frac{{ - 1}}{2} . \label{eqn:I2}
\end{align}
Plugging these back into Eq.~\eqref{eqn:Ttruncated}, the functional trace $T$ evaluated at the matching scale $\mu=M$ is given by
\begin{equation}
T\left( {{A_1},{A_2},\mu  = M} \right) \supset \frac{i}{{{{\left( {4\pi } \right)}^2}}}\int {{d^4}x{\rm{tr}}\left[ {\left( {{A_1}{A_2}} \right) + \frac{1}{{{M^2}}}\frac{1}{2}\left( {D{A_1}} \right)\left( {D{A_2}} \right)} \right]} . \label{eqn:Tresult}
\end{equation}

\subsubsection{Trace Evaluation Example 2: $T_0$}

As a second example, consider the functional trace $T_0$ defined as
\begin{equation}
{T_0}\left( {{A_1},{A_2},{A_3}} \right) \equiv {\rm{Tr}}\left( {\frac{1}{{ - {D^2} - {m^2}}}{A_1}\frac{1}{{ - {D^2} - {m^2}}}{A_2}\frac{{ - {D^2}}}{{ - {D^2} - {M^2}}}{A_3}} \right) .
\end{equation}
Again, we first follow the prescription of Eq.~\eqref{eqn:gen_prescription} to get
\begin{equation}
{T_0} = \int {{d^4}x\int {\frac{{{d^4}p}}{{{{\left( {2\pi } \right)}^4}}}{\rm{tr}}\left[ {\frac{1}{{{{\left( {iD - p} \right)}^2} - {m^2}}}{A_1}\frac{1}{{{{\left( {iD - p} \right)}^2} - {m^2}}}{A_2}\frac{{{{\left( {iD - p} \right)}^2}}}{{{{\left( {iD - p} \right)}^2} - {M^2}}}{A_3}} \right]} } . \label{eqn:T0shifted}
\end{equation}
Next is the ``CDE step'':
\begin{align}
\frac{{{{\left( {iD - p} \right)}^2}}}{{{{\left( {iD - p} \right)}^2} - {M^2}}} &= \frac{{{p^2}}}{{{p^2} - {M^2}}} + \frac{{{M^2}}}{{{{\left( {{p^2} - {M^2}} \right)}^2}}}2i{p^\mu }{D_\mu } + \frac{{\left( {1 - 4/d} \right){M^2}{p^2} - {M^4}}}{{{{\left( {{p^2} - {M^2}} \right)}^3}}}{D^2} . \label{eqn:CDE1M}
\end{align}
Now using Eqs.~\eqref{eqn:CDE0m} and~\eqref{eqn:CDE1M} in Eq.~\eqref{eqn:T0shifted} and keeping only zeroth power in $D$, we get
\begin{equation}
{T_0}\left( {{A_1},{A_2},{A_3}} \right) \supset \int {{d^4}x \cdot {\rm{tr}}\left( {{A_1}{A_2}{A_3}} \right)\int {\frac{{{d^4}p}}{{{{\left( {2\pi } \right)}^4}}}\frac{{{p^2}}}{{\left( {{p^2} - {M^2}} \right){{\left( {{p^2} - {m^2}} \right)}^2}}}} } . \label{eqn:T0truncated}
\end{equation}
The momentum integral is
\begin{equation}
I \equiv \int {\frac{{{d^4}p}}{{{{\left( {2\pi } \right)}^4}}}\frac{{{p^2}}}{{\left( {{p^2} - {M^2}} \right){{\left( {{p^2} - {m^2}} \right)}^2}}}}  = \frac{i}{{{{\left( {4\pi } \right)}^2}}}\left[ {\ln \frac{{{\mu ^2}}}{{{m^2}}} + \frac{{{M^2}}}{{{M^2} - {m^2}}}\left( {\frac{{{M^2}}}{{{M^2} - {m^2}}}\ln \frac{{{m^2}}}{{{M^2}}} + 1} \right)} \right] ,
\end{equation}
Taking the limit $m^2/M^2 \to 0$
\begin{equation}
I = \frac{i}{{{{\left( {4\pi } \right)}^2}}}\left( {\ln \frac{{{\mu ^2}}}{{{m^2}}} + \ln \frac{{{m^2}}}{{{M^2}}} + 1} \right) = \frac{i}{{{{\left( {4\pi } \right)}^2}}}\left( {\ln \frac{{{\mu ^2}}}{{{M^2}}} + 1} \right) . \label{eqn:I}
\end{equation}
Plugging this into Eq.~\eqref{eqn:T0truncated}, the functional trace $T_0$ evaluated at $\mu=M$ is given by
\begin{equation}
{T_0}\left( {{A_1},{A_2},{A_3},\mu  = M} \right) \supset \frac{i}{{{{\left( {4\pi } \right)}^2}}}\int {{d^4}x \cdot {\rm{tr}}\left( {{A_1}{A_2}{A_3}} \right)} . \label{eqn:T0result}
\end{equation}

\subsubsection{Trace Evaluation Example 3: $T_1$}

As a final example, we evaluate the functional trace $T_1$ defined as
\begin{equation}
{T_1}\left( {{A_1},{A_2}} \right) \equiv \frac{1}{{{M^2}}}{\rm{Tr}}\left( {\frac{1}{{ - {D^2} - {m^2}}}{A_1}\frac{{{D^4}}}{{ - {D^2} - {M^2}}}{A_2}} \right) .
\end{equation}
This time let us put all the steps together and move more smoothly. The entire evaluation is:
\begin{align}
{T_1} &= \frac{1}{{{M^2}}}{\rm{Tr}}\left( {\frac{1}{{ - {D^2} - {m^2}}}{A_1}\frac{{{D^4}}}{{ - {D^2} - {M^2}}}{A_2}} \right) \nonumber \\
 &= \frac{1}{{{M^2}}}\int {{d^4}x\int {\frac{{{d^4}p}}{{{{\left( {2\pi } \right)}^4}}}{\rm{tr}}\left[ {\frac{1}{{{{\left( {iD - p} \right)}^2} - {m^2}}}{A_1}\frac{{{{\left( {iD - p} \right)}^4}}}{{{{\left( {iD - p} \right)}^2} - {M^2}}}{A_2}} \right]} } \nonumber \\
 &\supset \frac{1}{{{M^2}}}\int {{d^4}x\int {\frac{{{d^4}p}}{{{{\left( {2\pi } \right)}^4}}}{\rm{tr}}\left\{ \begin{array}{l}
\left[ {\frac{1}{{{p^2} - {m^2}}} + \frac{1}{{{{\left( {{p^2} - {m^2}} \right)}^2}}}2ipD + \frac{{\left( {1 - 4/d} \right){p^2} - {m^2}}}{{{{\left( {{p^2} - {m^2}} \right)}^3}}}{D^2}} \right]{A_1}\\
 \times \left[ {\frac{{{p^4}}}{{{p^2} - {M^2}}} + \frac{{ - {p^4} + 2{M^2}{p^2}}}{{{{\left( {{p^2} - {M^2}} \right)}^2}}}2ipD + \frac{{ - {p^6} + 3{M^2}{p^4} - \left( {2 + 4/d} \right){M^4}{p^2}}}{{{{\left( {{p^2} - {M^2}} \right)}^3}}}{D^2}} \right]{A_2}
\end{array} \right\}} } \nonumber \\
 &\supset \frac{1}{{{M^2}}}\int {{d^4}x\int {\frac{{{d^4}p}}{{{{\left( {2\pi } \right)}^4}}}{\rm{tr}}\left[ \begin{array}{l}
\frac{{{p^4}}}{{\left( {{p^2} - {M^2}} \right)\left( {{p^2} - {m^2}} \right)}}{A_1}{A_2}\\
 + \frac{{{p^6} - 3{M^2}{p^4} + 3{M^4}{p^2}}}{{{{\left( {{p^2} - {M^2}} \right)}^3}\left( {{p^2} - {m^2}} \right)}}\left( {D{A_1}} \right)\left( {D{A_2}} \right)
\end{array} \right]} } \nonumber \\
 &= \int {{d^4}x \cdot {\rm{tr}}\left[ {{I_3}{M^2}\left( {{A_1}{A_2}} \right) + {I_4}\left( {D{A_1}} \right)\left( {D{A_2}} \right)} \right]} \nonumber \\
 &= \frac{i}{{{{\left( {4\pi } \right)}^2}}}\int {{d^4}x \cdot {\rm{tr}}\left[ {{M^2}\left( {{A_1}{A_2}} \right) + \frac{1}{2}\left( {D{A_1}} \right)\left( {D{A_2}} \right)} \right]} .
\end{align}
In the above, we have kept terms up to $D^2$ in the ``CDE step''. We have also used the fact that under the limit $m^2/M^2\to0$, the momentum integrals $I_3$ and $I_4$ are
\begin{align}
{I_3} &\equiv \frac{1}{{{M^4}}}\int {\frac{{{d^4}p}}{{{{\left( {2\pi } \right)}^4}}}\frac{{{p^4}}}{{\left( {{p^2} - {M^2}} \right)\left( {{p^2} - {m^2}} \right)}}}  = \frac{i}{{{{\left( {4\pi } \right)}^2}}} , \\
{I_4} &\equiv \frac{1}{{{M^2}}}\int {\frac{{{d^4}p}}{{{{\left( {2\pi } \right)}^4}}}\frac{{{p^6} - 3{M^2}{p^4} + 3{M^4}{p^2}}}{{{{\left( {{p^2} - {M^2}} \right)}^3}\left( {{p^2} - {m^2}} \right)}}}  = \frac{i}{{{{\left( {4\pi } \right)}^2}}}\frac{1}{2} .
\end{align}

We hope that the general evaluation technique is clear from the three examples above. This is a systematic prescription that one can use to work out any functional trace in the form of Eq.~\eqref{eqn:TraceGeneral_app}.

%% file: app_DeriveResolve.tex
In this appendix, we give the derivation of taking the functional determinant
\begin{equation}
\log\det \left( { - {{\left. {\frac{{{\delta ^2}{S_{{\rm{UV}}}}\left[ {\phi ,\Phi } \right]}}{{\delta {{\left( {\phi ,\Phi } \right)}^2}}}} \right|}_{\Phi  = {\Phi _c}\left[ \phi  \right]}}} \right) , \label{eqn:mixeddet_app}
\end{equation}
and bringing it to the form of Eq.~\eqref{eqn:determinantsresolved}. 
The basic idea is to break down this big functional determinant into smaller ones and then apply some straightforward manipulations of the functional derivative.

The functional derivative matrix in Eq.~\eqref{eqn:mixeddet} has the following $2\times2$ form
\begin{equation}
- {\left. {\frac{{{\delta ^2}{S_{{\rm{UV}}}}\left[ {\phi ,\Phi } \right]}}{{\delta {{\left( {\phi ,\Phi } \right)}^2}}}} \right|_{\Phi  = {\Phi _c}\left[ \phi  \right]}} = \left( {\begin{array}{*{20}{c}}
{{{\left. { - \frac{{{\delta ^2}{S_{{\rm{UV}}}}\left[ {\phi ,\Phi } \right]}}{{\delta {\phi ^2}}}} \right|}_{{\Phi _c}}}}&{{{\left. { - \frac{{{\delta ^2}{S_{{\rm{UV}}}}\left[ {\phi ,\Phi } \right]}}{{\delta \phi \delta \Phi }}} \right|}_{{\Phi _c}}}}\\
{ - {{\left. {\frac{{{\delta ^2}{S_{{\rm{UV}}}}\left[ {\phi ,\Phi } \right]}}{{\delta \phi \delta \Phi }}} \right|}_{{\Phi _c}}}}&{ - {{\left. {\frac{{{\delta ^2}{S_{{\rm{UV}}}}\left[ {\phi ,\Phi } \right]}}{{\delta {\Phi ^2}}}} \right|}_{{\Phi _c}}}}
\end{array}} \right) \equiv \left( {\begin{array}{*{20}{c}}
a&b\\
c&d
\end{array}} \right) ,
\end{equation}
whose determinant should then follow as
\begin{equation}
\det \left( {\begin{array}{*{20}{c}}
a&b\\
c&d
\end{array}} \right) = \det \left( {\begin{array}{*{20}{c}}
{a - b{d^{ - 1}}c}&0\\
c&d
\end{array}} \right) = \det \left( d \right)\det \left( {a - b{d^{ - 1}}c} \right) . \label{eqn:detda}
\end{equation}
We make the identifications that
\begin{align}
d &=  - {\left. {\frac{{{\delta ^2}{S_{{\rm{UV}}}}\left[ {\phi ,\Phi } \right]}}{{\delta {\Phi ^2}}}} \right|_{{\Phi _c}}} , \label{eqn:detd} \\
a - b{d^{ - 1}}c &= {\left. { - \frac{{{\delta ^2}{S_{{\rm{UV}}}}\left[ {\phi ,\Phi } \right]}}{{\delta {\phi ^2}}}} \right|_{{\Phi _c}}} + {\left. {\frac{{{\delta ^2}{S_{{\rm{UV}}}}\left[ {\phi ,\Phi } \right]}}{{\delta \phi \delta \Phi }}} \right|_{{\Phi _c}}}{\left( {{{\left. {\frac{{{\delta ^2}{S_{{\rm{UV}}}}\left[ {\phi ,\Phi } \right]}}{{\delta {\Phi ^2}}}} \right|}_{{\Phi _c}}}} \right)^{ - 1}}{\left. {\frac{{{\delta ^2}{S_{{\rm{UV}}}}\left[ {\phi ,\Phi } \right]}}{{\delta \phi \delta \Phi }}} \right|_{{\Phi _c}}} . \label{eqn:deta}
\end{align}
Our task is to simplify Eq.~\eqref{eqn:deta}. Making use of the definition of $\Phi_c[\phi]$,
\begin{equation}
0 = {\left. {\frac{{\delta {S_{{\rm{UV}}}}\left[ {\phi ,\Phi } \right]}}{{\delta \Phi }}} \right|_{{\Phi _c}}} ,
\end{equation}
and the chain rule,
\begin{equation}
\frac{\delta }{{\delta \phi }} = \left. \frac{\delta }{{\delta \phi }}\right|_{\Ph_c} + \frac{{\delta {\Phi _c}\left[ \phi  \right]}}{{\delta \phi }}\frac{\delta }{{\delta {\Phi _c}\left[ \phi  \right]}} ,
\end{equation}
we get
\begin{equation}
0 = \frac{\delta }{{\delta \phi }}\left( {{{\left. {\frac{{\delta {S_{{\rm{UV}}}}\left[ {\phi ,\Phi } \right]}}{{\delta \Phi }}} \right|}_{{\Phi _c}}}} \right) = {\left. {\frac{{{\delta ^2}{S_{{\rm{UV}}}}\left[ {\phi ,\Phi } \right]}}{{\delta \phi \delta \Phi }}} \right|_{{\Phi _c}}} + {\left. {\frac{{\delta {\Phi _c}\left[ \phi  \right]}}{{\delta \phi }}\frac{{{\delta ^2}{S_{{\rm{UV}}}}\left[ {\phi ,\Phi } \right]}}{{\delta {\Phi ^2}}}} \right|_{{\Phi _c}}} ,
\end{equation}
which gives
\begin{equation}
\frac{{\delta {\Phi _c}\left[ \phi  \right]}}{{\delta \phi }} =  - {\left. {\frac{{{\delta ^2}{S_{{\rm{UV}}}}\left[ {\phi ,\Phi } \right]}}{{\delta \phi \delta \Phi }}} \right|_{{\Phi _c}}}{\left( {{{\left. {\frac{{{\delta ^2}{S_{{\rm{UV}}}}\left[ {\phi ,\Phi } \right]}}{{\delta {\Phi ^2}}}} \right|}_{{\Phi _c}}}} \right)^{ - 1}} .
\end{equation}
This expression is nothing but the usual formula for ``the derivative of an implicit function''. However, it helps us a lot because using it in Eq.~\eqref{eqn:deta} we get
\begin{equation}
a - b{d^{ - 1}}c = {\left. { - \frac{{{\delta ^2}{S_{{\rm{UV}}}}\left[ {\phi ,\Phi } \right]}}{{\delta {\phi ^2}}}} \right|_{{\Phi _c}}} - \frac{{\delta {\Phi _c}\left[ \phi  \right]}}{{\delta \phi }}{\left. {\frac{{{\delta ^2}{S_{{\rm{UV}}}}\left[ {\phi ,\Phi } \right]}}{{\delta \phi \delta \Phi }}} \right|_{{\Phi _c}}} =  - \frac{{{\delta ^2}{S_{{\rm{UV}}}}\left[ {\phi ,{\Phi _c}\left[ \phi  \right]} \right]}}{{\delta {\phi ^2}}} .
\end{equation}
Now using Eq.~\eqref{eqn:detda}, we obtain a resolved form of Eq.~\eqref{eqn:mixeddet_app}:
\begin{equation}
\log \det \left( { - {{\left. {\frac{{{\delta ^2}{S_{{\rm{UV}}}}\left[ {\phi ,\Phi } \right]}}{{\delta {{\left( {\phi ,\Phi } \right)}^2}}}} \right|}_{{\Phi _c}}}} \right) = \log\det \left( { - {{\left. {\frac{{{\delta ^2}{S_{{\rm{UV}}}}\left[ {\phi ,\Phi } \right]}}{{\delta {\Phi ^2}}}} \right|}_{{\Phi _c}}}} \right) + \log\det \left( { - \frac{{{\delta ^2}{S_{{\rm{UV}}}}\left[ {\phi ,{\Phi _c}\left[ \phi  \right]} \right]}}{{\delta {\phi ^2}}}} \right) . \label{eqn:determinantsresolved_app}
\end{equation}
which is nothing but Eq.~\eqref{eqn:determinantsresolved}.

%% file: app_Wilsonian.tex
In this appendix, we elaborate more on the local vs. non-local mismatch discussed in section~\ref{subsec:DirectGeneral}. We learned in section~\ref{subsec:DirectGeneral} that the non-local action $S_\text{eff}[\phi]$ obtained by integrating out the heavy field $\Phi$ in the path integral
\begin{equation}
{e^{i{S_{{\rm{eff}}}}\left[ \phi  \right]}} = \int {D\Phi {e^{i{S_{{\rm{UV}}}}\left[ {\phi ,\Phi } \right]}}} ,
\end{equation}
is completely equivalent to the UV theory in regarding to the low energy physics of $\phi$, namely that
\begin{equation}
{\Gamma _{{\rm{eff}}}}\left[ \phi  \right] = {\Gamma _{{\rm{L,UV}}}}\left[ \phi  \right].
\end{equation}
On the other hand, the ``local counterpart'' to $S_\text{eff}[\phi]$, which is simply an expansion of it into a sum of local operators, would not give the same 1PI effective action. To better distinguish the non-local $S_\text{eff}[\phi]$ and its local counterpart, if we use the notation $S_\text{eff, non-local}[\phi]\equiv S_\text{eff}[\phi]$, and $S_\text{eff, local}[\phi]$ for the local counterpart of $S_\text{eff, non-local}[\phi]$, then this local vs. non-local difference can be expressed as
\begin{equation}
{\Gamma _{{\text{eff, local}}}}\left[ \phi  \right] \ne {\Gamma _{{\text{eff, non-local}}}}\left[ \phi  \right] = {\Gamma _{{\rm{L,UV}}}}\left[ \phi  \right] .
\end{equation}
It is exactly due to this mismatch that the mixed one-loop piece of matching, \textit{i.e.} $c_{i,\text{mixed}}^{(1)}$, is nonzero.

On the other hand, we also learned that this mismatch $\Gamma_\text{eff, local}[\phi] \ne \Gamma_\text{eff, non-local}[\phi]$ is a result of using dimensional regularization. Specifically, using $\Gamma_\text{eff, local}[\phi]$ amounts to expanding the propagator $\frac{1}{p^2-M^2}$ inside the momentum integral (see the discussion around Eq.~\eqref{eqn:dimregmismatch}), which is illegitimate under dimensional regularization. However, this mismatch can be avoided by using other regularization scheme, such as a hard cutoff $p^2 < \Lambda_0^2$ in the Euclidean space:
\begin{eqnarray}
  \int_0^{\Lambda_0} {\cal D}\phi{\cal D}\Phi e^{-S_{\rm UV}(\phi,\Phi;\Lambda_0)}
  &=& \int_0^{\Lambda_1}\int_{\Lambda_1}^{\Lambda_0}
  {\cal D}\phi{\cal D}\Phi e^{-S_{\rm UV}(\phi,\Phi;\Lambda_0)} \nonumber \\
  &=& \int_0^{\Lambda_1}
  {\cal D}\phi{\cal D}\Phi e^{-S_{\rm UV}(\phi,\Phi; \Lambda_1)} \nonumber \\
  &=& \int_0^{\Lambda_1}
  {\cal D}\phi e^{-S_{\rm eff} (\phi; \Lambda_1)}\ .
\end{eqnarray}
In the very last step, we merely integrate out the heavy field $\Phi$ of mass $M$, given that the cutoff $\Lambda_1 < M$ at this stage.
Then $S_\text{eff} (\phi; \Lambda_1)$ should make the same physical predictions as the UV theory, even after it is expanded into a sum of local operators. This $S_\text{eff} (\phi; \Lambda_1)$ defined with a hard cutoff regularization scheme is the Wilsonian effective action. We demonstrate this point in this appendix.

Let us come back to the example model considered in section~\ref{subsec:ToyScalar},
\begin{equation}
{{\cal L}_{{\text{UV}}}}\left( {\phi ,\Phi } \right) = \frac{1}{2}\Phi \left( { - {\partial ^2} - {M^2}} \right)\Phi  - \frac{\lambda }{{3!}}\Phi {\phi ^3} + \frac{1}{2}\phi \left( { - {\partial ^2} - {m^2}} \right)\phi  - \frac{\kappa }{{4!}}{\phi ^4} .
\end{equation}
Here, $\phi$ is the light field of mass $m$, while $\Phi$ is the heavy field of mass $M \gg m$.  We assume $\kappa \approx \frac{\lambda^2}{(4\pi)^2}$ (similar to the assumption $\kappa \approx \frac{e^2}{(4\pi)^2}$ in the Coleman--Weiberg theory). This assumption is not necessary, but is there only for the purpose of ignoring $\kappa^2$ correction to $\kappa$. Our main consideration is the interplay between the heavy-light loop at $O(\lambda^2)$ that renormalizes $\kappa$ after $\Phi$ is integrated out.

Let us first look at the computation in the UV theory. When the momentum slices $\Lambda_1 <  p < \Lambda_0$ are integrated out, the coupling
$\kappa$ needs to be changed as
\begin{eqnarray}
  \lefteqn{
  \kappa(\Lambda_1)
  = \kappa(\Lambda_0)
      - 3\lambda^2 \int_{\Lambda_1}^{\Lambda_0}
  \frac{d^4 p}{(2\pi)^4} \frac{1}{p^2+M^2}
      \frac{1}{p^2+m^2} }\nonumber \\
  &&= \kappa(\Lambda_0) - \frac{3\lambda^2}{(4\pi)^2} \frac{1}{M^2-m^2}
      \left( M^2 \log \frac{\Lambda_0^2+M^2}{\Lambda_1^{2}+M^2}
      - m^2 \log \frac{\Lambda_0^2+m^2}{\Lambda_1^{2}+m^2} \right).
\end{eqnarray}
In particular when $M, m \ll \Lambda_1, \Lambda_0$, it reduces to the usual logarithmic running
\begin{eqnarray}
  \kappa(\Lambda_1)
  &=& \kappa(\Lambda_0) - \frac{3\lambda^2}{(4\pi)^2}
      \log \frac{\Lambda_0^2}{\Lambda_1^{2}}\ .
\end{eqnarray}
Yet our main interest is when the all the momenta above $M$ are integrated out so that $\Lambda_1 < M$.

We can further integrate out momentum slices to go to even lower $\Lambda_2 < \Lambda_1 < M$.  The change in the coupling is obviously
\begin{equation}
  \kappa(\Lambda_2)
  = \kappa(\Lambda_1) - \frac{3\lambda^2}{(4\pi)^2} \frac{1}{M^2-m^2}
      \left( M^2 \log \frac{\Lambda_1^2+M^2}{\Lambda_2^{2}+M^2}
      - m^2 \log \frac{\Lambda_1^2+m^2}{\Lambda_2^{2}+m^2} \right). \label{eqn:kappacutoff}
\end{equation}
Note that this expression allows for a Taylor expansion in $\frac{\Lambda_{1,2}^2}{M^2}$ within the radius of convergence $\Lambda_{1,2}^2 < M^2$.

The question is whether this result can be reproduced by the IR theory with the local Lagrangian $\mathcal{L}_\text{eff, local}(\phi)$ after $\Phi$ is integrated out. The answer is yes as long as $\Lambda_1 < M$. Integrating out $\Phi$ results in the $\mathcal{L}_\text{eff, local}(\phi)$ with the effective local operators
\begin{equation}
  {\cal L}_\text{eff, local} (\phi) = \frac{1}{2} \phi (-\partial^2-m^2) \phi - \frac{\kappa}{4!} \phi^4
  + \sum\limits_{n=0}^\infty \frac{\lambda^2}{72M^2} \phi^3 \left(\frac{-\partial^2}{M^2}\right)^n \phi^3 .
\end{equation}
Contracting a $\phi$ before the derivative and another $\phi$ after the derivative in the effective operators results in the renormalization of $\kappa$ in the IR theory,\footnote{Other contractions result in different operators such as $\phi \partial^2 \phi^3$, not of our interest here.}
\begin{eqnarray}
  \kappa(\Lambda_2)
  &=& \kappa(\Lambda_1) - 3\lambda^2 \int_{\Lambda_2}^{\Lambda_1}
  \frac{d^4 p}{(2\pi)^4} \frac{1}{M^2} \sum_{n=0}^\infty
  \left(\frac{-p^2}{M^2}\right)^n
      \frac{1}{p^2+m^2} \nonumber \\
  &=& \kappa(\Lambda_1)
      - 3\lambda^2 \frac{1}{(4\pi)^2} \int_{\Lambda_2}^{\Lambda_1}
      \frac{d p^2}{(2\pi)^4} \sum_{n=0}^\infty \frac{1}{M^{2n+2}}
      \frac{(-1)^n (p^2)^{n+1}}{p^2+m^2} \nonumber \\
  &=& \kappa(\Lambda_1)
      - \frac{3\lambda^2}{(4\pi)^2} \sum_{n=0}^\infty \frac{m^{2n+2}}{M^{2n+2}}
      (B_{-\Lambda_1^2/m^2}(n+2,0) - B_{-\Lambda_2^2/m^2}(n+2,0) ).
      \nonumber \\
\end{eqnarray}
Here $B_z(p,q)$ is the incomplete Beta function $ B_z(p,q) = \int_0^z t^{p-1} (1-t)^{q-1} dt $ (not to be confused with the beta function of the running coupling constant). For large $n$,
\begin{equation}
 B_{-\Lambda^2/m^2}(n+2,0) \approx \frac{-1}{n+1} \left( \frac{-\Lambda^2}{m^2} \right)^{n+1} ,
\end{equation}
and hence the sum over $n$ converges for $\Lambda_{1,2} < M$. Therefore, one can interchange the order of the sum over $n$ and the integration over $p^2$. We see that the IR theory with $\Phi$ integrated out reproduces the correct result in the cutoff-dependence of the coupling $\kappa$.

However, this is not the case in the $\overline{\rm MS}$ scheme. As a function of the renormalization scale $\mu$ in $d=4-2\epsilon$ dimensions, the UV theory gives
\begin{equation}
  \kappa(\mu_1) = \kappa(\mu_0)
  - \frac{3\lambda^2 (\mu_0^{2\epsilon} - \mu_1^{2\epsilon})}{(4\pi)^2}
  \frac{\Gamma(\epsilon)}{1-\epsilon} \frac{1}{M^2-m^2}
  ((M^2)^{1-\epsilon} - (m^2)^{1-\epsilon}) , \label{eqn:MSbarUV}
\end{equation}
which obviously describes the same running as in Eq.~\eqref{eqn:kappacutoff}. But the IR theory does not reproduce this result,
\begin{align}
  \kappa(\mu_1) - \kappa(\mu_0) &= -3\lambda^2 (\mu_0^{2\epsilon} - \mu_1^{2\epsilon})
      \int \frac{d^{d}p}{(2\pi)^d}
      \frac{1}{M^2} \sum_{n=0}^\infty
      \left(\frac{-p^2}{M^2}\right)^n \frac{1}{p^2+m^2} \nonumber \\
  &= -3\lambda^2 (\mu_0^{2\epsilon} - \mu_1^{2\epsilon})
      \int \frac{\pi^{d/2}}{\Gamma(d/2)}
      \frac{(p^2)^{1-\epsilon} d p^2}{(2\pi)^d}
      \frac{1}{M^2} \sum_{n=0}^\infty
      \left(\frac{-p^2}{M^2}\right)^n \frac{1}{p^2+m^2}  \nonumber \\
  &= -\frac{3\lambda^2 (\mu_0^{2\epsilon} - \mu_1^{2\epsilon})}
      {(4\pi)^{2-\epsilon}\Gamma(2-\epsilon)}
      \frac{1}{M^2} \sum_{n=0}^\infty \frac{1}{M^{2n}}
      \int d p^2
      \frac{(-1)^n (p^2)^{n+1-\epsilon}}{p^2+m^2}
      \nonumber \\
  &= -\frac{3\lambda^2 (\mu_0^{2\epsilon} - \mu_1^{2\epsilon})}
      {(4\pi)^{2-\epsilon}\Gamma(2-\epsilon)}
      \frac{1}{M^2} \sum_{n=0}^\infty (-1)^n
      \frac{(m^2)^{n+1-\epsilon} }{M^{2n}}
      \frac{\Gamma(n+2-\epsilon) \Gamma(-n-1+\epsilon)}{\Gamma(1)}\nonumber \\
  &= -\frac{3\lambda^2 (\mu_0^{2\epsilon} - \mu_1^{2\epsilon})}
      {(4\pi)^{2-\epsilon}\Gamma(2-\epsilon)}
      \frac{(m^2)^{1-\epsilon}}{M^2} \sum_{n=0}^\infty
      \left( \frac{m^2}{M^2} \right)^{n}
      \frac{(-1)^n\pi}{\sin (2-\epsilon+n)\pi} \nonumber \\
  &= -\frac{3\lambda^2 (\mu_0^{2\epsilon} - \mu_1^{2\epsilon})}
      {(4\pi)^{2-\epsilon}\Gamma(2-\epsilon)}
      \frac{(m^2)^{1-\epsilon}}{M^2} \sum_{n=0}^\infty
      \left( \frac{m^2}{M^2} \right)^{n}
      \frac{- \pi}{\sin \epsilon\pi} \nonumber \\
  &= -\frac{3\lambda^2 (\mu_0^{2\epsilon} - \mu_1^{2\epsilon})}
      {(4\pi)^{2-\epsilon}\Gamma(2-\epsilon)}
      \frac{(m^2)^{1-\epsilon}}{M^2}
      \frac{1}{1-\frac{m^2}{M^2}}
      \Gamma(-1+\epsilon) \Gamma(2-\epsilon) \nonumber \\
  &= \frac{3\lambda^2 (\mu_0^{2\epsilon} - \mu_1^{2\epsilon})}
      {(4\pi)^{2-\epsilon}}
      \frac{(m^2)^{1-\epsilon}}{M^2-m^2}
      \frac{\Gamma(\epsilon)}{1-\epsilon} .
\end{align}
Compared with Eq.~\eqref{eqn:MSbarUV}, we see that the piece proportional to $(M^2)^{1-\epsilon}$ is not reproduced. The reason is very simple. In the IR theory, the dependence on $M^2$ is always in integer powers because it comes from the local operators after the expansion in the inverse power of $M^2$. The momentum integral no longer knows anything about $M^2$. Therefore, it can never reproduce a fractional power $(M^2)^{1-\epsilon}$ in the $\overline{\rm MS}$ scheme. This missing piece is the $c_{i,\text{mixed}}^{(1)}$.

Note that we defined the Wilsonian effective action using a hard cutoff regularization scheme, but one can also define it using the Gaussian cutoff. In this case, however, the loop integral involves momenta above the cutoff even though its contribution is supposed to be Gaussian-suppressed. This causes the IR theory to be an asymptotic expansion in $1/M$, which does not converge, but provides a good approximation for $\Lambda_1 \ll M$. Again one can confirm that the IR theory reproduces the result in the UV theory, not as a Taylor expansion but rather as an asymptotic expansion.

%% file: app_TripletScalar.tex
This appendix contains some supplementary calculation details for our Triplet Scalar Model example discussed in Section~\ref{subsec:TripletScalar}. Specifically, we encounter four functional traces $S_1$, $S_2$, $S_3$, and $S_K$ in Eqs.~\eqref{eqn:S123} and~\eqref{eqn:SK}, but did not show the details of evaluating them. Let us list out some of the steps here. Before the actual evaluating steps, it is useful to prepare a list of functional traces that are involved. These functional traces were evaluated using the CDE technique described in Appendix~\ref{app:Trace}.
\begin{align}
{T_1}\left( {{A_1},{A_2}} \right) &\equiv \frac{1}{{{M^2}}}{\rm{Tr}}\left( {\frac{1}{{ - {D^2} - {m^2}}}{A_1}\frac{{{D^4}}}{{ - {D^2} - {M^2}}}{A_2}} \right) \nonumber\\
 &\supset \frac{i}{{{{\left( {4\pi } \right)}^2}}}\int {{d^4}x \cdot {\rm{tr}}\left[ {{M^2}\left( {{A_1}{A_2}} \right) + \frac{1}{2}\left( {D{A_1}} \right)\left( {D{A_2}} \right)} \right]} \label{eqn:T1} \\
{T_2}\left( {{A_1},{A_2},{A_3}} \right) &\equiv {\rm{Tr}}\left( {\frac{1}{{ - {D^2} - {m^2}}}{A_1}\frac{1}{{ - {D^2} - {m^2}}}{A_2}\frac{{{D^4}}}{{ - {D^2} - {M^2}}}{A_3}} \right) \nonumber \\
 &\supset \frac{i}{{{{\left( {4\pi } \right)}^2}}}\int {{d^4}x \cdot {\rm{tr}}\left[ \begin{array}{l}
{M^2}\left( {{A_1}{A_2}{A_3}} \right) + \frac{1}{2}\left( {D{A_1}} \right)\left( {D{A_2}} \right){A_3}\\
 + \frac{1}{2}\left( {D{A_1}} \right){A_2}\left( {D{A_3}} \right) + \frac{5}{2}{A_1}\left( {D{A_2}} \right)\left( {D{A_3}} \right)
\end{array} \right]} \label{eqn:T2} \\
{T_3}\left( {{A_1},{A_2},{A_3}} \right) &\equiv {\rm{Tr}}\left( {\frac{1}{{ - {D^2} - {m^2}}}{A_1}\frac{1}{{ - {D^2} - {M^2}}}{A_2}\frac{{{D^4}}}{{ - {D^2} - {M^2}}}{A_3}} \right) \nonumber \\
 &\supset \frac{i}{{{{\left( {4\pi } \right)}^2}}}\int {{d^4}x \cdot {\rm{tr}}\left[ \begin{array}{l}
\frac{1}{6}\left( {D{A_1}} \right)\left( {D{A_2}} \right){A_3} + \frac{1}{2}\left( {D{A_1}} \right){A_2}\left( {D{A_3}} \right)\\
 + \frac{2}{3}{A_1}\left( {D{A_2}} \right)\left( {D{A_3}} \right)
\end{array} \right]} \label{eqn:T3} \\
{T_4}\left( {{A_1},{A_2},{A_3}} \right) &\equiv \frac{1}{{{M^4}}}{\rm{Tr}}\left( {\frac{1}{{ - {D^2} - {m^2}}}{A_1}\frac{{{D^4}}}{{ - {D^2} - {M^2}}}{A_2}\frac{{{D^4}}}{{ - {D^2} - {M^2}}}{A_3}} \right) \nonumber \\
 &\supset \frac{i}{{{{\left( {4\pi } \right)}^2}}}\int {{d^4}x \cdot {\rm{tr}}\left[ \begin{array}{l}
\frac{7}{6}\left( {D{A_1}} \right)\left( {D{A_2}} \right){A_3} + \frac{3}{2}\left( {D{A_1}} \right){A_2}\left( {D{A_3}} \right)\\
 + \frac{7}{6}{A_1}\left( {D{A_2}} \right)\left( {D{A_3}} \right)
\end{array} \right]} \label{eqn:T4} \\
{T_5}\left( {{A_1},{A_2},{A_3},{A_4}} \right) &\equiv {\rm{Tr}}\left( {\frac{1}{{ - {D^2} - {m^2}}}{A_1}\frac{{ - {D^2}}}{{ - {D^2} - {M^2}}}{A_2}\frac{1}{{ - {D^2} - {m^2}}}{A_3}\frac{{{D^4}}}{{ - {D^2} - {M^2}}}{A_4}} \right) \nonumber \\
 &\supset \frac{i}{{{{\left( {4\pi } \right)}^2}}}\int {{d^4}x \cdot {\rm{tr}}\left[ \begin{array}{l}
\frac{1}{6}\left( {D{A_1}} \right)\left( {D{A_2}} \right){A_3}{A_4} + \frac{1}{6}\left( {D{A_1}} \right){A_2}\left( {D{A_3}} \right){A_4}\\
 + \frac{1}{2}\left( {D{A_1}} \right){A_2}{A_3}\left( {D{A_4}} \right) + \frac{1}{2}{A_1}\left( {D{A_2}} \right)\left( {D{A_3}} \right){A_4}\\
 + \frac{1}{6}{A_1}\left( {D{A_2}} \right){A_3}\left( {D{A_4}} \right) + \frac{2}{3}{A_1}{A_2}\left( {D{A_3}} \right)\left( {D{A_4}} \right)
\end{array} \right]} \label{eqn:T5} \\
{T_6}\left( {{A_1},{A_2},{A_3},{A_4}} \right) &\equiv \frac{1}{{{M^2}}}{\rm{Tr}}\left( {\frac{1}{{ - {D^2} - {m^2}}}{A_1}\frac{{{D^4}}}{{ - {D^2} - {M^2}}}{A_2}\frac{1}{{ - {D^2} - {m^2}}}{A_3}\frac{{{D^4}}}{{ - {D^2} - {M^2}}}{A_4}} \right) \nonumber \\
 &\supset \frac{i}{{{{\left( {4\pi } \right)}^2}}}\int {{d^4}x \cdot {\rm{tr}}\left[ \begin{array}{l}
\frac{7}{6}\left( {D{A_1}} \right)\left( {D{A_2}} \right){A_3}{A_4} + \frac{1}{6}\left( {D{A_1}} \right){A_2}\left( {D{A_3}} \right){A_4}\\
 + \left( {D{A_1}} \right){A_2}{A_3}\left( {D{A_4}} \right) + {A_1}\left( {D{A_2}} \right)\left( {D{A_3}} \right){A_4}\\
 + \frac{1}{6}{A_1}\left( {D{A_2}} \right){A_3}\left( {D{A_4}} \right) + \frac{7}{6}{A_1}{A_2}\left( {D{A_3}} \right)\left( {D{A_4}} \right)
\end{array} \right]} \label{eqn:T6}
\end{align}

Now let us start with $S_1$. The procedure goes as follows:
\begin{align}
{S_1} &\equiv \frac{i}{2}{\rm{Tr}}{\left[ { - 2\lambda {\kappa ^2}\frac{1}{{ - {D^2} - {m^2}}}\left( {\begin{array}{*{20}{c}}
{{a_1}}&{{b_1}}\\
{b_1^*}&{a_1^T}
\end{array}} \right)\frac{1}{{ - {D^2} - {m^2}}}\left( {\begin{array}{*{20}{c}}
{{a_2}}&{{b_2}}\\
{b_2^*}&{a_2^T}
\end{array}} \right)} \right]_{\rm{d}}} \nonumber \\
 &=  - i\lambda {\kappa ^2}{\rm{Tr}}{\left[ {\frac{1}{{ - {D^2} - {m^2}}}{a_1}\frac{1}{{ - {D^2} - {m^2}}}{a_2} + \frac{1}{{ - {D^2} - {m^2}}}{b_1}\frac{1}{{ - {D^2} - {m^2}}}b_2^* + c.c.} \right]_{\rm{d}}} \nonumber \\
 &=  - i\lambda {\kappa ^2}{\rm{Tr}}{\left\{ \begin{array}{l}
\frac{1}{{ - {D^2} - {m^2}}}\left( {{{\left| H \right|}^2} + H{H^\dag }} \right)\frac{1}{{ - {D^2} - {m^2}}}\left[ \begin{array}{l}
{t^a}{\left( {\frac{1}{{ - {D^2} - {M^2}}}{H^\dag }{t^a}H} \right)_x}\\
 + {t^a}H\frac{1}{{ - {D^2} - {M^2}}}{H^\dag }{t^a}
\end{array} \right] + c.c.\\
 + \frac{1}{{ - {D^2} - {m^2}}}H{H^T}\frac{1}{{ - {D^2} - {m^2}}}{t^{a*}}{H^*}\frac{1}{{ - {D^2} - {M^2}}}{H^\dag }{t^a} + c.c.
\end{array} \right\}_{\rm{d}}} \nonumber \\
 &= \frac{{ - i\lambda {\kappa ^2}}}{{{M^4}}}\text{Tr}\left[ \begin{array}{l}
\frac{1}{{ - {D^2} - {m^2}}}\left( {{{\left| H \right|}^2} + H{H^\dag }} \right)\frac{1}{{ - {D^2} - {m^2}}}\left( {{t^a}H} \right)\frac{{{D^4}}}{{ - {D^2} - {M^2}}}\left( {{H^\dag }{t^a}} \right) + c.c.\\
 + \frac{1}{{ - {D^2} - {m^2}}}\left( {H{H^T}} \right)\frac{1}{{ - {D^2} - {m^2}}}\left( {{t^{a*}}{H^*}} \right)\frac{{{D^4}}}{{ - {D^2} - {M^2}}}\left( {{H^\dag }{t^a}} \right) + c.c.
\end{array} \right] \nonumber \\
 &= \frac{{ - i\lambda {\kappa ^2}}}{{{M^4}}}\left[ {{T_2}\left( {{{\left| H \right|}^2} + H{H^\dag },{t^a}H,{H^\dag }{t^a}} \right) + {T_2}\left( {H{H^T},{t^{a*}}{H^*},{H^\dag }{t^a}} \right) + c.c.} \right] \nonumber \\
 &= \frac{{\lambda {\kappa ^2}}}{{{M^4}}}\frac{1}{{{{\left( {4\pi } \right)}^2}}}\int {{d^4}x\left[ {\frac{{13}}{8}{{\left( {{D_\mu }{{\left| H \right|}^2}} \right)}^2} - \frac{3}{2}{{\left| {{H^\dag }{D_\mu }H} \right|}^2} + \frac{{25}}{4}{{\left| H \right|}^2}{{\left| {{D_\mu }H} \right|}^2}} \right]} . \label{eqn:S1app}
\end{align}

Let us describe what we have done in the above six lines. We started with the definition of $S_1$ (\textit{i.e.} Eq.~\eqref{eqn:S1}) in the first line, and multiplied the matrices out to obtain the second line. In the third line, we plugged in the expression of $a_1$, $b_1$, $a_2$, and $b_2$. 
Then we identified and dropped the ``local counterparts'' in the fourth line, according to the splitting in Eqs.~\eqref{eqn:TSPsplit}. In the fifth line, the result is written in terms of the functional traces defined in Eqs.~\eqref{eqn:T1}-\eqref{eqn:T6}. In the last line, we used the prepared list of evaluated functional traces to write the result in terms of effective operators. Clearly, $S_2$, $S_3$, and $S_K$ can be evaluated with the same procedure. The detailed steps are given below.

\begin{align}
{S_2} &\equiv \frac{i}{2}{\rm{Tr}}{\left[ { - 8{\kappa ^4}\frac{1}{{ - {D^2} - {m^2}}}\left( {\begin{array}{*{20}{c}}
{{a_2}}&{{b_2}}\\
{b_2^*}&{a_2^T}
\end{array}} \right)\frac{1}{{ - {D^2} - {m^2}}}\left( {\begin{array}{*{20}{c}}
{{a_2}}&{{b_2}}\\
{b_2^*}&{a_2^T}
\end{array}} \right)} \right]_{\rm{d}}} \nonumber \\
 &=  - 4i{\kappa ^4}{\rm{Tr}}{\left[ {\frac{1}{{ - {D^2} - {m^2}}}{a_2}\frac{1}{{ - {D^2} - {m^2}}}{a_2} + \frac{1}{{ - {D^2} - {m^2}}}{b_2}\frac{1}{{ - {D^2} - {m^2}}}b_2^* + c.c.} \right]_{\rm{d}}} \nonumber \\
 &=  - 4i{\kappa ^4}{\rm{Tr}}{\left[ \begin{array}{l}
\frac{1}{{ - {D^2} - {m^2}}}{t^a}{\left( {\frac{1}{{ - {D^2} - {M^2}}}{H^\dag }{t^a}H} \right)_x}\frac{1}{{ - {D^2} - {m^2}}}{t^b}{\left( {\frac{1}{{ - {D^2} - {M^2}}}{H^\dag }{t^b}H} \right)_x} + c.c.\\
 + 2\frac{1}{{ - {D^2} - {m^2}}}{t^a}{\left( {\frac{1}{{ - {D^2} - {M^2}}}{H^\dag }{t^a}H} \right)_x}\frac{1}{{ - {D^2} - {m^2}}}{t^b}H\frac{1}{{ - {D^2} - {M^2}}}{H^\dag }{t^b} + c.c.\\
 + \frac{1}{{ - {D^2} - {m^2}}}{t^a}H\frac{1}{{ - {D^2} - {M^2}}}{H^\dag }{t^a}\frac{1}{{ - {D^2} - {m^2}}}{t^b}H\frac{1}{{ - {D^2} - {M^2}}}{H^\dag }{t^b} + c.c.\\
 + \frac{1}{{ - {D^2} - {m^2}}}{t^a}H\frac{1}{{ - {D^2} - {M^2}}}{H^T}{t^{a*}}\frac{1}{{ - {D^2} - {m^2}}}{t^{b*}}{H^*}\frac{1}{{ - {D^2} - {M^2}}}{H^\dag }{t^b} + c.c.
\end{array} \right]_{\rm{d}}} \nonumber \\
 &= \frac{{ - 4i{\kappa ^4}}}{{{M^6}}}{\rm{Tr}}\left\{ \begin{array}{l}
 - 2\frac{1}{{ - {D^2} - {m^2}}}\left[ \begin{array}{l}
{t^a}{H^\dag }{t^a}H + {t^a}H{H^\dag }{t^a}\\
 - \frac{1}{{{M^2}}}{t^a}{\left( {{D^2}{H^\dag }{t^a}H} \right)_x}
\end{array} \right]\frac{1}{{ - {D^2} - {m^2}}}{t^b}H\frac{{{D^4}}}{{ - {D^2} - {M^2}}}{H^\dag }{t^b} + c.c.\\
 + 2\frac{1}{{ - {D^2} - {m^2}}}{t^a}H\frac{{ - {D^2}}}{{ - {D^2} - {M^2}}}{H^\dag }{t^a}\frac{1}{{ - {D^2} - {m^2}}}{t^b}H\frac{{{D^4}}}{{ - {D^2} - {M^2}}}{H^\dag }{t^b} + c.c.\\
 - \frac{1}{{{M^2}}}\frac{1}{{ - {D^2} - {m^2}}}{t^a}H\frac{{{D^4}}}{{ - {D^2} - {M^2}}}{H^\dag }{t^a}\frac{1}{{ - {D^2} - {m^2}}}{t^b}H\frac{{{D^4}}}{{ - {D^2} - {M^2}}}{H^\dag }{t^b} + c.c.\\
 - 2\left[ {\frac{1}{{ - {D^2} - {m^2}}}{t^a}H{H^T}{t^{a*}}\frac{1}{{ - {D^2} - {m^2}}}{t^{b*}}{H^*}\frac{{{D^4}}}{{ - {D^2} - {M^2}}}{H^\dag }{t^b} + c.c.} \right]\\
 + 2\left[ {\frac{1}{{ - {D^2} - {m^2}}}{t^a}H\frac{{ - {D^2}}}{{ - {D^2} - {M^2}}}{H^T}{t^{a*}}\frac{1}{{ - {D^2} - {m^2}}}{t^{b*}}{H^*}\frac{{{D^4}}}{{ - {D^2} - {M^2}}}{H^\dag }{t^b} + c.c.} \right]\\
 - \frac{1}{{{M^2}}}\frac{1}{{ - {D^2} - {m^2}}}{t^a}H\frac{{{D^4}}}{{ - {D^2} - {M^2}}}{H^T}{t^{a*}}\frac{1}{{ - {D^2} - {m^2}}}{t^{b*}}{H^*}\frac{{{D^4}}}{{ - {D^2} - {M^2}}}{H^\dag }{t^b} + c.c.
\end{array} \right\} \nonumber \\
 &= \frac{{ - 4i{\kappa ^4}}}{{{M^6}}}\left\{ \begin{array}{l}
 - 2{T_2}\left( {{t^a}{H^\dag }{t^a}H + {t^a}H{H^\dag }{t^a} - \frac{1}{{{M^2}}}{t^a}{D^2}\left( {{H^\dag }{t^a}H} \right),{t^b}H,{H^\dag }{t^b}} \right) + c.c.\\
 + 2{T_5}\left( {{t^a}H,{H^\dag }{t^a},{t^b}H,{H^\dag }{t^b}} \right) - {T_6}\left( {{t^a}H,{H^\dag }{t^a},{t^b}H,{H^\dag }{t^b}} \right) + c.c.\\
 - 2\left[ {{T_2}\left( {{t^a}H{H^T}{t^{a*}},{t^{b*}}{H^*},{H^\dag }{t^b}} \right) + c.c.} \right]\\
 + 2\left[ {{T_5}\left( {{t^a}H,{H^T}{t^{a*}},{t^{b*}}{H^*},{H^\dag }{t^b}} \right) + c.c.} \right]\\
 - {T_6}\left( {{t^a}H,{H^T}{t^{a*}},{t^{b*}}{H^*},{H^\dag }{t^b}} \right) + c.c.
\end{array} \right\} \nonumber \\
 &= \frac{{{\kappa ^4}}}{{{M^6}}}\frac{1}{{{{\left( {4\pi } \right)}^2}}}\int {{d^4}x\left[ { - 2{{\left( {{D_\mu }{{\left| H \right|}^2}} \right)}^2} - {{\left| {{H^\dag }{D_\mu }H} \right|}^2} - \frac{{21}}{2}{{\left| H \right|}^2}{{\left| {{D_\mu }H} \right|}^2}} \right]} . \label{eqn:S2app}
\end{align}

\begin{align}
{S_3} &\equiv \frac{i}{2}{\rm{Tr}}{\left[ { - \frac{{8{\kappa ^2}\eta }}{{ - {D^2} - {m^2}}}\left( {\begin{array}{*{20}{c}}
{{a_3}}&{{b_3}}\\
{b_3^*}&{a_3^T}
\end{array}} \right)} \right]_{\rm{d}}} \nonumber \\
 &=  - 8i{\kappa ^2}\eta {\rm{Tr}}{\left[ {\frac{1}{{ - {D^2} - {m^2}}}{a_3}} \right]_{\rm{r}}} \nonumber \\
 &=  - 8i{\kappa ^2}\eta {\rm{Tr}}{\left\{ \begin{array}{l}
\frac{1}{{ - {D^2} - {m^2}}}{t^a}{\left( {\frac{1}{{ - {D^2} - {M^2}}}{{\left| H \right|}^2}\frac{1}{{ - {D^2} - {M^2}}}{H^\dag }{t^a}H} \right)_x}\\
 + \frac{1}{2}\frac{1}{{ - {D^2} - {m^2}}}{\left( {\frac{1}{{ - {D^2} - {M^2}}}{H^\dag }{t^a}H} \right)_x}{\left( {\frac{1}{{ - {D^2} - {M^2}}}{H^\dag }{t^a}H} \right)_x}\\
 + \left[ {\frac{1}{{ - {D^2} - {m^2}}}{t^a}H\frac{1}{{ - {D^2} - {M^2}}}{H^\dag }{{\left( {\frac{1}{{ - {D^2} - {M^2}}}{H^\dag }{t^a}H} \right)}_x} + c.c.} \right]\\
 + \frac{1}{{ - {D^2} - {m^2}}}{t^a}H\frac{1}{{ - {D^2} - {M^2}}}{\left| H \right|^2}\frac{1}{{ - {D^2} - {M^2}}}{H^\dag }{t^a}
\end{array} \right\}_{\rm{d}}} \nonumber \\
 &= \frac{{ - 8i{\kappa ^2}\eta }}{{{M^4}}}{\rm{Tr}}\left\{ \begin{array}{l}
\frac{1}{{ - {D^2} - {m^2}}}{t^a}H\frac{{{D^4}}}{{ - {D^2} - {M^2}}}{H^\dag }\left[ { - \frac{1}{{{M^2}}}{H^\dag }{t^a}H + \frac{1}{{{M^4}}}{{\left( {{D^2}{H^\dag }{t^a}H} \right)}_x}} \right] + c.c.\\
 + \frac{1}{{ - {D^2} - {m^2}}}{t^a}H\frac{1}{{ - {D^2} - {M^2}}}{\left| H \right|^2}\frac{{{D^4}}}{{ - {D^2} - {M^2}}}{H^\dag }{t^a} + c.c.\\
 - \frac{1}{{{M^4}}}\frac{1}{{ - {D^2} - {m^2}}}{t^a}H\frac{{{D^4}}}{{ - {D^2} - {M^2}}}{\left| H \right|^2}\frac{{{D^4}}}{{ - {D^2} - {M^2}}}{H^\dag }{t^a}
\end{array} \right\} \nonumber \\
 &= \frac{{ - 8i{\kappa ^2}\eta }}{{{M^4}}}\left\{ \begin{array}{l}
\left[ {{T_1}\left( {{t^a}H,{H^\dag }\left[ { - {H^\dag }{t^a}H + \frac{1}{{{M^2}}}{D^2}\left( {{H^\dag }{t^a}H} \right)} \right]} \right) + c.c.} \right]\\
 + \left[ {{T_3}\left( {{t^a}H,{{\left| H \right|}^2},{H^\dag }{t^a}} \right) + c.c.} \right] - {T_4}\left( {{t^a}H,{{\left| H \right|}^2},{H^\dag }{t^a}} \right)
\end{array} \right\} \nonumber \\
 &= \frac{{{\kappa ^2}\eta }}{{{M^4}}}\frac{1}{{{{\left( {4\pi } \right)}^2}}}\int {{d^4}x\left[ { - 7{{\left( {{D_\mu }{{\left| H \right|}^2}} \right)}^2} + 16{{\left| {{H^\dag }{D_\mu }H} \right|}^2} - 21{{\left| H \right|}^2}{{\left| {{D_\mu }H} \right|}^2}} \right]} . \label{eqn:S3app}
\end{align}

\begin{align}
{S_K} &\equiv  - 2i{\kappa ^2}{\rm{Tr}}{\left[ {\frac{1}{{ - {D^2} - {m^2}}}\left( {\begin{array}{*{20}{c}}
{{a_2}}&{{b_2}}\\
{b_2^*}&{a_2^T}
\end{array}} \right)} \right]_{\rm{d}}} \nonumber \\
 &=  - 4i{\kappa ^2}{\rm{Tr}}{\left[ {\frac{1}{{ - {D^2} - {m^2}}}{a_2}} \right]_{\rm{d}}} \nonumber \\
 &=  - 4i{\kappa ^2}{\rm{Tr}}{\left[ {\frac{1}{{ - {D^2} - {m^2}}}{t^a}{{\left( {\frac{1}{{ - {D^2} - {M^2}}}{H^\dag }{t^a}H} \right)}_x} + \frac{1}{{ - {D^2} - {m^2}}}{t^a}H\frac{1}{{ - {D^2} - {M^2}}}{H^\dag }{t^a}} \right]_{\rm{d}}} \nonumber \\
 &=  - \frac{{4i{\kappa ^2}}}{{{M^4}}}{\rm{Tr}}\left[ {\frac{1}{{ - {D^2} - {m^2}}}{t^a}H\frac{{{D^4}}}{{ - {D^2} - {M^2}}}{H^\dag }{t^a}} \right] \nonumber \\
 &=  - \frac{{4i{\kappa ^2}}}{{{M^2}}}{T_1}\left( {{t^a}H,{H^\dag }{t^a}} \right) \nonumber \\
 &= \frac{{{\kappa ^2}}}{{{M^2}}}\frac{1}{{{{\left( {4\pi } \right)}^2}}}\int {{d^4}x\left[ {\frac{3}{2}{{\left| {DH} \right|}^2}} \right]} . \label{eqn:SKapp}
\end{align}